\documentclass[fleqn,usenatbib]{mnras}
\usepackage[T1]{fontenc}
\usepackage{ae,aecompl}
\usepackage{array}
\usepackage{todonotes}
\usepackage{float}

\usepackage{graphicx}	% Including figure files
\usepackage{amsmath}	% Advanced maths commands
\usepackage{amssymb}	% Extra maths symbols

\title{Nuclear Burning in Collapsar Accretion Disks}

\author[Y.~Zenati, D.~M.~Siegel, B.~D.~Metzger \& H.~B.~Perets]{
Yossef Zenati$^{1}$, 
Daniel M.~Siegel$^{2,3}$,
Brian D.~Metzger$^{4}$,
and
Hagai B.~Perets$^{1}$
\\
$^{1}$Physics Department, Technion - Israel Institute of Technology,
Haifa 3200004, Israel\\
$^{2}$Perimeter Institute for Theoretical Physics, Waterloo, Ontario, Canada, N2L 2Y5\\
$^3$Department of Physics, University of Guelph, Guelph, Ontario, Canada, N1G 2W1\\
$^4$Department of Physics, Columbia University, New York, USA
}

\date{Accepted XXX. Received YYY; in original form ZZZ}
 
\pubyear{2020}

\begin{document}
\label{firstpage}
\pagerange{\pageref{firstpage}--\pageref{lastpage}}
\maketitle

% Abstract of the paper
\begin{abstract}
The core collapse of massive, rapidly-rotating stars are thought to be the progenitors of long-duration gamma-ray bursts (GRB) and their associated hyper-energetic supernovae (SNe).  At early times after the collapse, relatively low angular momentum material from the infalling stellar envelope will circularize into an accretion disk located just outside the black hole horizon, resulting in high accretion rates necessary to power a GRB jet.  Temperatures in the disk midplane at these small radii are sufficiently high to dissociate nuclei, while outflows from the disk can be neutron-rich and may synthesize $r$-process nuclei.  However, at later times, and for high progenitor angular momentum, the outer layers of the stellar envelope can circularize at larger radii $\gtrsim 10^{7}$ cm, where nuclear reactions can take place in the disk midplane (e.g.~$^{4}$He + $^{16}$O $\rightarrow$ $^{20}$Ne + $\gamma$).  Here we explore the effects of nuclear burning on collapsar accretion disks and their outflows by means of hydrodynamical $\alpha$-viscosity torus simulations coupled to a 19-isotope nuclear reaction network, which are designed to mimic the late infall epochs in collapsar evolution when the viscous time of the torus has become comparable to the envelope fall-back time.  Our results address several key questions, such as the conditions for quiescent burning and accretion versus detonation and the generation of $^{56}$Ni in disk outflows, which we show could contribute significantly to powering GRB supernovae. Being located in the slowest, innermost layers of the ejecta, the latter could provide the radioactive heating source necessary to make the spectral signatures of $r$-process elements visible in late-time GRB-SNe spectra.      

\end{abstract}

%\newpage

\section{Introduction}

The core collapse of rapidly-rotating stars, which are stripped of their outer hydrogen (and, potentially, also helium) envelopes, are considered to be the main progenitor channel for hyper-energetic supernovae (broad-lined Type I supernovae; SNe Ic-BL) and long-duration gamma-ray bursts (GRB) (e.g.~\citealt{Woosley&Bloom06,Nagataki+18}).  The exact mechanism by which such stars are endowed with rapid rotation at the time of their deaths remains a matter of debate.  Such high angular final momentum is unlikely to characterize single-star evolution (e.g.~\citealt{Ma&Fuller19}), pointing to some form of binary interaction (e.g.~\citealt{Cantiello+07}) or highly efficient mixing process during the main sequence leading to chemically-homogeneous evolution (e.g.~\citealt{Maeder87}).  The rarity of long GRBs and SNe Ic-BL, and the typically low metallicities of their host galaxies, provide potentially important clues (e.g.~\citealt{Modjaz+19}) for their origins. 

%\newpage

Most work on the core collapse of rapidly-rotating stars has focused on the collapse phase and the supernova explosion (e.g.~\citealt{Sekiguchi&Shibata11,Mosta+14,Takiwaki+16,Gilkis+19}), or on the early accretion phase onto the central compact object (e.g.~\citealt{Dessart+08,MacFadyen&Woosley99,Lindner+10}).  

The central engine powering the ultra-relativistic GRB jet could be either a millisecond-period magnetar (e.g.~\citealt{Metzger+11}) and/or a hyper-accreting black hole (e.g.~\citealt{MacFadyen&Woosley99}).  In the latter case, the accretion rate can be sufficiently high $\gtrsim 10^{-3}-10^{-1}M_{\odot}$ s$^{-1}$ for the disk midplane to cool efficiently via neutrino emission (e.g.~\citealt{Narayan+01,Chen&Beloborodov07}) and thus to become neutron-rich through weak interactions on radial scales $\lesssim$ tens of gravitational radii (e.g.~\citealt{Beloborodov03}).  Such hyper-accreting disks are also known to generate powerful outflows (e.g.~\citealt{Kohri+05}), which under these conditions of high temperatures and densities will result in the formation of heavy nuclei.  Recent general-relativistic magneto-hydrodynamical (GRMHD) simulations of the inner regions of GRB accretion disks indicate that the outflows can retain a low electron fraction $Y_e \ll 0.5$ (\citealt{Siegel+19,Miller+19}), thus synthesizing heavy $r$-process nuclei with mass number $A \gtrsim 130$.  Collapsars may therefore rival neutron star mergers  as sources of $r$-process production in the Universe, particularly at low metallicity \cite{Siegel+19}.  

The SNe Ic-BL observed to accompany GRBs not only possess high kinetic energies $\gtrsim 10^{52}$ erg, they are also more optically-luminous than ordinary Type Ibc supernova.  In particular, the mass of radioactive $^{56}$Ni required to power Ic-BL light curves is typically $\approx 0.3-0.5M_{\odot}$ (e.g.~\citealt{Drout+11,Cano16}), compared to $\lesssim 0.03-0.1M_{\odot}$ in ordinary Type Ib/c SNe (e.g.~\citealt{Anderson19}).\footnote{In fact, \citet{Ertl+19} find that the $^{56}$Ni yields generated from the explosion of stripped-envelope progenitor models are insufficient to explain the peak luminosities of $\sim$ half of ordinary Type Ib/c, instead arguing for possible enhancement of the supernova luminosity by a magnetar engine (\citealt{Maeda+07,Kasen&Bildsten10,Woosley10}).}  As in the majority of ordinary core collapse supernovae, $^{56}$Ni may be formed via shock-heating of the inner layers of the infalling progenitor star core.  However, explaining $\approx 0.3-0.5M_{\odot}$ of $^{56}$Ni requires not only a highly-energetic explosion ($\gtrsim 10^{52}$ erg, but also one which this energy is released promptly following collapse $\lesssim 1$ s (e.g.~in comparison to the duration of the GRB jet $\gtrsim 10$ s), in order that enough shocked core material achieve high enough temperatures to reach nuclear statistical equilibrium (e.g.~\citealt{Barnes+18}).  Given the need for such a prompt, energetic explosion, it is not clear whether enough fall-back would occur to form a black hole at all, possibly instead favoring a magnetar engine \citep{Metzger+11}.    

On the other hand, it is possible that the $^{56}$Ni responsible for lighting up GRB-SNe is not generated by shock heating the infalling star. \citet{MacFadyen&Woosley99} suggested $^{56}$Ni is created in outflows from the black hole accretion disk.  However, $^{56}$Ni synthesis requires material with a high electron fraction $Y_e \gtrsim 0.5$, which recent GRMHD simulations disfavor, instead finding $Y_e < 0.5$ and $r$-process production in outflows from the inner regions of collapsar disks \citep{Siegel+19,Miller+19}.  Higher $Y_e \gtrsim 0.5$ outflows capable of generating $^{56}$Ni may arise at earlier times (higher accretion rates), when the hot proto-neutron star is still present (e.g.~\citealt{Dessart+08}) and neutrino irradiation can play a larger role in raising the $Y_e$ of the disk winds (e.g.~\citealt{Surman+11}).  

This paper explores a different mechanism for nucleosynthesis in collapsar disks, one that occurs at later times in their evolution than is usually simulated.  Although the rotation profiles of massive stars at the time of core collapse is notoriously uncertain (e.g.~\citealt{Heger+00}), the specific angular momentum $j(r)$ is generally an increasing function of stellar radius $r$.  As a result, infalling envelope circularizes at increasingly large distances around the black hole with time after the explosion (e.g.~\citealt{Kumar+08}).  At early times, matter circularizes at small radii, where the high virial temperatures $\gtrsim 10^{9}-10^{10}$ K result in nuclear dissociation and neutronization.  During these early epochs, the timescale for local angular momentum transport (``viscous'' timescale, $t_{\rm visc}$) is generally much shorter than the mass-infall time ($\sim$ envelope free-fall time $t_{\rm ff}$), in which case matter is funneled through the disk onto the black hole at roughly the same rate it is fed by fall-back accretion. Though not the focus of this study, fallback accretion disk could also occur under other circumstances such as the $\mu$TDEs - the tidal disruption/collision of stars and planets by/with stellar black holes \citep{Per+16}.  Our results could therefore be also relevant to these type of transients. 

At later times, matter will tend to circularize further and further from the black hole, where the viscous timescale increases and may become comparable to the free-fall time of the outer layers of the progenitor star.  In the limit $t_{\rm visc} \gg t_{\rm ff}$, the black hole feeding rate is controlled by the viscous time instead of the free-fall time, and one can (to first approximation) evolve the disk starting from a torus of fixed mass and angular momentum.  The model of such a ``viscously-evolving isolated torus'' for the late-time evolution of collapsar disk was proposed to explain the unexpected behavior of some GRB X-ray afterglows \citep{Metzger+08,Kumar+08,Cannizzo+11}.  

The midplane temperature of the accretion flow is also lower $\lesssim 10^{8}$ K at these larger radii, such that material in the disk will initially retain the nuclear composition of the progenitor star envelope from which it was built.  However, as this unprocessed matter accretes onto the central black hole, its temperature will become high enough at small radii to ignite nuclear burning in the midplane \citep{Metzger12}.  The effects of energy generation from nuclear burning during the collapse of rotating stars was first studied by \citet{Bodenheimer&Woosley83}, who found with axisymmetric hydrodynamical simulations that ``hang-up'' due to the centrifugal barrier and oxygen burning could lead to a weak explosion (even absent a central explosion).  \citet{Kushnir&Katz15} argued that nuclear burning during the collapse phase of a rapidly rotating star could power a supernova explosion through a similar mechanism.\footnote{\citet{Kushnir&Katz15} proposed thermonuclear explosion of rotating massive stars as a generic explanation for all core collapse supernovae.  However, the inference based on the discoveries by LIGO/Virgo that most black holes are rotating slowly make such rapidly spinning progenitor models less tenable (e.g.~\citealt{Ma&Fuller19}).}  

This paper explores the effects of nuclear burning on the late-time accretion flows generated by the core collapse of rapidly-rotating stripped-envelope stars.  However, rather than simulating the entire collapse phase (which is not computationally feasible at the required resolution needed to follow the nuclear burning) we instead focus on a simplified model in which we treat the accretion flow starting from an equilibrium torus.  This is a reasonable description of the system properties at sufficiently late times that $t_{\rm visc} \gtrsim t_{\rm ff}$.  The composition, mass, and energy of the initial torus are motivated based on stellar progenitor models \citep{Heger+00}.  Given its larger associated uncertainty, we allow ourselves sizable freedom in specifying the progenitor angular momentum profile, exploring the sensitivity of our results to its overall normalization.  The goal of our analysis is to understand the conditions for quiescent accretion, versus explosive burning (detonation), and to assess the role of large-angular momentum disk outflows as a source of comparatively late-time nucleosynthesis in collapsars.   
This paper is organized as follows.  

%\newpage

\section{Methods: Collapsar Disks with Nuclear Burning}

\subsection{Description of Numerical Code}
\label{sec:code}

We simulate the evolution of the black hole (BH) accretion torus using the publicly available FLASH code \citep{2000ApJS..131..273F}.  We employ the unsplit ${\rm PPM}$ solver of FLASH in ${\rm 2D}$ axisymmetric cylindrical coordinates $[{\bar\rho},z]$ on a grid of fiducial size $(10^{10}$ cm$)\times (9 \times 10^{10}$ cm) using adaptive mesh
refinement.  We solve the equations of mass, momentum, energy, and chemical species conservation,
\begin{eqnarray}
\frac{\partial\rho}{\partial t}+\nabla\cdot\left(\rho\mathbf{v}\right)&=&0,  \\
\frac{d\mathbf v}{dt}&=&\mathbf{f}_{c}-\frac{1}{\rho}\nabla p + \nabla\phi,  \\
\rho \frac{dl_z}{dt} &=& \bar\rho (\nabla\cdot\mathbf{T})_\phi \\
 \rho\frac{d e_{\rm int}}{d t}+p\nabla\cdot\mathbf v &=& \frac{1}{\rho \nu}\mathbf{T}:\mathbf{T}+\rho(\dot{Q}_{\rm nuc} - \dot{Q}_\nu), \\
 \frac{\partial \mathbf{X}}{\partial t} &=& \dot{\mathbf{X}}, \label{eq:chemical_species_evolution} \\
 \nabla^{2}\phi&=& 4 \pi G \rho + \nabla^{2}\phi_{\rm BH}, \\
\mathbf{f}_{c}&=&\frac{l_{z}^{2}}{\bar\rho^{3}}\hat{{\bar\rho}}.
\end{eqnarray}
We have included sources terms for gravity, shear viscosity, nuclear reactions, and neutrino cooling as described in detail below. Here, $\mathbf{f}_{c}$ is an implicit centrifugal source term, where $l_z$ is the z-component of the specific angular momentum. Variables have their standard meaning: $\rho$, $\mathbf{v}$, $p$, $e_{\rm int}$, $\nu$, $\mathbf{T}$, $\phi$, and $\mathbf{X}=\{X_i\}$ denote, respectively, fluid density, poloidal velocity, total pressure, specific internal energy, fluid viscosity, viscous stress tensor for azimuthal shear, gravitational potential, and mass fractions of the isotopes $X_i$, with $\sum_i X_i = 1$. Furthermore, $d/dt = \partial/\partial t + \mathbf{v}\cdot\nabla$, $\phi_{\rm BH}$ denotes the black hole potential, and $
\dot{Q}_{\rm nuc}$ and $\dot{Q}_{\rm \nu}$ represent the specific nuclear heating rate due to nuclear reactions and the specific cooling rate due to neutrino emission.

We employ the Helmholtz equation of state (EOS) in FLASH (\citealt{2000ApJS..126..501T}). The Helmholtz EOS includes contributions from partially degenerate electrons and positrons, radiation, and non-degenerate ions.  It employs a look-up table scheme for high performance. The most important aspect of the Helmholtz EOS is its ability to handle thermodynamic states where radiation dominates, and under conditions of very high pressure.  The contributions of both nuclear reaction (see below) and neutrino cooling \citep{1989ApJ...346..847C,1991ApJ...376..234H} are included in the internal energy evolution, though the latter does not play an appreciable role for the range of disk radii that we simulate.  

Self-gravity is included as a multipole expansion of up to multipole $l_{\rm max}=12-18$ using the new FLASH multipole solver, to which we add a point-mass gravitational potential, $\phi_{\rm BH}$, to account for gravity of the BH. The potential $\phi_{\rm BH}$ is modeled as a non-spinning pseudo-Newtonian point-mass. The mass is set to the sum of the layers of the progenitor star interior to those which form our initial torus, as further described in Sec.~\ref{sec:initial_models}. The BH spin can be safely ignored, as our disk material circularizes at large radii from the BH where deviations from a spinning BH potential are minute.

Since our calculations are axisymmetric and do not include magnetic fields, we cannot self-consistently account for angular momentum transport due to the magneto-rotational instability or non-axisymmetric instabilties (e.g. associated with self-gravity).  Instead, we make the common approximation of modeling the viscosity using the $\alpha$-viscosity parameterization of \citet{Shakura&Sunyaev73}, for which the kinematic viscosity is taken to be
\begin{equation} 
    \nu_{\alpha}=\alpha c_{s}^{2}/\Omega_{\rm K},  \label{eq:nualpha}
\end{equation}
where $\Omega_{\rm K} = (GM_{\rm enc}/r^{3})^{1/2}$ is the Keplerian frequency given the enclosed mass $M_{\rm enc}$ and $c_{s}$ is the sound speed.  In our fiducial models we take $\alpha = 0.1$ for the dimensionless viscosity coefficient, but we explore the dependence of our simulation results on different values of $\alpha$ (see Sec.~\ref{sec:viscosity}).

Nuclear burning is included in the simulations following a similar approach to that employed by us and other authors in the field of thermonuclear SNe (e.g. \citealt{Mea+09,Zen+19a,Zen+19b}). The nuclear network used is the FLASH $\alpha$-chain network of 19 isotopes, which provides the source terms $\dot{\mathbf{X}}$ in Eq.~\eqref{eq:chemical_species_evolution} and adequately captures the energy generated
during nuclear burning $\dot{Q}_{\rm nuc}$ \citep{2000ApJS..126..501T}. 

Unphysical early detonations are avoided by switching on the source terms due to nuclear burning and viscosity only after a brief initial relaxation phase. Furthermore, in order to prevent the generation of unphysical early detonations arising from insufficient numerical resolution, we apply a limiter to the burning following \citet{2013ApJ...778L..37K}, restricting energy injection into a grid cell to $\dot{Q}_{\rm nuc} < 0.1 e_{\rm int} t_{\rm s}^{-1}$, where $t_{\rm s}$ is the sound crossing time of the cell. Multiple simulations were run with increasing resolution until convergence was achieved in the properties of nuclear burning.  A resolution of $6-10$ km was found to be sufficient to achieve $< 5\%$ convergence in energy (cf.~Sec.~\ref{sec:viscosity}, Fig.~\ref{fig:reso}). Detonations are handled by the reactive hydrodynamics solver in FLASH without the need for a front tracker, which is possible since unresolved Chapman--Jouguet
(CJ) detonations retain the correct jump conditions and propagation
speeds. Numerical stability is maintained by preventing nuclear burning
within the shock. This is necessary because shocks are artificially
spread out over a few zones by the ${\rm PPM}$ hydrodynamics solver,
which can lead to nonphysical burning within shocks that can destabilise
the burning front \citep{1989BAAS...21.1209F}.

\subsection{Initial Torus Models}
\label{sec:initial_models}

\begin{figure}[h]
\centering
\includegraphics[width=0.99\linewidth]{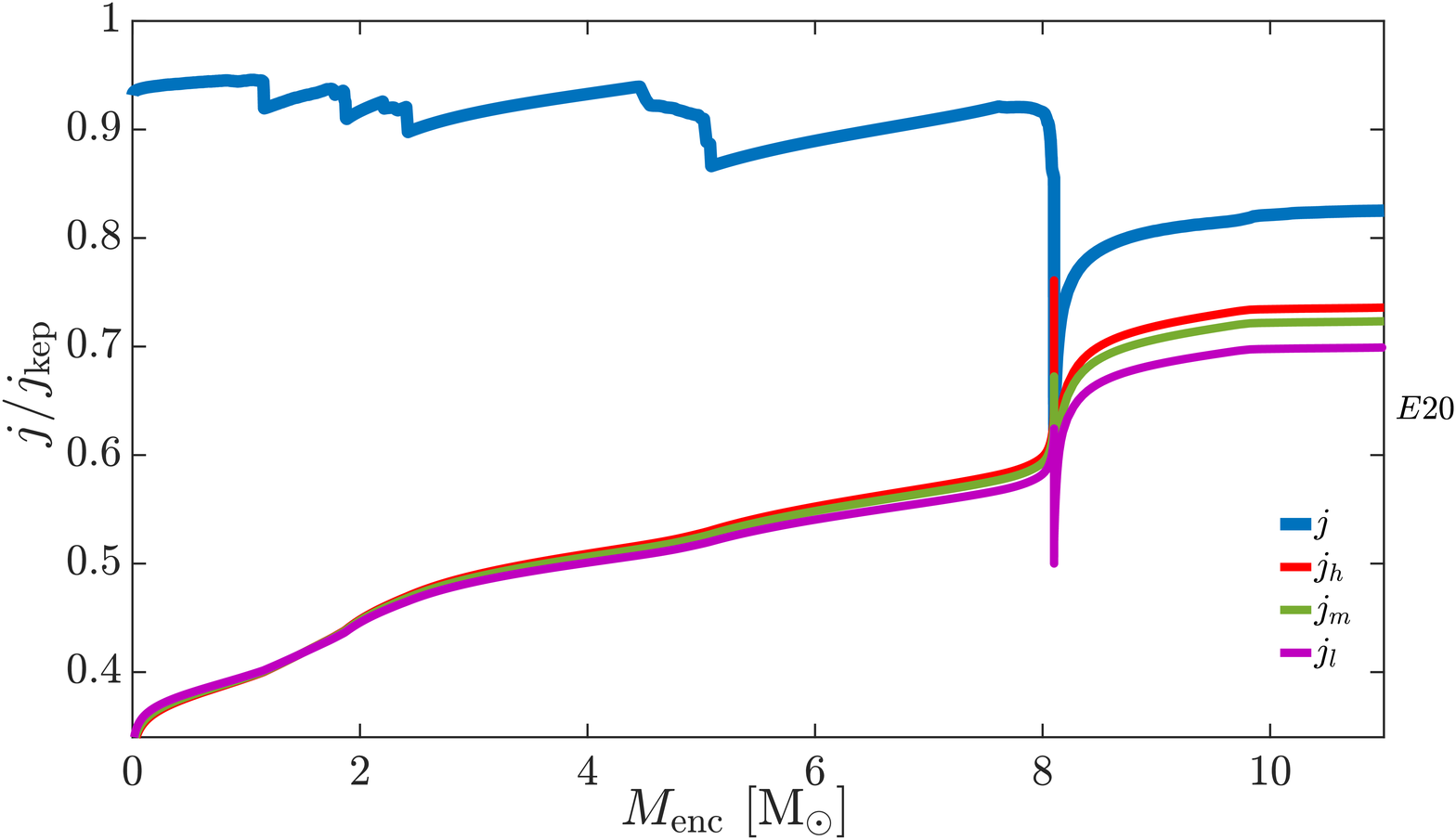}
\includegraphics[width=0.99\linewidth]{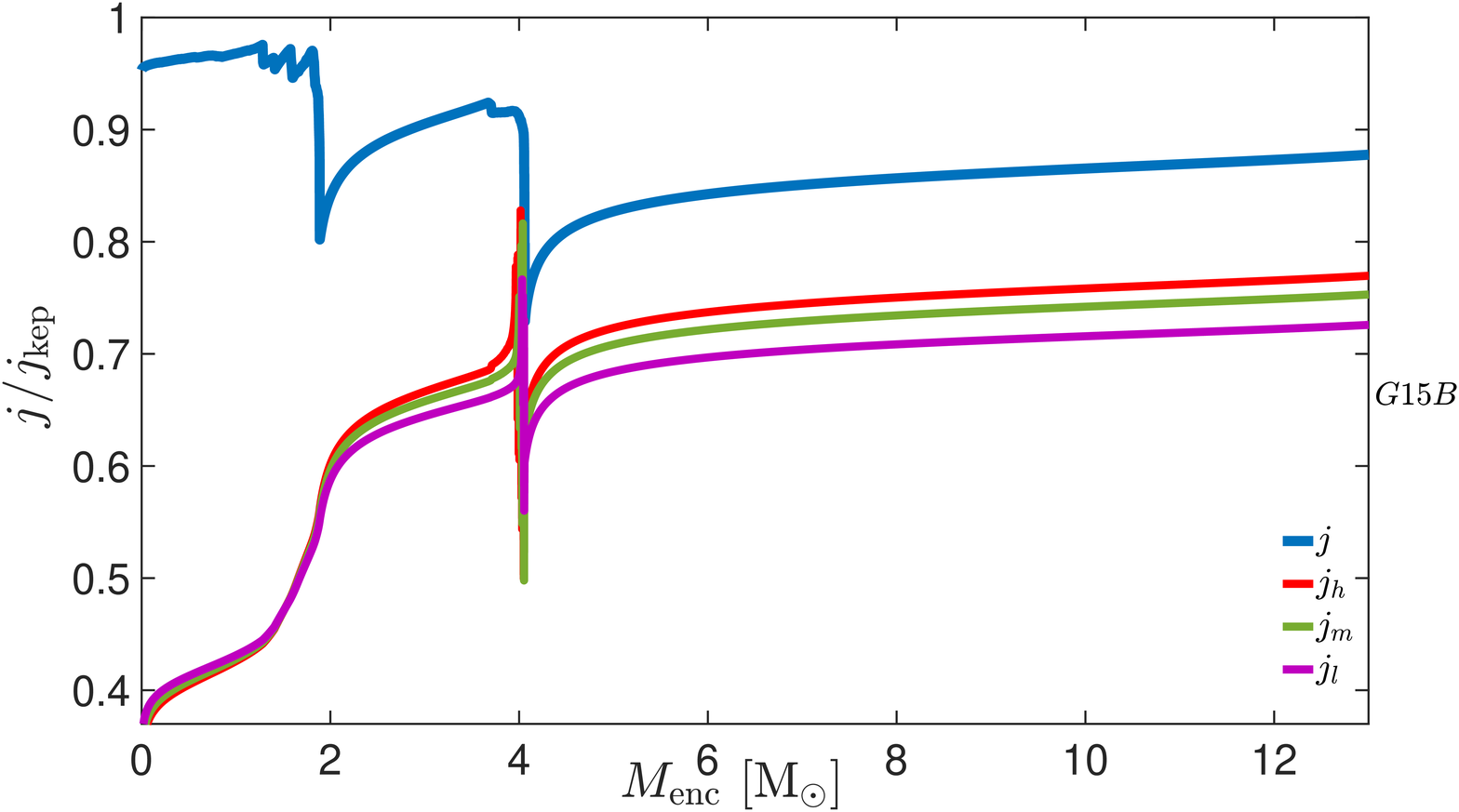}
\includegraphics[width=0.99\linewidth]{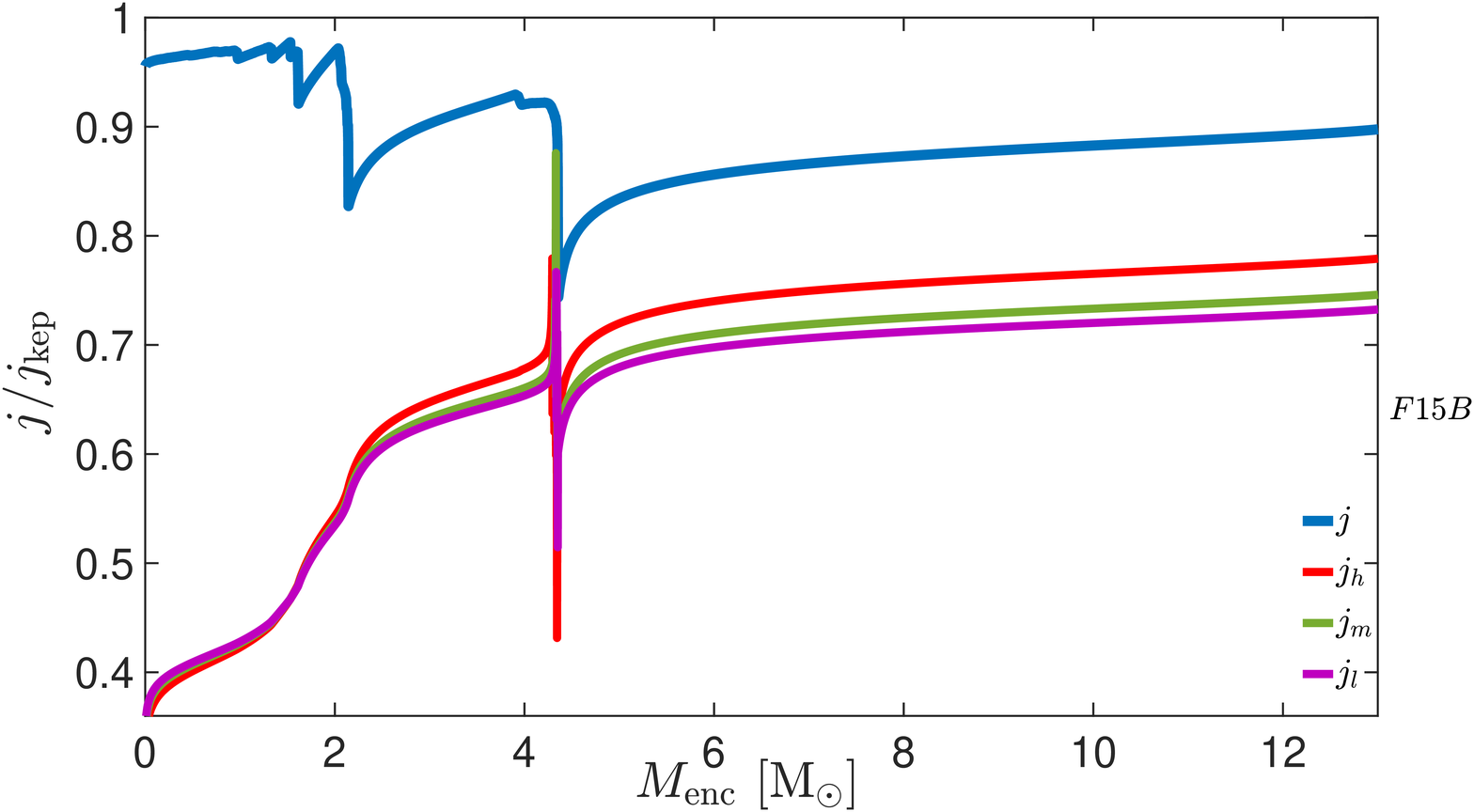}
\caption{Specific angular momentum profiles of the E20, G15B, and F15B progenitor models of \citet{Heger+00}, normalized by the Keplerian angular momentum $j_{kep}=(GM_{\rm enc}r)^{1/2}$ (blue solid line). Shown are also the modified profiles $j_{\rm l}$, $j_{\rm m}$, $j_{\rm h}$ with small, medium, and large surface angular momentum, respectively (see Eq.~\eqref{eq:j_profile}, Fig.~\ref{fig:visc_fftimescale}, and the text for details).}
\label{fig:j_jkep_Menc}
\end{figure}

Initial data for our collapsar accretion disks are constructed by modeling the collapse of presupernova stellar models similar to previous work \citep{Kumar+08}. We consider a range of progenitor models: E20, G15B, and F15B from \citet{Heger+00}; Table \ref{tab:modeltab} provides the mass, stellar radius and surface angular velocity for each model. The radial angular velocity profiles provided in these models assume rigid rotation on spherical shells, $\Omega(r,\theta)=\Omega(r)$, i.e.,
\begin{equation}
    j(r,\theta) = j(r) \sin(\theta),    
\end{equation}
where $\Omega$ denotes angular velocity, and $j$ denotes the specific angular momentum. Figure \ref{fig:j_jkep_Menc} shows $j$ as a function of enclosed mass $M_{\rm enc}$ for each progenitor model.

\begin{table}
\begin{centering}
\begin{tabular}{ccccccc}
\hline
\#  &  {$M_{*}[M_{\odot}]$} &  {$R_{*}$[cm]} & {$\Omega/ \Omega_{\rm BU}$}\tabularnewline %& {$\rm SBH[M_{\odot}]$}\tabularnewline
\hline 
E20 & $11.02$ & $1.65\times 10^{14} $ & $0.83$ \tabularnewline% & $7.3$ \tabularnewline
%\hline 
G15B & $13.47$ & $3.85\times 10^{13}$ & $0.88$ \tabularnewline%& $5.2$ \tabularnewline
%\hline 
F15B & $12.90$ & $4.32\times 10^{13}$ & $0.89$ \tabularnewline%& $3.2$ \tabularnewline
\hline 
\end{tabular}
\par\end{centering}
\caption{Initial masses $M_\star$, stellar radii $R_\star$, surface angular velocity in units of the break-up limit $\Omega_{\rm BU}=(GM_{\star}/r^{3})^{1/2}$ of progenitor models from \citet{Heger+00} considered here.}
\label{tab:modeltab}
\end{table}

For each layer of a given progenitor star, one may define the free-fall timescale for matter to reach the central compact object or accretion disk,
\begin{equation}
	t_{\rm ff}(r) = \left(\frac{r^{3}}{GM_{\rm enc}}\right)^{1/2}. \label{eq:tsff}
\end{equation}
Each layer will circularize outside the black hole at a cylindrical radius, which is approximately given by angular momentum conservation, 
\begin{equation}
    r_{\rm circ}(r) = \frac{j(r)^{2}}{2GM_{\rm enc}}. \label{eq:r_circ}
\end{equation}
Once matter is in the disk, accretion will proceed on the local viscous timescale evaluated at $r_{\rm circ}$,
\begin{eqnarray}
	t_{\rm visc}(r) \mskip-10mu&=& \mskip-10mu  \frac{r^2_{\rm circ}(r)}{\nu_{\alpha}} \approx \frac{1}{\alpha}\left(\frac{r_{\rm circ}^{3}}{GM_{\rm enc}}\right)^{1/2}\left(\frac{H_{0}}{r_{\rm circ}}\right)^{-2} \label{eq:tvisc} \\
	\approx&&\mskip-50mu  20\,{\rm s} \left(\frac{\alpha}{0.1}\right)^{-1}\mskip-5mu\left(\frac{r_{\rm circ}}{10^8{\rm cm}}\right)^{\frac{3}{2}}\mskip-5mu \left(\frac{M_{\rm enc}}{5M_\odot}\right)^{-\frac{1}{2}}\mskip-5mu \left(\frac{H_0/r_{\rm circ}}{0.15}\right)^{-2}, \nonumber
%&\sim& 3400{\rm\,s} \left(\frac{0.1}{\alpha}\right)\left(\frac{R_{eq}}{10^{9.4}\rm cm}\right)^{3/2}
%\left(\frac{8M_{\sun}}{M_{\rm c}}\right)^{1/2}\left(\frac{H_{0}}{0.5R_{eq}}\right)^{-2}\nonumber\\
\end{eqnarray}
where the kinematic viscosity $\nu_{\alpha}$, defined in Eq.~\eqref{eq:nualpha}, is evaluated at the corresponding circularization radius $r_{\rm circ}$. In what follows, we take $\alpha = 0.1$ as a fiducial value. Furthermore, we have evaluated the above expression for characteristic initial parameters of the resulting tori we consider (see below, Tab.~\ref{tab:models}), such as a typical radius of the disk $\!\sim 10^{8}$\,cm, black-hole mass of $\approx\!5\,M_\odot$, and initial scale-height of the torus of $H_0 \equiv C_{\rm s}/\Omega_{\rm K} \approx 0.15 r_{\rm circ}$.

\begin{figure}[h]
\centering

\includegraphics[width=0.99\linewidth]{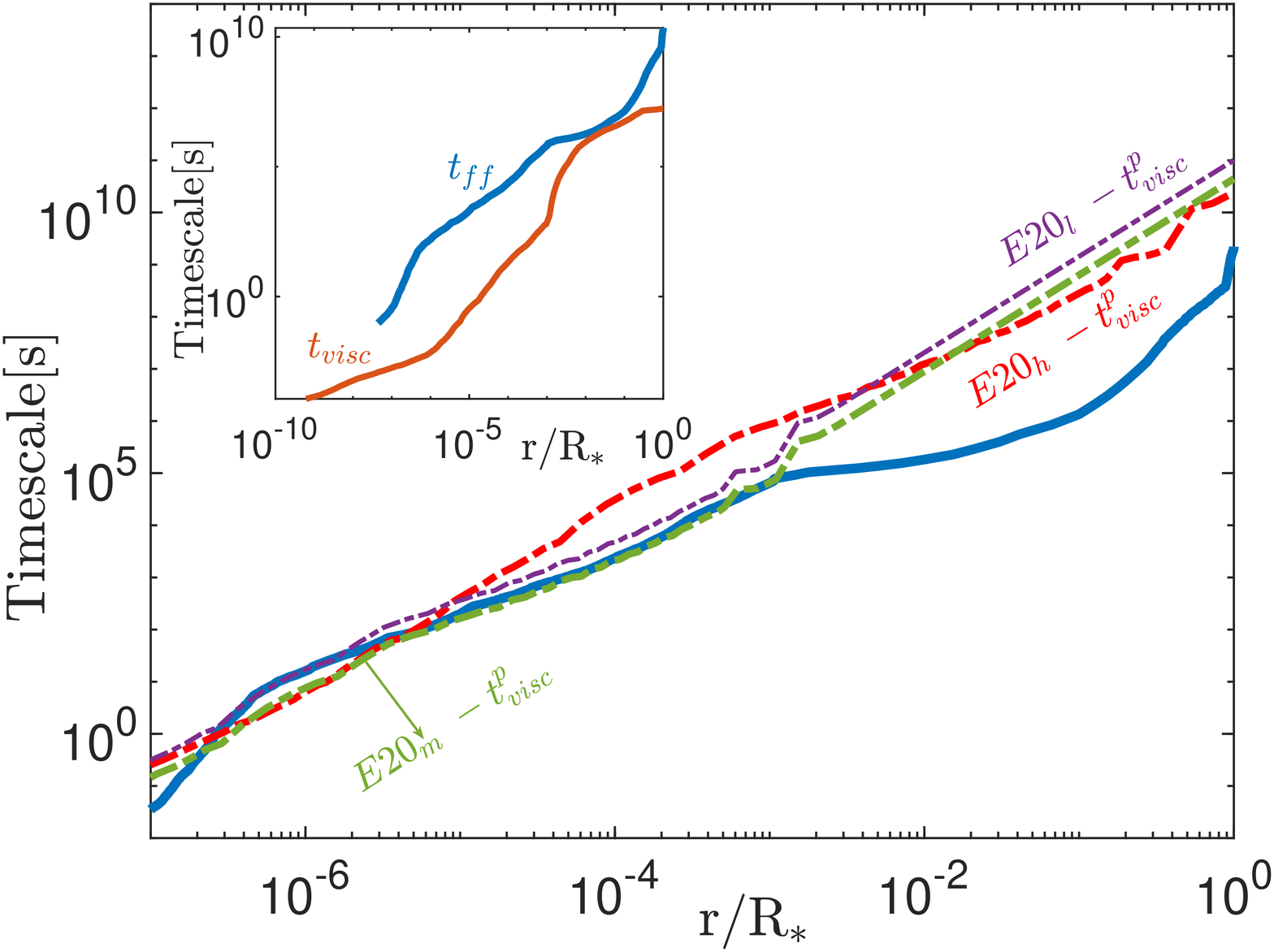}
\includegraphics[width=0.99\linewidth]{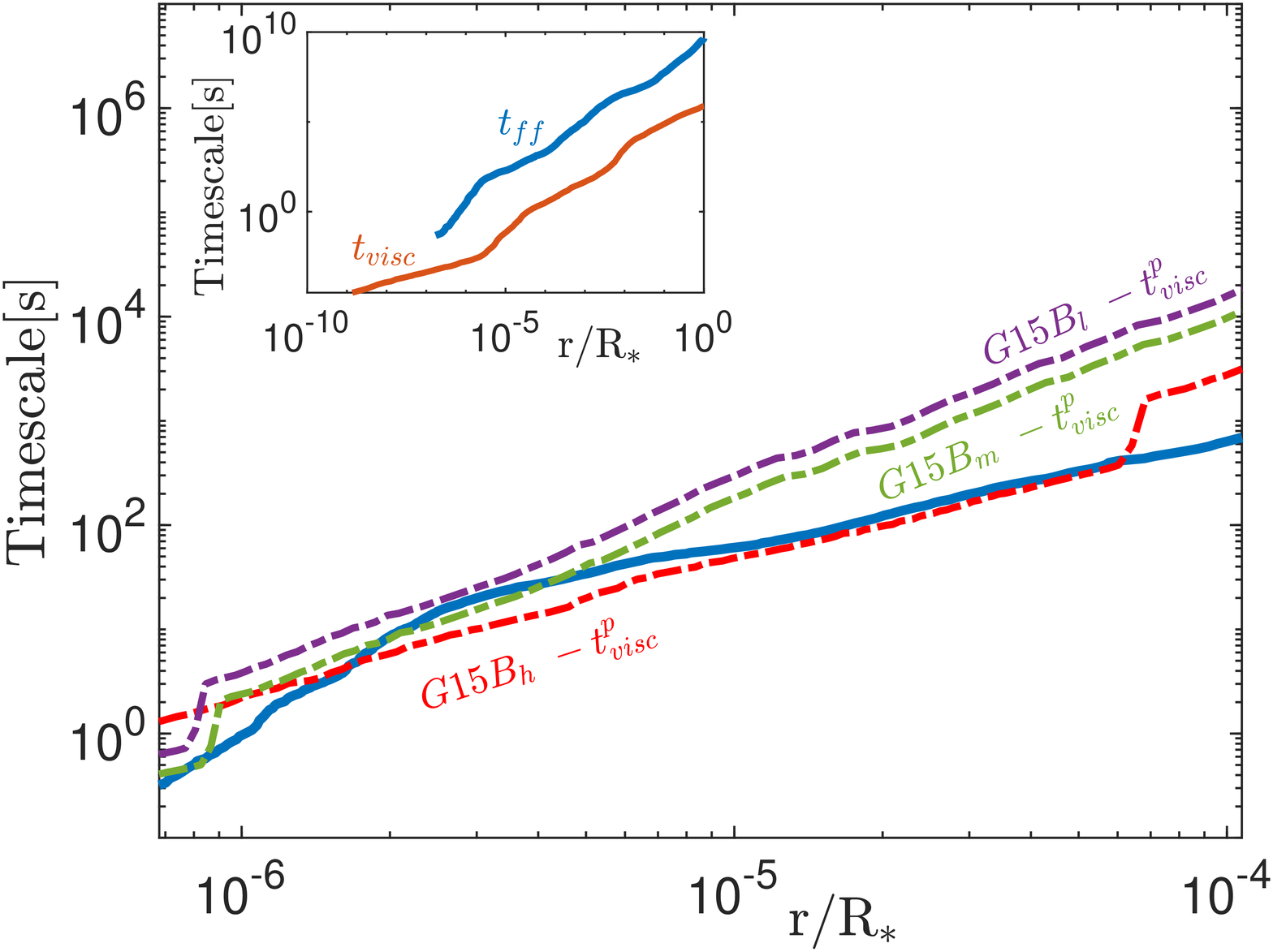}
\includegraphics[width=0.99\linewidth]{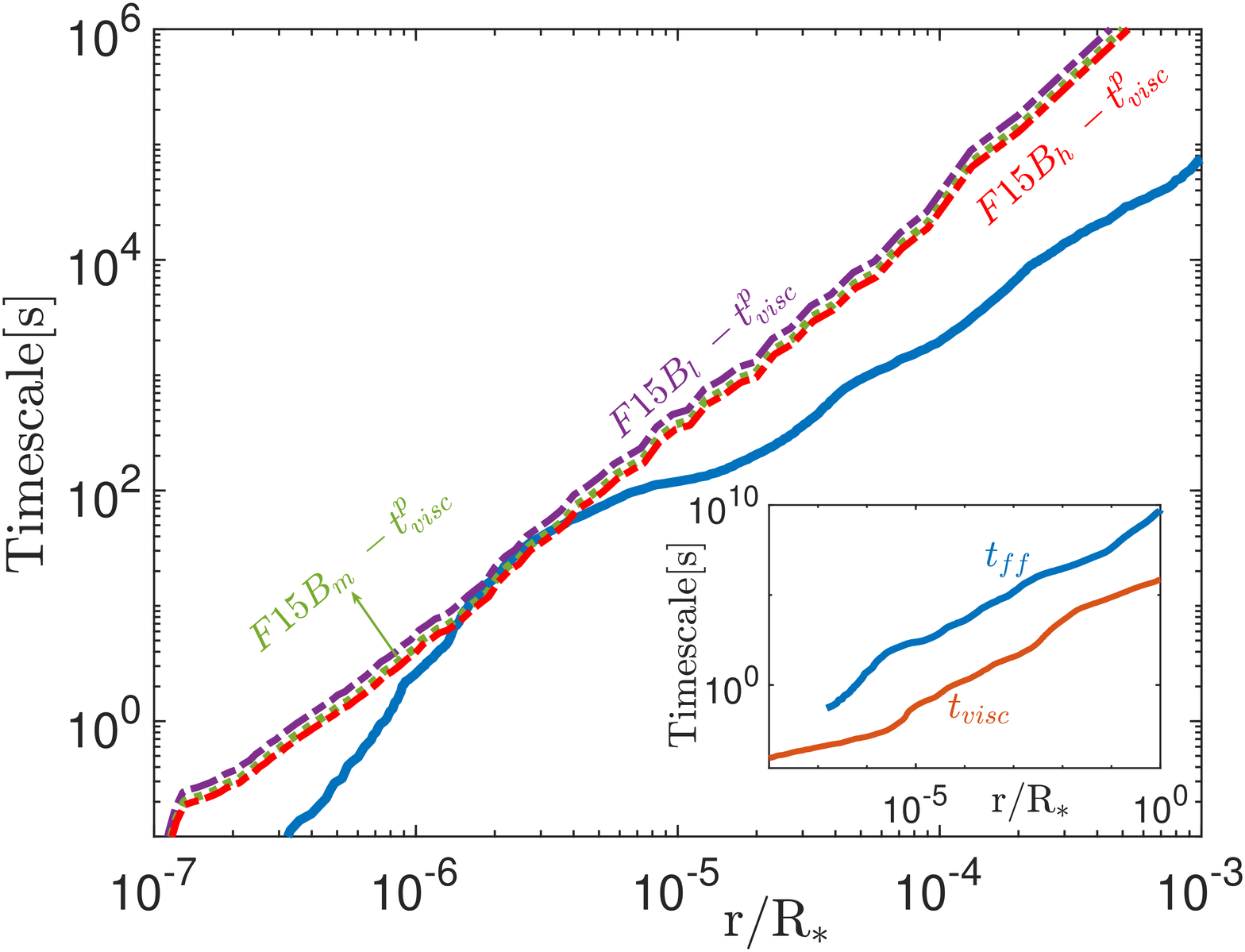}
\caption{Comparison of the free-fall timescale $t_{\rm ff}$ (Eq.~\eqref{eq:tsff}; blue solid lines) of material in the progenitor stars E20 (top), G15B (center) and F15B (bottom) of \citet{Heger+00} to the corresponding viscous timescales $t_{\rm visc}$ (Eq.~\eqref{eq:tvisc}) of that matter when circularized in the resulting accretion disk. Insets show $t_{\rm visc}$ computed with the original angular momentum profiles of \citet{Heger+00}, while the dashed-dotted lines show $t_{\rm visc}$ obtained by adopting the modified specific angular momentum profiles (Eq.~\eqref{eq:j_profile}) with small (`l'; purple), medium (`m'; green), and large (`h'; red) surface angular momentum (see the text for details). Quantities are plot as a function of stellar radius coordinate $r$ normalized to the progenitor surface radius $R_\star$.}
\label{fig:visc_fftimescale}
\end{figure}

\begin{figure}[h]
\centering
\includegraphics[width=0.99\linewidth]{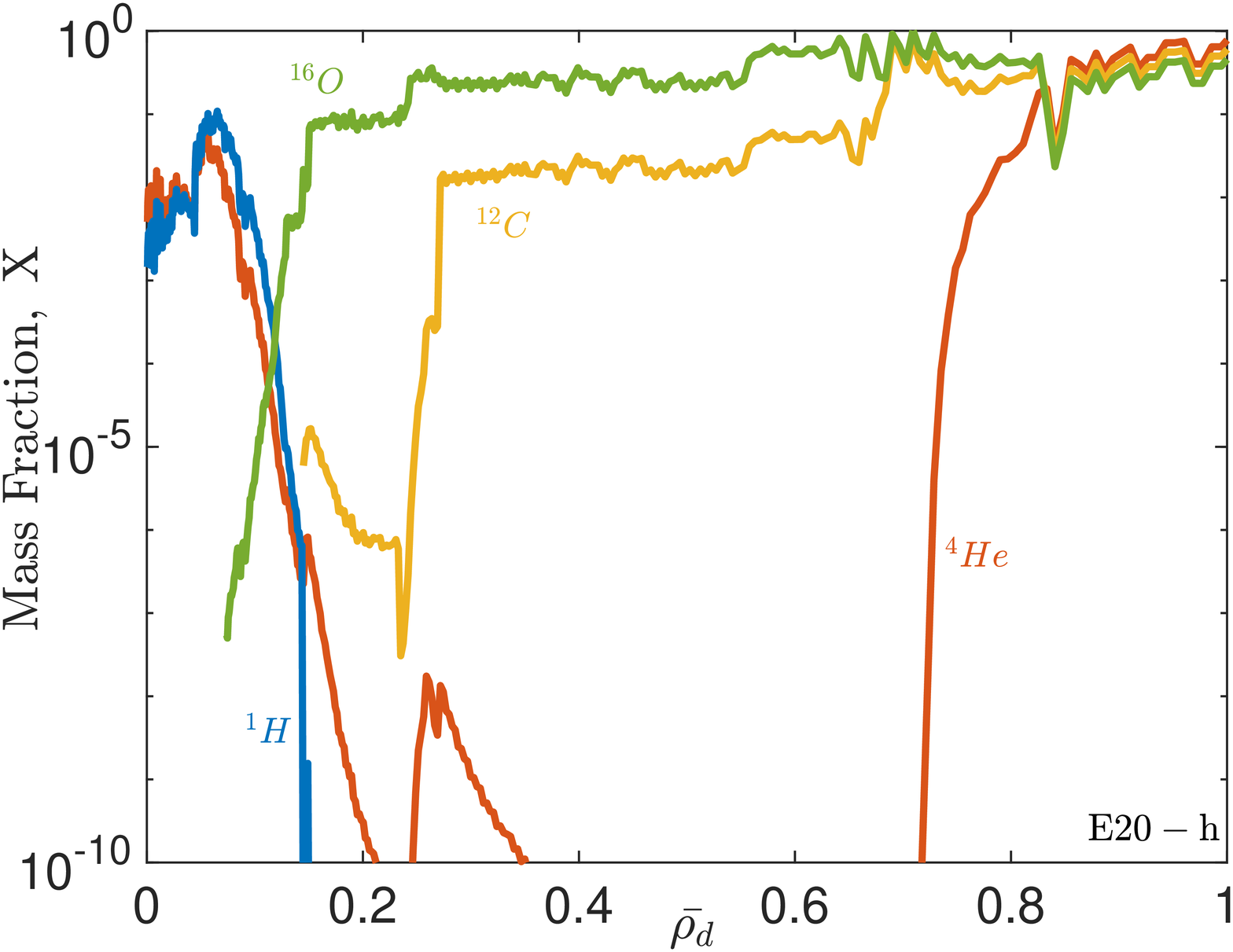}
\includegraphics[width=0.99\linewidth]{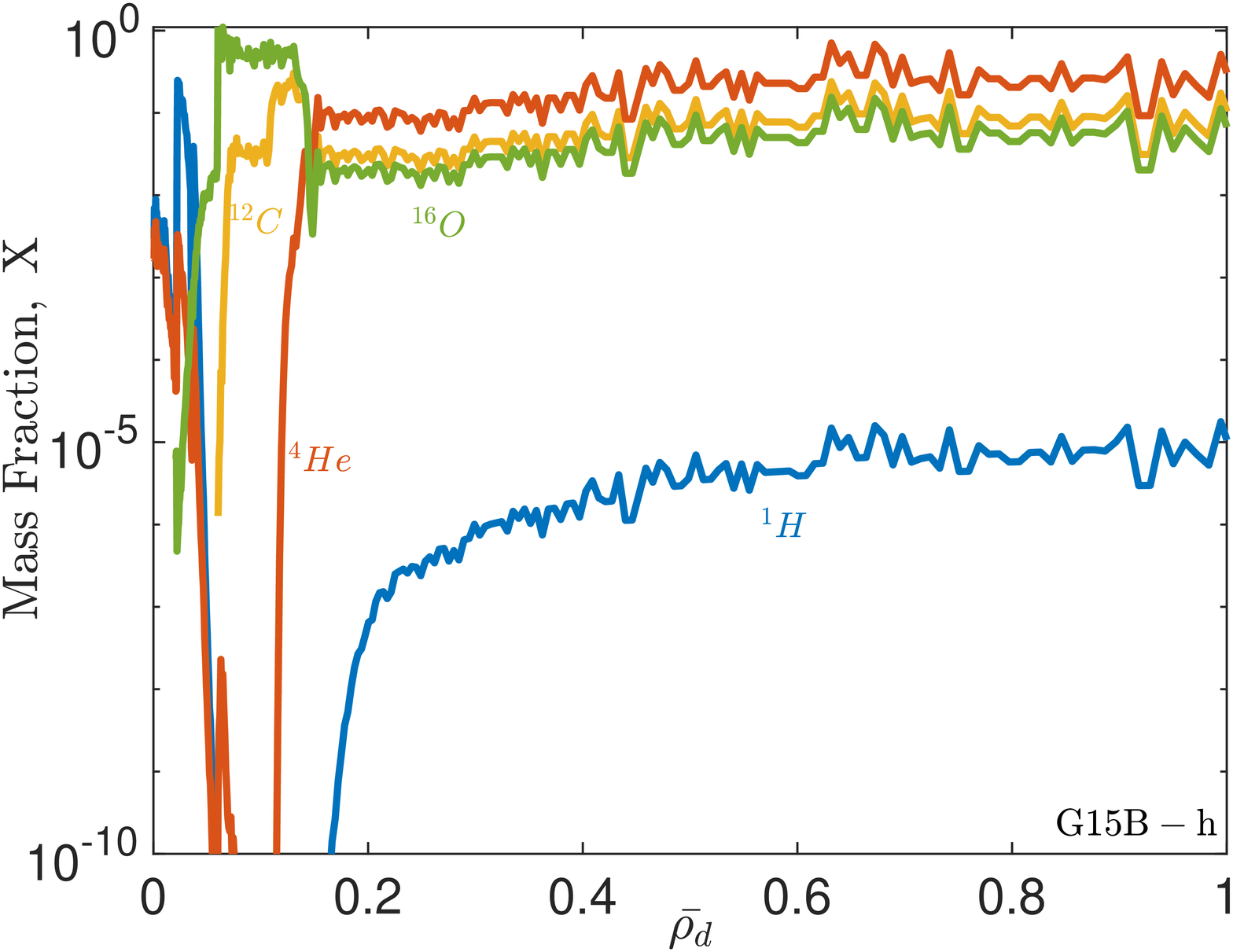}
\includegraphics[width=0.99\linewidth]{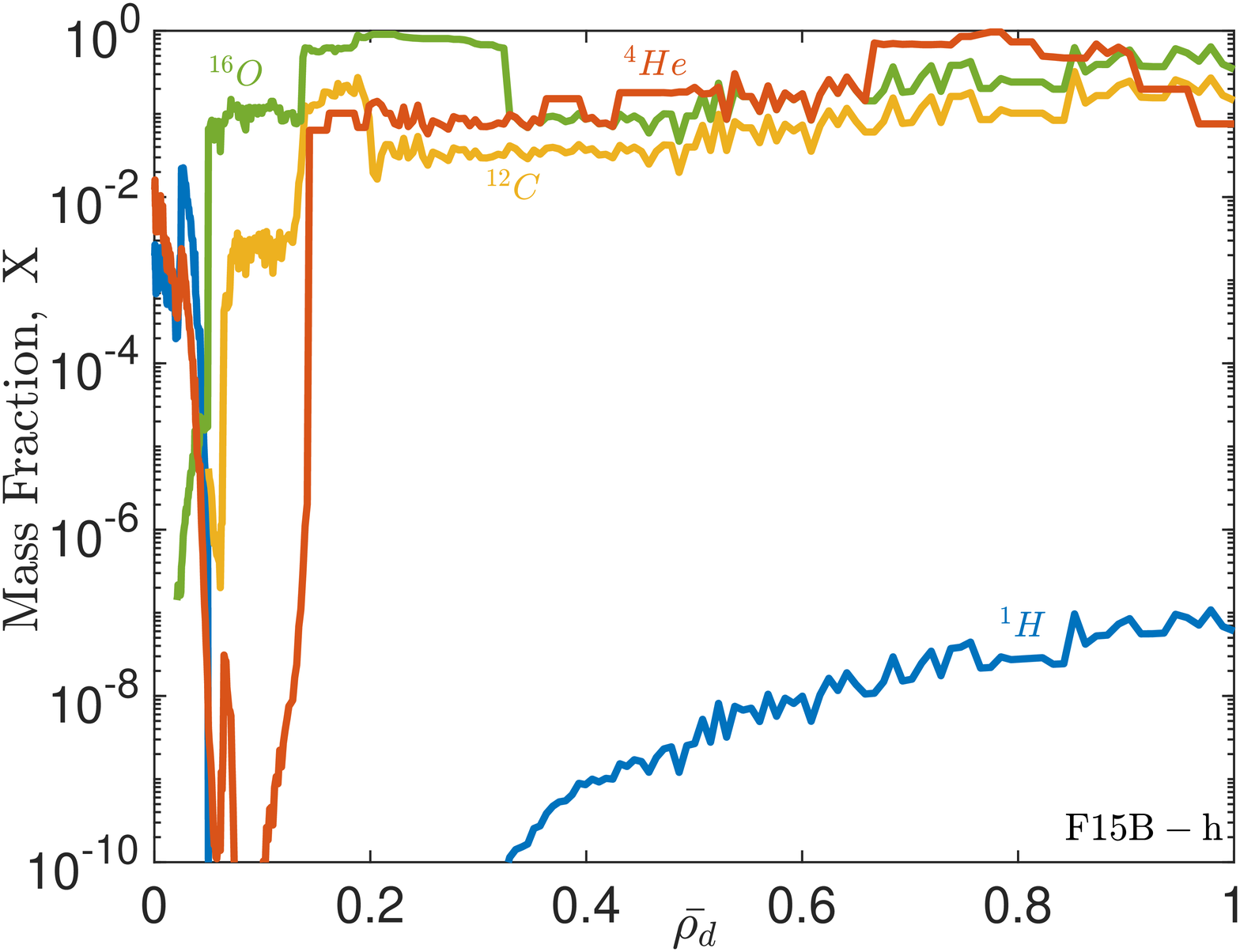}
\caption{Initial compositions of the collapsar accretion disks. Shown are mass fractions $X_i$ of various elements as a function of position within the torus $\bar{\rho}_{\rm d}= (\bar{\rho}-r_{\rm circ, in})/(r_{\rm circ,out}-r_{\rm circ,in})$ for disks resulting from the progenitor models E20, G15B, and F15B of \citet{Heger+00} with modified large surface angular momentum (`h') as defined in Eq.~\eqref{eq:j_profile} (see the text and Tab.~\ref{tab:models} for details).}
\label{fig:Massfraction_Mass_disc}
\end{figure}

At early times (small radii), one generally expects that $t_{\rm visc} \ll t_{\rm ff}$, in which case matter will be fed into the BH at roughly the same rate as it is supplied by the infalling star.  However, at later times (larger radii), $t_{\rm visc}$ increases faster than $t_{\rm ff}$, such that once $t_{\rm visc} > t_{\rm ff}$ one may treat the subsequent evolution as approximately that of a torus of fixed mass.

For all progenitor models considered here, we observe that $t_{\rm ff} \gtrsim t_{\rm visc}$ at all radii (see insets of Fig.~\ref{fig:visc_fftimescale}).  However, as discussed earlier, the process of angular momentum transport in massive stars comes with large uncertainty. This motivates us to consider a more general, parameterized angular momentum profile that will give rise to $t_{\rm ff} \sim t_{\rm visc}$ at some radius. Somewhat arbitrarily, we alter $j(r)$ from the original form given by \citet{Heger+00} to instead take on a power-law form,
\begin{equation} 
	j(r) = j(R_{\star})\left(\frac{r}{R_{\star}}\right)^{p}, \label{eq:j_profile}
\end{equation}
where the power-law index $p$ is taken to be the smallest value such that $t_{\rm ff}$ intersects $t_{\rm visc}$ at some radius, which we define as $r_{\rm in}$. In what follows, we denote by $r_{\rm circ,in}=r_{\rm circ}(r_{\rm in})$ the corresponding circularization radius of matter originating from the stellar layer at $r_{\rm in}$. Essentially, this rescaling procedure of the angular momentum profile reduces the amount of angular momentum, specifically in the core of the progenitor star, which is in line with recent findings of enhanced angular momentum transport in massive stars \citep{Ma&Fuller19}.

For each of the original progenitor models, we consider three qualitatively different scenarios, one with small (`l'), one with intermediate (`m'), and one with large (`h') surface angular momentum and hence different intersection values of $r_\text{in}$.  Each of these models, expressed in terms of the surface rotation rate $\Omega$ relative to the break-up value $\Omega_{\rm BU} = (GM_{\star}/R_{\star}^{3})^{1/2}$ and their assumed $p$ value, are listed in Table \ref{tab:models}. Figure \ref{fig:j_jkep_Menc} shows the angular momentum profiles $j(r)$ of these rescaled models, while Fig.~\ref{fig:visc_fftimescale} shows the resulting profiles of $t_{\rm ff}$ versus $t_{\rm visc}$.

In order to predict the properties of the resulting accretion disks---their radial extent, mass, and initial composition profiles, in particular,---we adapt the method of \citet{Kumar+08} and compute the circularization radius and associated circularization time $t_{\rm circ}$ (cf.~their Eq.~(6)) of each radial shell once $r>r_{\rm in}$. All material from shells $r<r_{\rm in}$ defines the central black-hole mass $M_{\rm BH}$, while we ignore shells with $r > r_{\rm out} \sim 10^9$\,cm, as their corresponding $t_{\rm circ}(r)<t_{\rm ff}(r)$, i.e., the formation timescale of the accretion disk, becomes comparable to the evolution timescale $t_{\rm visc}$ of the disk, which is typically determined by the innermost part of the disk. Furthermore, the composition of these outer layers is not of interest in terms of nuclear burning and would not qualitatively change the detonation behavior discussed in Sec.~\ref{sec:results}. Material residing between $r_{\rm in}$ and $r_{\rm out}$ sets the mass $M_{\rm d}$ of the resulting accretion disk.  The corresponding $r_{\rm in}$ and angular momentum profile of each model determines the compactness of the accretion disk. We note that different rotation profiles give rise to different values of $r_{\rm in}$, which thus lead to different compositions of the resulting accretion disks. Figure \ref{fig:Massfraction_Mass_disc} shows the composition of the resulting disk models with large (`h') surface angular momentum as a function of radius within the disk. Table~\ref{tab:models} lists properties of the accretion disks.

\begin{table*}
\begin{centering}
\begin{tabular}{cccccccccccccccc}
\hline 
Model & ${\rm M_{BH}}$  & $\rho_{\rm max,d}$ & $T_{\rm max,d}$ & $\Omega$ & $p$ & $r_{\rm circ,in-out}$ & $R_0$ & $d$ & $M_{\rm d}$ & ${\rm ^{1}H}$ & ${\rm ^{4}He}$ & ${\rm ^{12}C}$ & ${\rm ^{16}O}$\tabularnewline

- & $[M_{\odot}]$ & $[{\rm g/cm}^3]$ & $[10^8{\rm K}]$ & $[\Omega_{\rm BU}]$ & - & $[{\rm cm}]$ & $[10^8{\rm cm}]$ & - & $[M_{\odot}]$ & $[\%]$ & $[\%]$ & $[\%]$ & $[\%]$\tabularnewline
\hline 

E20-{h} & $6.22$ & $1.5\times10^{8}$ & $4.0$ & $0.735$ & $0.36$ &  $9\times10^7 - 1.3\times10^9$ & $1.14$ & $1.5$ & $3.53$ & $2$ & $22$  & $20$  & $56$ \tabularnewline
%\hline 
G15B-{h} & $5.36$ & $4.6\times10^{8}$ & $3.0$ & $0.767$ & $0.25$ & $9\times10^7 - 1.1\times10^9$ & $1.22$ & $1.45$ & $1.39$ & $3$ & $43$  & $26$  & $28$\tabularnewline
%\hline 
F15B-{h} & $3.42$ & $3.6\times10^{8}$ & $3.0$ & $0.776$ & $0.31$ & $8.5\times10^7 - 1.17\times10^9$ & $1.04$ & $1.45$ & $1.15$ & $7$ & $32$  & $27$  & $34$\tabularnewline
\hline 
%\hline 
E20-{l} & $6.29$ & $6.2\times10^{5}$ & $2.0$ & $0.698$ & $1.22$ & $9\times10^7 - 1.2\times10^9$ & $1.46$ & $1.5$ & $3.50$ & $3$ & $17$  & $25$  & $55$\tabularnewline
%\hline 
G15B-{l} & $5.89$ & $5.3\times10^{5}$ & $1.7$ & $0.722$ & $1.35$ & $8\times10^7 - 1.1\times10^9$ & $1.21$ & $1.5$ & $1.35$ & $5$ & $37$  & $23$  & $35$\tabularnewline
%\hline 
F15B-{l} & $4.51$  & $6.2\times10^{5}$ & $1.2$ & $0.731$ & $1.16$ & $7.5\times10^7 - 1.2\times10^9$ & $1.34$ & $1.5$ & $1.14$ & $4$ & $37$  & $26$  & $33$\tabularnewline
\hline 
%\hline 
E20-{m} & $6.23$  & $2.8\times10^{6}$ & $1.4$ & $0.723$ & $0.50$ & $5\times10^7 - 1.6\times10^9$ & $1.63$ & $1.48$ & $3.52$ & $3$ & $19$  & $27$  & $51$\tabularnewline
%\hline 
G15B-{m} & $5.38$  & $5.0\times10^{6}$ & $1.6$ & $0.748$ & $0.47$ & $6\times10^7 - 1.3\times10^9$ & $1.52$ & $1.63$ & $1.33$ & $7$ & $28$  & $27$  & $38$\tabularnewline
%\hline 
F15B-{m} & $4.47$ & $4.0\times10^{6}$ & $2.1$ & $0.744$ & $0.48$ & $7\times10^7 - 1.4\times10^9$ & $1.22$ & $1.5$ & $1.13$ & $4$ & $36$  & $24$  & $36$\tabularnewline
\hline 
\end{tabular}
\par\end{centering}
\caption{Initial parameters of the collapsar accretion disk models considered here. Form left to right: black-hole mass, maximum density and temperature of the disk, surface angular momentum (in units of the break-up limit) and angular momentum power-law coefficient of the stellar progenitor model, circularization radius of stellar mass shells contributing to torus formation, radius $R_0$ of the location of maximum density in the torus, torus $d$ parameter, torus mass $M_{\rm d}$, and mass fractions of various elements.}
\label{tab:models}
\end{table*}

The initial conditions for our hydrodynamic disk simulations are equilibrium tori that roughly resemble the properties predicted by the fallback model described above. The total disk mass $M_{\rm d}$ (cf.~Tab.~\ref{tab:models}) and radial profiles in composition (cf.~Fig.~\ref{fig:Massfraction_Mass_disc}) as calculated by the fallback model have been imposed on torus solutions. Following \citet{1999MNRAS.310.1002S} and \citet{2013ApJ...763..108F}, the torus density is normalized to its maximum value, $\rho_{\rm max}$, which fixes the polytropic constant in terms of the adiabatic index, $\gamma$, and the torus ``distortion parameter'', $d$.  The latter is a measure of the internal energy content of the torus which is closely related to its vertical scale-height \citep{PapaloizouPringle,1999MNRAS.310.1002S,2013ApJ...763..108F}. The initial torus radii and maximum density are chosen to closely resemble the fallback solutions.
The uncertainty in $j(r)$ and the simplicity of the fallback model impede a more precise determination of the initial torus properties.

The initial internal energy content of the torus and thus the value of $d$ are chosen as follows. The matter forming the collapsar disks falls in from much larger radii on nearly parabolic (zero-energy) orbits.  Furthermore, the material is sufficiently dense and infall sufficiently rapid that matter does not have time to cool during the process of disk formation.  Therefore, we expect that the total initial energy of the torus, which includes the kinetic, gravitational, and internal energies,
\begin{equation}
	e_{\rm tot} = \frac{1}{2} \left[v_{\bar{\rho}}^2 + \frac{l_z^2}{\bar{\rho}^2}\right] -\frac{GM_{\rm BH}}{r} - E_{\rm grav,d} + e_{\rm int}, \label{eq:etot}
\end{equation}
to be approximately zero, $e_{\rm tot} \simeq 0$. Here, $E_{\rm grav,d}$ is the self-gravity of the disk. It is thus the condition $e_{\rm tot} = 0$ that determines the disk distortion parameter $d$.

 In practical terms, tori solutions are initialized first by choosing approximate initial values for the torus structure as described in Appendix \ref{app:tori_solutions}, 
 from which we then derive the actual disk structure by relaxation using the full Helmholtz EOS in FLASH. Our torus solutions include self-gravity and lead to a self-consistent rotation profile. We relax the initial torus data to an equilibrium disk 
 until the relative acceleration is $<10^{-16}$ and the structure has converged to within a relative accuracy of $10^{-5}$ in hydrodynamic variables between consecutive time steps. For numerical stability during evolution, we surround the equilibrium torus with an atmosphere of low constant density and temperature ($\rho_{\rm atm} \lesssim 10^{-2} {\rm g\,cm}^{-3}, T_{\rm atm} \lesssim 10^4 {\rm K}$).

\subsection{Disk relaxation and simulation diagnostics}
\label{sec:diagnostics}

All simulation runs include an initial relaxation phase of typically $\approx\!160\,\mathrm{s}$, during which the initial data are evolved under the influence of gravity, but without nuclear reactions and the viscous source term. Once the initial acceleration is zero to machine precision for all grid points, we switch on the source terms due to viscosity and nuclear reactions. In what follows, we renormalize time labels with respect to the fully relaxed torus state and refer to this state as $t=0$. 

Mass outflow from the accretion disk in form of viscously driven winds or detonations is monitored throughout the simulations. We refer to {\it outflow} with mass $M_{\rm out}$ as all material crossing the outer boundary of the simulation grid. A fraction of this outflow is unbound and will eventually be ejected from the system. We define such {\it ejecta} with mass $M_{\rm ej}$ as outflow material that additionally has positive energy,
\begin{equation}
    |E_{\rm grav}| - \left(\frac{1}{2} \left[v_{\bar\rho}^{2} + \frac{l_{z}^{2}}{\bar\rho^{2}}\right] + |e_{\rm int}|\right) > 0.
\end{equation}
We cross-check this criterion by additionally evaluating the Bernoulli function across the grid,
\begin{equation}
    B=\frac{1}{2} \left[v_{\bar\rho}^{2} + \frac{l_{z}^{2}}{\bar\rho^{2}}\right]+e_{\rm int} + \frac{p}{\rho} +\phi,
\end{equation}
and requiring $B>0$. Both criteria allow for the conversion of internal energy into kinetic energy through adiabatic expansion and typically agree to a relative precision of $<10^{-5}$ in all our simulations, precluding ambiguity between the two criteria. 

All disk models whether detonating or not give rise to viscous accretion onto the black hole. Accreted mass onto the BH, denoted by $M_{\rm acc}$, is defined as all matter falling through a cylindrical radius $\bar{\rho}_{\rm acc}$ of $2\times 10^7\,{\rm cm}$ (for all `h' and `m' models) and $6\times 10^7\,{\rm cm}$ (for all `l' models). We extrapolate the accretion rate beyond the timescales simulated here, by fitting a power-law to the late-time trend of the accretion rate. This yields the approximate total amount of accreted material $M_{{\rm acc},\infty}$ resulting from a particular BH--disk system. Additionally, this allows us to estimate the total amount of material $M_{{\rm out},\infty}\equiv M_{\rm d} - M_{{\rm acc},\infty} - M_{\rm ej}$ that remains on the grid at the end of the simulation and that will eventually be unbound in viscous outflows.

The importance of nuclear reactions is monitored by computing the specific nuclear energy on the grid, defined by
\begin{equation}
    e_{\rm nuc} = \dot{Q}_{\rm nuc}\Delta t,
\end{equation}
where $\Delta t$ is the current time step and $\dot{Q}_{\rm nuc}$ is the total energy generation rate through nuclear reactions per unit mass (cf.~Sec.~\ref{sec:code}). We monitor the nuclear burning timescale, defined as
\begin{equation}
	t_{\rm burn} = \Big|\frac{X_i}{\dot{X}_i}\Big| = \Big|\frac{Y_i}{\dot{Y}_i}\Big|, \label{eq:t_burn}
\end{equation}
Here, $X_i$ denotes the mass fraction of fuel or burning products and $Y_i$ the corresponding number fractions, specifically of $^{4}$He, $^{12}$C, and $^{28}$Si. In order to become dynamically important, nuclear burning must occur on the orbital/dynamical timescale,
\begin{equation}
	t_{\rm burn} \lesssim t_{\rm dyn} = \sqrt{\frac{\bar{\rho}^3}{G M_{\rm BH}}}. \label{eq:t_dyn}
\end{equation}

Furthermore, we monitor the ignition timescale, which we define similarly to \citet{Woosley2004} as
\begin{eqnarray}
	t_{\rm ign} &=& \left(\frac{1}{\dot{Q}_{\rm nuc}}\frac{d\dot{Q}_{\rm nuc}}{dt}\right)^{-1} \approx \left(\frac{1}{\dot{Q}_{\rm nuc}}\frac{\partial \dot{Q}_{\rm nuc}}{\partial T}\frac{dT}{dt}\right)^{-1} \\
	&=& \frac{c_v T}{\beta \dot{Q}_{\rm nuc}} \approx \frac{e_{\rm int}}{\beta \dot{Q}_{\rm nuc}}, \label{eq:t_ign}
\end{eqnarray}
where the last identity holds for an approximately polytropic gas. We have assumed that $\dot{Q}_{\rm nuc} \propto \rho^\alpha T^\beta$ can be approximated by a power law in temperature with an effective exponent $\beta$, and we have used the fact that changes in the specific heat are given by $\dot{q} = \dot{Q}_{\rm nuc}$ and $dq = c_v dT$ at constant volume. For a given isolated volume element, $t_{\rm ign}$ represents the time needed to increase the initial temperature to a temperature `run away', such that nuclear reactions become instantaneous. We compute $t_{\rm ign}$ at every grid point by extracting $c_v$, $\dot{Q}_{\rm nuc}$, and $\beta$ using the full reaction network and EOS information. We define the onset of a detonation event by
\begin{equation}
	t_{\rm burn} \approx t_{\rm ign}, \label{eq:detonation_criterion}
\end{equation}
where $t_{\rm burn}$ denotes the timescale computed from either mass fraction of the aforementioned species.

\section{Results}
\label{sec:results}

Initial data for all collapsar accretion disks simulated here are obtained using the methods prescribed in Secs.~\ref{sec:initial_models} and \ref{sec:diagnostics} (see Tab.~\ref{tab:models} for a list of models considered here and corresponding initial parameters). Characteristic timescales of these accretion disks are reported in Tab.~\ref{tab:viscositytab}. Due to the computational cost of resolving the required time and length scales, we do not follow the actual collapse and the formation of the disk. It is therefore possible that the infalling material has already been susceptible to nuclear burning even prior to disk formation. In such cases, the formulation of our prescribed disks is physically inconsistent, and we would expect these disks to immediately show significant nuclear burning, and possibly give rise to prompt nuclear detonations. In more consistent cases, one would expect nuclear burning to be delayed, and begin only following the evolution of the disk and the onset of viscous heating, or not lead to any detonation, if the disk density and temperatures never give rise to the appropriate detonation conditions. 

Depending on the angular momentum profile of the progenitor star, we find three classes of collapsar accretion disks, based on the progenitor models described in Sec.~\ref{sec:initial_models}:
\begin{itemize}
     \item[$(i)$] Disks that lead to a relatively `prompt' detonation on a timescale $\lesssim t_{\rm visc}$, typically $\sim\!100 t_{\rm orb}$, where 
\begin{equation}
	t_{\rm orb}=2\pi \sqrt{R_0^3/G M_{\rm BH}}
\end{equation}
     is the orbital timescale,
\begin{equation}
     t_{\rm visc} = \frac{1}{\alpha}\sqrt{\frac{R_0^3}{G M_{\rm BH}}}\left(\frac{H}{\bar{\rho}}\right)^{-2}
\end{equation}
     is the viscous timescale, and $R_0$ is the radius at maximum density of the disk. These disks include the models labeled `h'.
    \item[$(ii)$] Disks that lead to delayed detonations on a timescale $\gtrsim (1-\mathrm{few}) t_{\rm visc}$. Models labeled `m' belong to this category of somewhat delayed and weaker detonations.
    \item[$(iii)$] Disks that do not show detonations and at most show mild nuclear burning; these are the models labeled `l'.
\end{itemize}

We do not find any immediate detonations on the orbital timescale of the disks in our models, indicating that none of our models may have been susceptible to nuclear burning even prior to formation and that they are thus formulated consistently. All models require at least a significant fraction of the viscous timescale to detonate. Furthermore, for consistency, our models have been set up such that the torus formation timescale (the circularization timescale $t_{\rm circ,out}$ of the outermost stellar layer contributing to disk formation) is less than the evolution timescale, $t_{\rm circ,out} \lesssim t_{\rm visc}$ (cf.~Sec.~\ref{sec:initial_models} and Tab.~\ref{tab:viscositytab} for a compilation of timescales).

In this section, we first discuss in detail the limiting cases $(i)$ and $(iii)$ (Secs.~\ref{sec:no_detonation} and \ref{sec:detonation}), before summarizing results of a survey of models throughout the collapsar accretion disk parameter space more broadly (Sec.~\ref{sec:survey_models}).

\subsection{Non-detonating disks}
\label{sec:no_detonation}

\begin{figure*}
\centering
\includegraphics[width=0.49\linewidth]{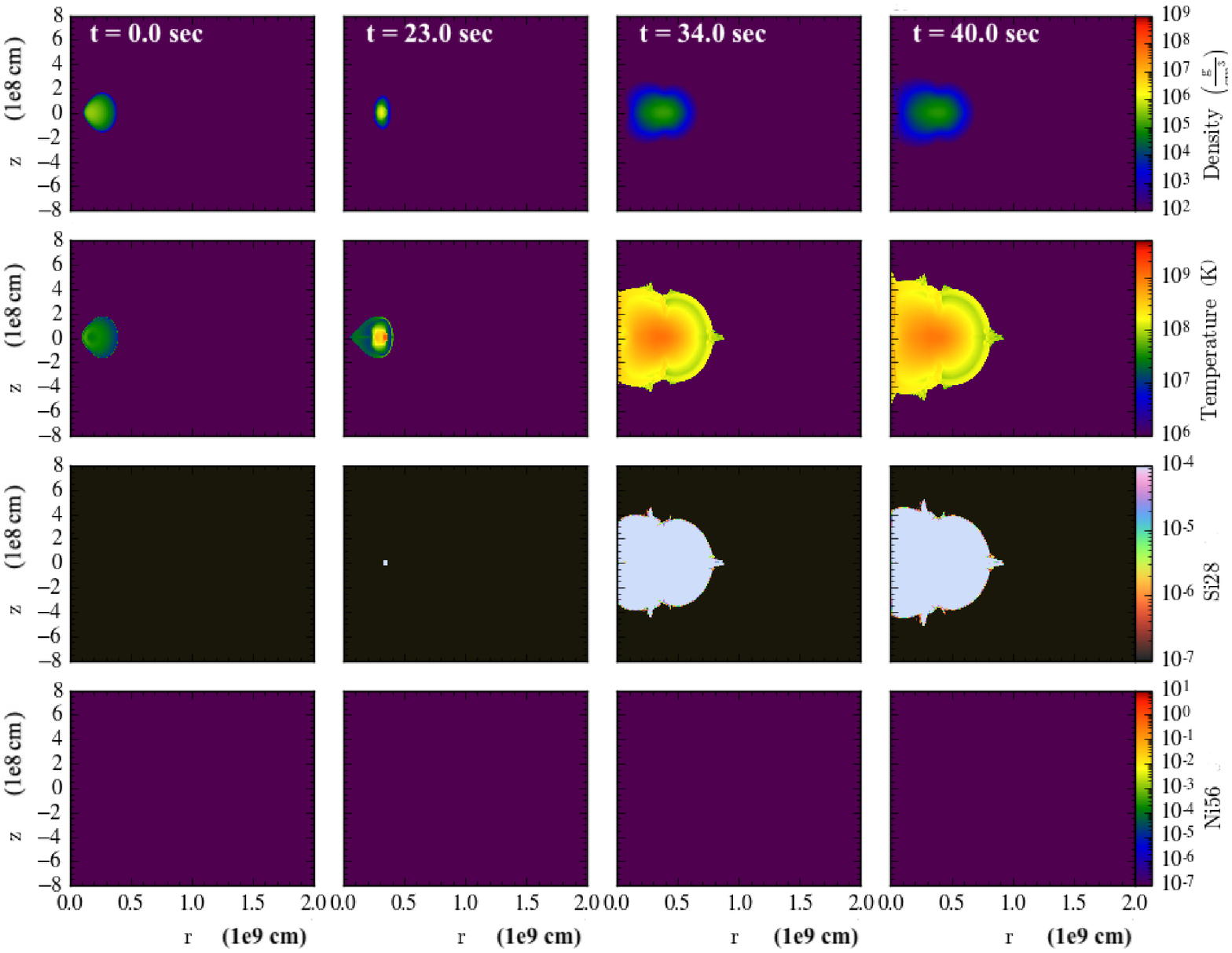}
\includegraphics[width=0.49\linewidth]{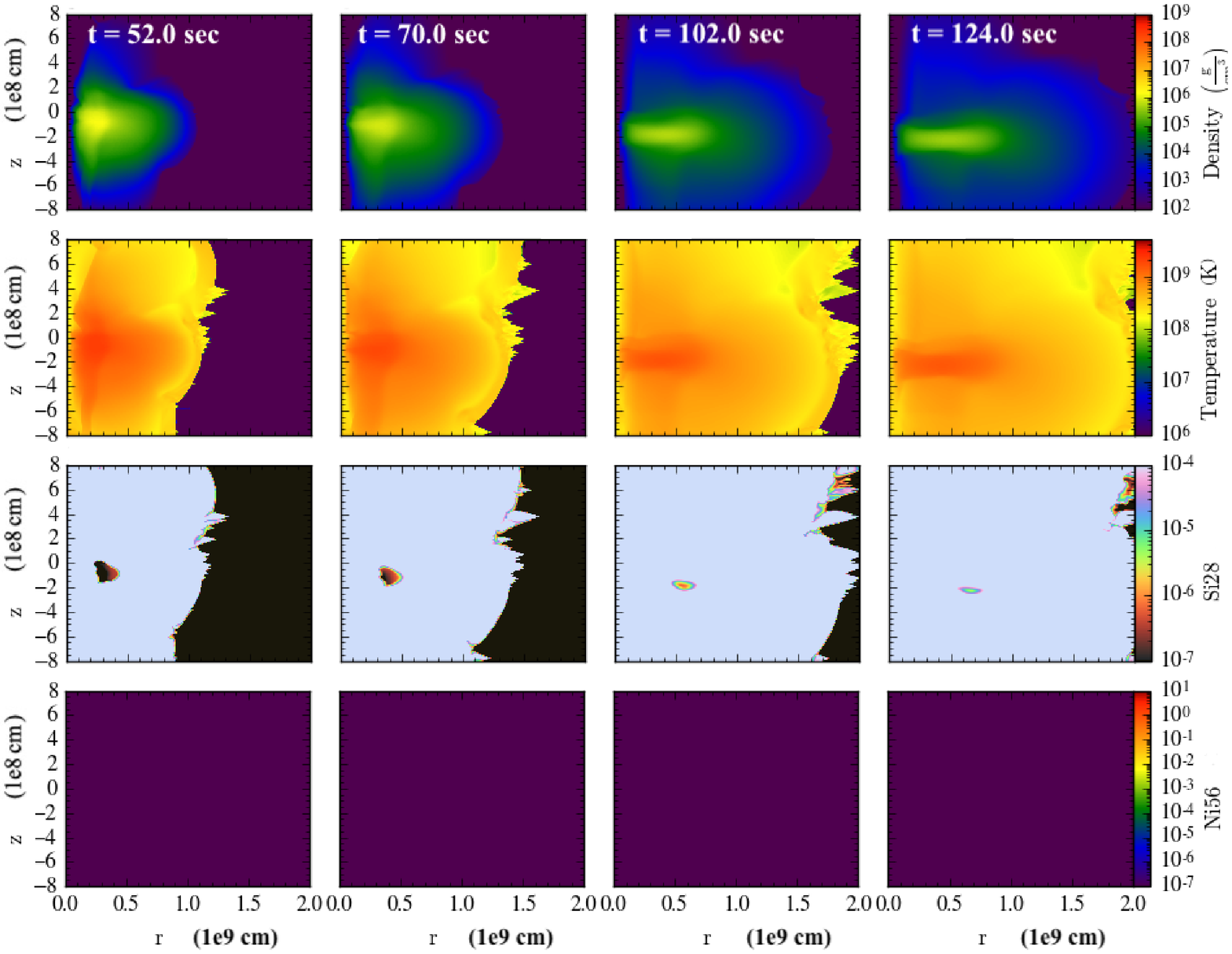}
\caption{Early evolution of model E20-l. Snapshots show density, temperature, and $^{28}$Si as well as $^{56}$Ni mass fractions in the meridional plane ($\bar\rho$,$z$) during the first few viscous timescales. Residual nuclear burning is evident from $^{28}$Si-production, but iron-group elements such as $^{56}$Ni are absent.
}
\label{fig:FLASHE20_l}
\end{figure*}

Non-detonating disks are defined as those models that do not show detonations during the evolution time, typically up to many viscous timescales; we refer to Tab.~\ref{tab:viscositytab} for a compilation of simulation, orbital, and viscous timescales of all simulation runs. Figure \ref{fig:FLASHE20_l} shows snapshots of several key quantities during the early evolution of model E20-l, which is representative for this category of collapsar disks.

\begin{figure}[h]
\centering
\includegraphics[width=0.99\linewidth]{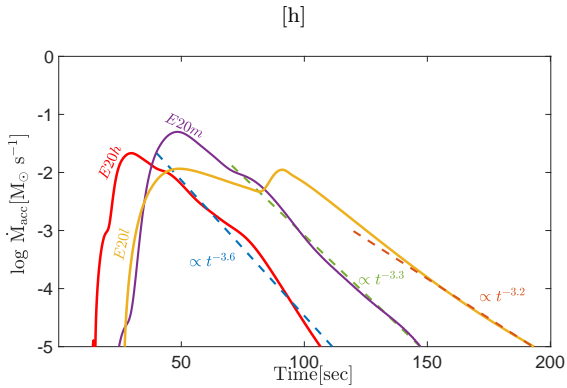}
\caption{Accretion rate onto the black hole for models E20-h (red), E20-l (yellow), and E20-m (purple), as well as power law fits to the late-time trend of E20-h and E20-m (dashed lines).}
\label{fig:Mdot_E20}
\end{figure}

\begin{figure}[h]
\centering
\includegraphics[width=0.95\linewidth]{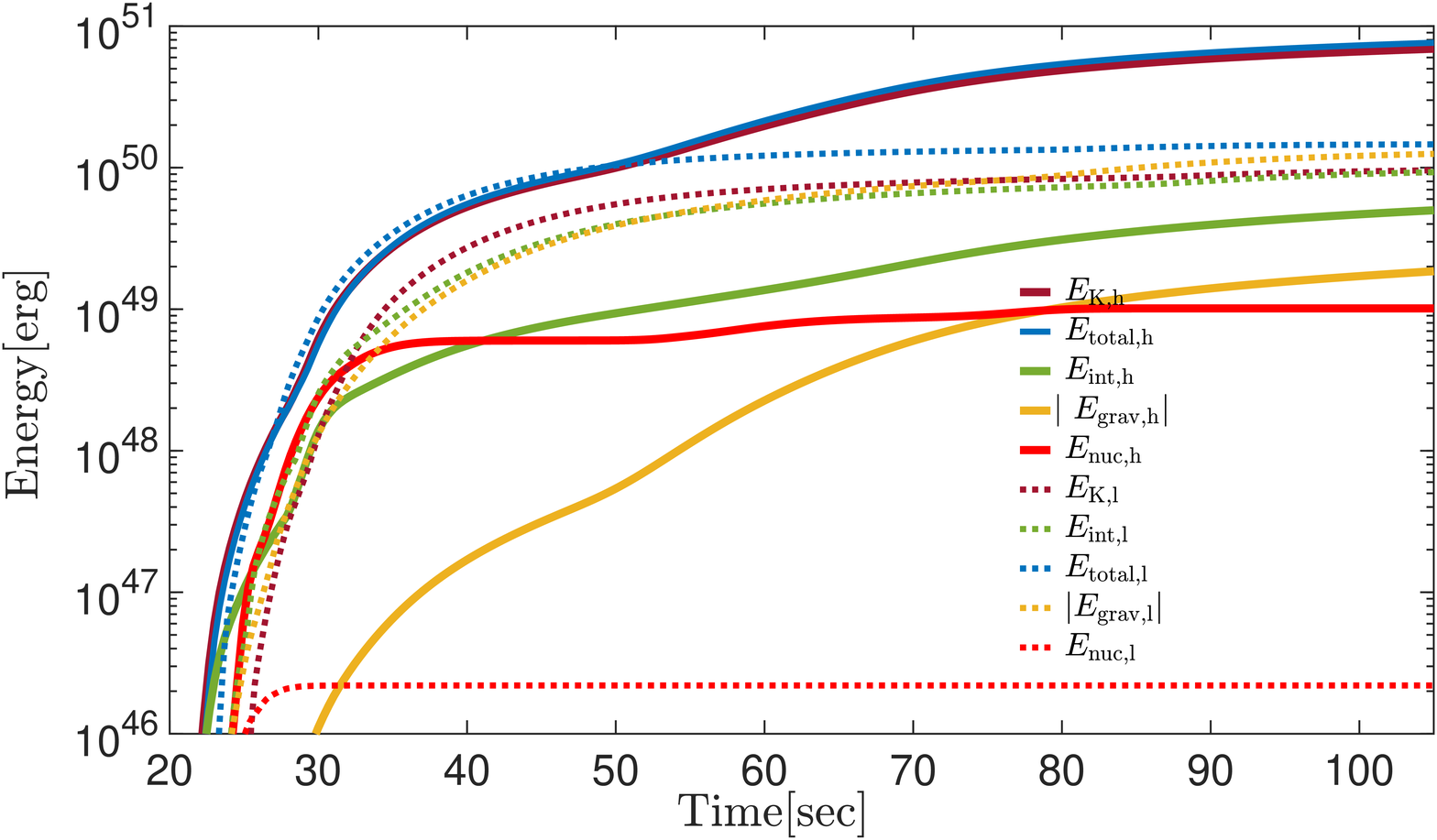}
\includegraphics[width=\linewidth]{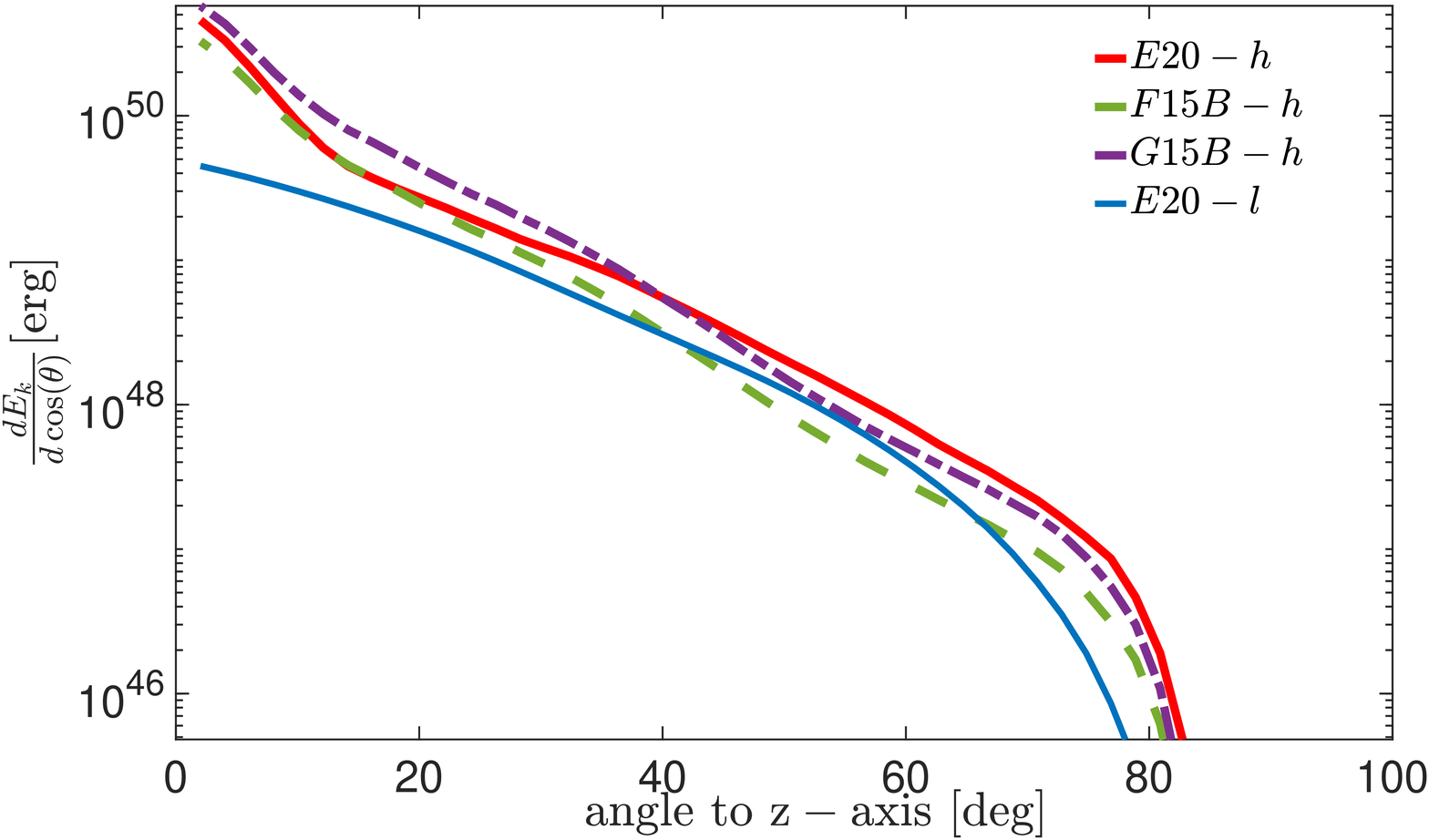}
\caption{Top: various energies across the numerical grid monitored for models E20-h and E20-l. The kinetic energy refers to ejecta only. Bottom: kinetic energy of ejecta as a function of polar angle for various models, indicating a significant increase of kinetic energy in equatorial directions for detonating models.}
\label{fig:energytotal_E20h}
\end{figure}

Initially, viscous accretion leads to an increase in internal energy of the disk as gravitational energy is being released and converted into heat of material viscously moving inward (cf.~Fig.~\ref{fig:energytotal_E20h}). We find an accretion rate peaking at $\dot{M}_{\rm acc}\simeq \times10^{-2}\,M_\odot\,{\rm s}^{-1}$ at $t\approx t_{\rm visc}\approx 40$\,s, before starting a power-law decline as a result of viscous spreading of the disk (cf.~Fig.~\ref{fig:Mdot_E20}). During the first viscous timescale residual nuclear burning takes place, which results in the production of intermediate-mass elements as, e.g., traced by $^{28}$Si; however, synthesis of iron-group elements such as $^{56}$Ni is absent (cf.~bottom two rows of Fig.~\ref{fig:FLASHE20_l}). The overall contribution of nuclear burning to the disk dynamics is not entirely negligible, typically $e_{\rm nuc}/e_{\rm int} \lesssim 0.1$ during the first viscous timescale, such that the disk starts to expand. However, energy release from nuclear burning is not sufficient to cause a global detonation---$e_{\rm nuc}$ remains orders of magnitude smaller than $e_{\rm tot}$ at all times. Nevertheless, initial expansion of the disk results in an eventual `recollapse' of the disk on itself after the burning period, which manifests as a bump in the accretion rate around $t\approx 2.5t_{\rm visc}\approx 100$\,s (Fig.~\ref{fig:Mdot_E20}).

\begin{figure}[h]
\centering
\includegraphics[width=0.99\linewidth]{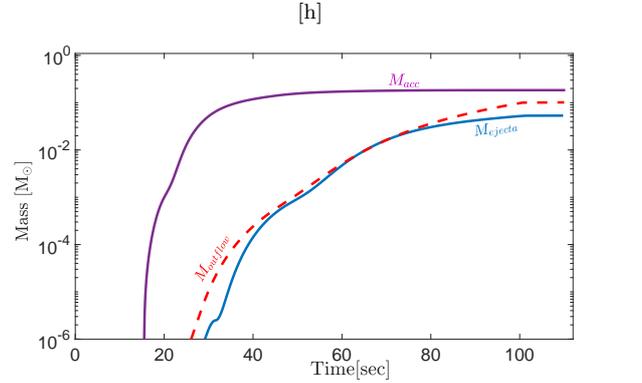}
\caption{Cumulative mass loss from the detonating collapsar accretion disk E20h: mass accreted onto the black hole ($M_{\rm acc}$), total mass in outflows (bound and unbound; $M_{\rm out}$), and total ejected mass (unbound material; $M_{\rm ej}$).}
\label{fig:CDF_massloss_E20h}
\end{figure}

In the absence of effective cooling, viscous heating drives vigorous convective turbulence (cf.~upper panels of Fig.~\ref{fig:FLASHE20_l}), which results in significant mass outflows from the disk. Material is ejected from the disk mostly in a wide funnel around the polar axis (cf.~Fig.~\ref{fig:energytotal_E20h}), with a tail of material in the equatorial direction. Most material in these early outflows has positive total energy and it is thus asymptotically unbound from the BH--disk system: about $\approx\!80\%$ of the total outflowing mass $M_{\rm out}$ during the simulated time window is unbound in our `l' models, which we refer to as $M_{\rm ej}$ (cf.~Tab.~\ref{tab:torusenergytab}). Although these ejecta outflows are rather fast, with typical velocity $v\sim 0.1-0.16$\,c, the kinetic energy of these early ejecta are not sufficient to unbind the remaining stellar envelope surrounding the BH-disk system (cf.~Tab.~\ref{tab:torusenergytab}).

As the disk spreads outward at $t\gtrsim t_{\rm visc}$ due to viscous angular momentum transport, the maximum density and temperature in the disk start to decrease rapidly; significant nuclear burning beyond the simulated time frame is thus not expected to occur for these accretion disks. This is also reflected by the flat trend in the cumulative nuclear energy for $t\gtrsim t_{\rm visc}$ shown in Fig.~\ref{fig:energytotal_E20h}. Another consequence of viscous spreading is that the cumulative accretion rate onto the BH converges rapidly (cf.~Fig.~\ref{fig:CDF_massloss_E20h}) and most of the remaining disk material will eventually be lost in outflows. As nuclear reactions are insignificant for the subsequent evolution of these models, the disk material is not further reprocessed and the composition of the total outflows reflects the composition of the early outflows simulated here. We find that significant amounts of intermediate-mass elements with atomic numbers $10\le Z \le 22$ are synthesized (Tab.~\ref{tab:torusenergytab}); typically, 30\%-60\% of the initial disk mass are converted into such intermediate-mass elements. However, outflows do not contain significant amounts of iron-group nuclei.

\begin{table*}
\begin{centering}
\begin{tabular}{cccccccccccc}
\hline
Model  &  $f_{\rm out}$ & $M_{\rm ej}$ & $M_{\rm acc,\infty}$ &  $M_{\rm out,\infty}$ & $E_{\rm K}$ &  $E_{\rm K, pol}$ & $E_{\rm grav,env} $ & $r_{\rm det}$ & $M_{\rm ej}^{\rm IGE}$ & $M_{\rm out,\infty}^{\rm IGE}$ & $M_{\rm out,\infty}^{\rm IME}$  \\
& $[10^{-2}M_{\rm d}]$ & $[10^{-2}M_{\odot}]$ & $[M_{\odot}]$ & $[M_{\odot}]$ & $[10^{51}{\rm erg}]$ & $[10^{51}{\rm erg}]$ & $[10^{51}{\rm erg}]$  & $[r_{\rm circ,in}]$ & $[M_{\odot}]$ & $[M_{\odot}]$ & $[M_{\odot}]$ \\
\hline 
E20-h & $2.8$ & $6.6$ & $0.218$ & $3.242$ & $0.5279$ & $0.1603$ & $0.0034$ & $ \lesssim 1.16$ & 9.22e-3 & 4.06e-1 & $1.13$\tabularnewline 

G15B-h & $6.0$ & $5.2$ & $0.145$ & $1.194$ & $0.6041$ & $0.1673$ & $0.0251$ & $\lesssim 1.01$ & 9.27e-3 & 5.84e-2 & $0.21$\tabularnewline
 
F15B-h & $5.8$ & $4.7$ &$0.157$ & $0.946$ & $0.4165$  & $0.1778$ & $0.0186$ & $\lesssim 1.02$ & 6.34e-3 & 5.41e-2 & $0.25$\tabularnewline

\hline 
E20-l & $0.069$ & $0.21$ & $0.104$ & $3.388$ & $0.0481$ & $0.0409$ &$0.0529$ & $-$ & 7.02e-8 & 4.52e-6 & $1.38$\tabularnewline

G15B-l & $0.086$ & $0.099$ & $0.011$ & $1.332$ & $0.0119$ & $0.0104$ & $0.0324$ & $-$ & 7.55e-8 & 2.48e-6 & $0.47$\tabularnewline

F15B-l & $0.078$ & $0.072$ & $0.009$ & $1.131$ & $0.0142$ & $0.0116$ & $0.0225$ & $-$ & 8.26e-8 & 4.08e-6 & $0.66$\tabularnewline

\hline 
E20-m & $5.3$ & $1.52$ & $0.195$ & $3.307$ & $0.0775$ & $0.1283$ & $0.0042$ & $\lesssim 3.1$ & 6.29e-3 & 4.15e-1 & $1.16$\tabularnewline

G15B-m & $1.5$ & $1.31$ & $0.126$ & $1.191$ & $0.1032$ & $0.0646$ & $0.0273$ & $\lesssim 2.5$ & 6.88e-3 & 7.62e-2 & $0.28$\tabularnewline
 
F15B-m & $1.4$ & $1.35$ & $0.077$ & $1.037$ & $0.0964$ & $0.1522$ & $0.0114$ & $\lesssim 2.8$ & 4.41e-3 & 5.77e-2  & $0.33$\tabularnewline
\hline 
\end{tabular}
\par\end{centering}
\caption{Detonation and outflow properties of the accretion disk models simulated here. Where applicable, quantities are extracted at the end of the simulation runs. From left to right: fraction of the initial disk mass in outflows, amount of unbound material (ejecta) from the BH-disk system, total mass eventually accreted onto the BH ($M_{{\rm acc},\infty}$; cf.~Sec.~\ref{sec:diagnostics}), total mass in outflows that will eventually be unbound ($M_{{\rm out},\infty}$; cf.~Sec.~\ref{sec:diagnostics}), total kinetic energy of the ejecta, fraction thereof in equatorial directions (polar angles $>15^\circ$), gravitational binding energies of surrounding stellar envelopes, locations at which detonations first arise in units of the initial inner torus radius, total mass of ejected iron-group elements (IGE) during simulation run, and total mass of iron-group and intermediate-mass elements (IME; atomic number $10\leq Z\leq 22$) in $M_{\rm out,\infty}$.
}
\label{tab:torusenergytab}
\end{table*}

\subsection{Detonating disks}
\label{sec:detonation}

\begin{figure*}
\centering
\includegraphics[width=0.49\linewidth]{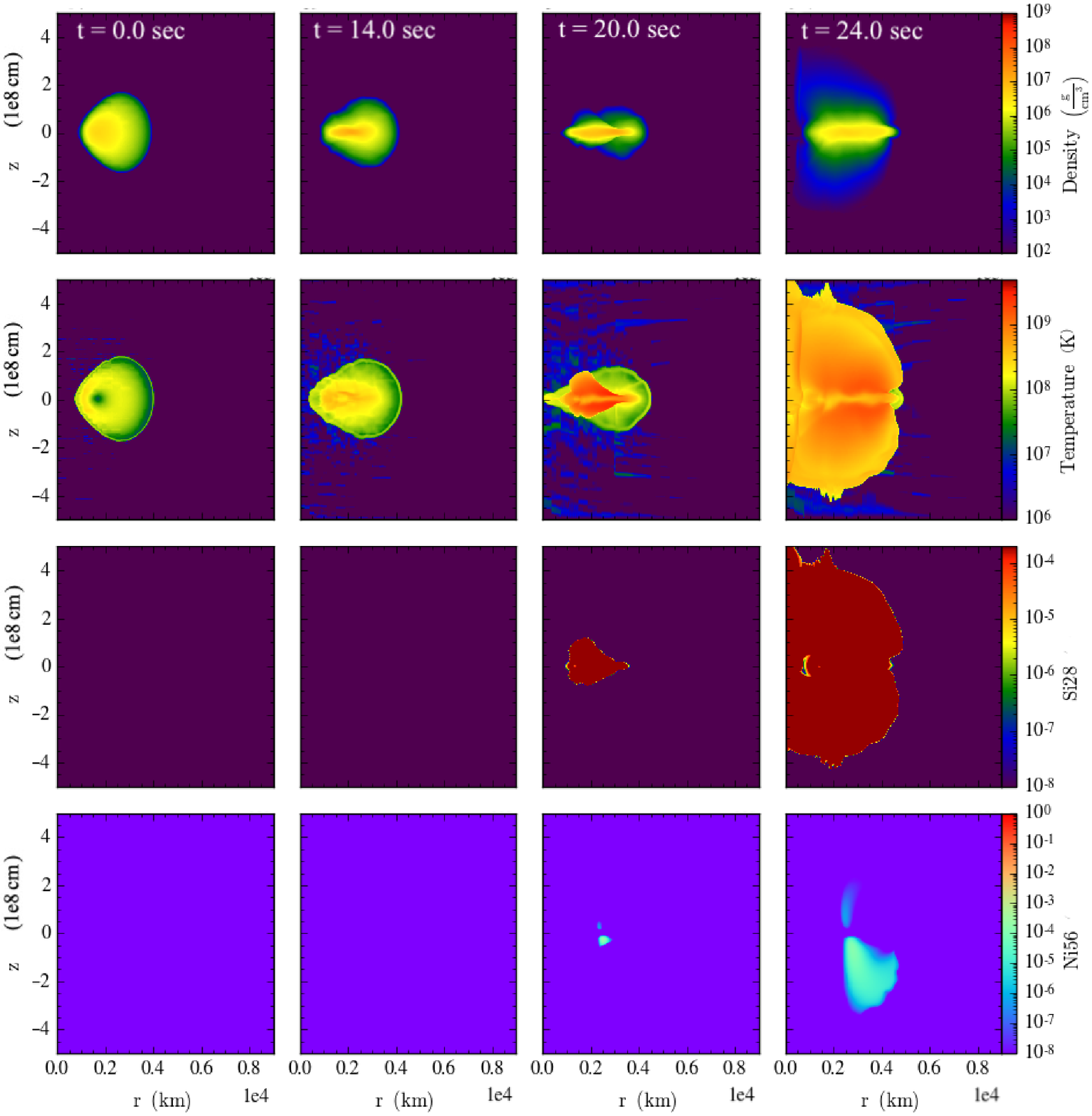}
\includegraphics[width=0.49\linewidth]{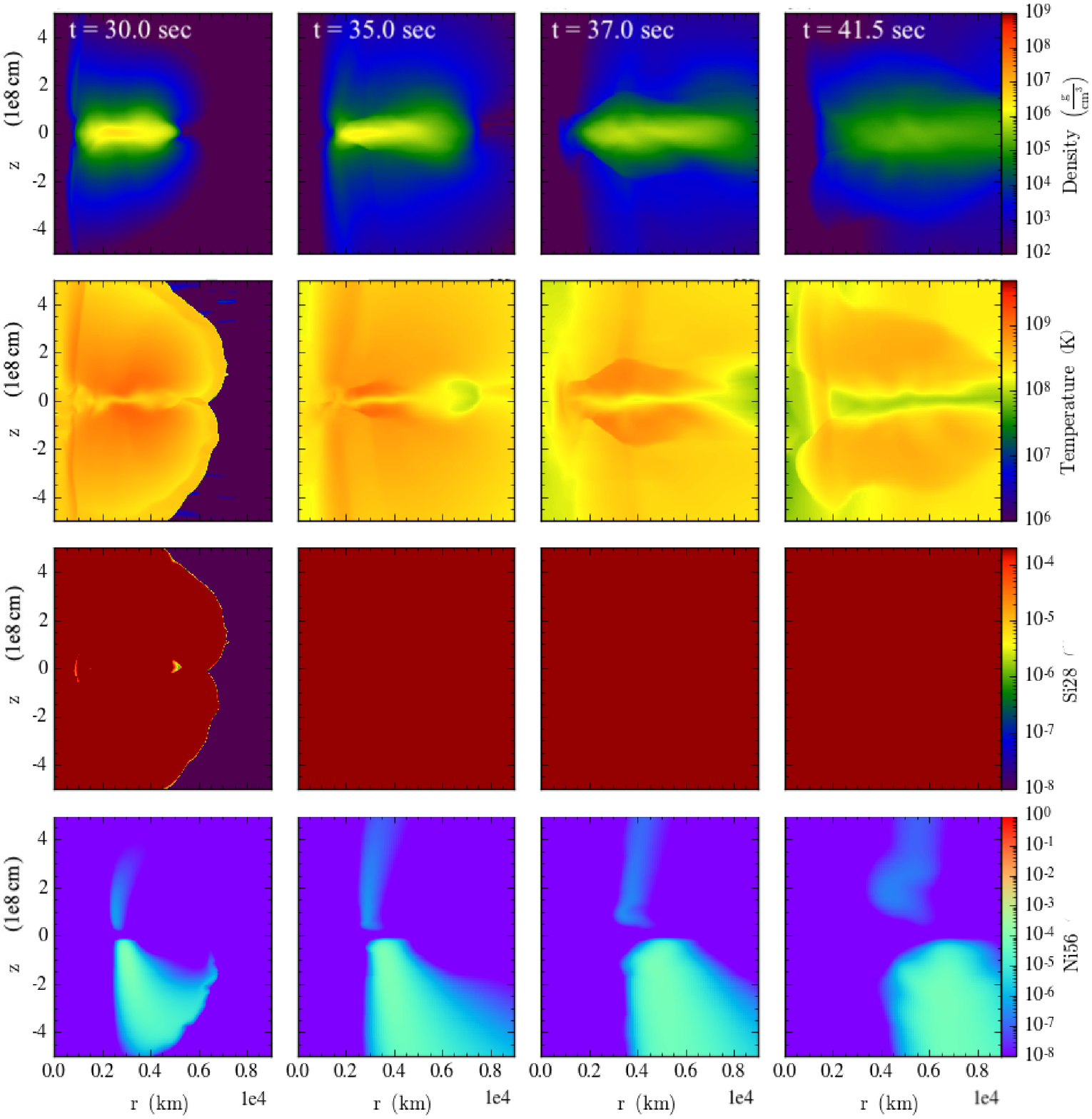}
\caption{Early evolution of model E20-h. Snapshots show density, temperature, and $^{28}$Si as well as $^{56}$Ni mass fractions in the meridional plane ($\bar\rho$,$z$) during the first $\lesssim 1.5$ viscous timescales. Production of $^{28}$Si and $^{56}$Ni indicate significant nuclear burning that develops into a global detonation of the accretion disk and leads to the production of iron-group nuclei.}
\label{fig:FLASHE20_h}
\end{figure*}

For models of the `h' category, we find that detonations typically occur on a timescale $t\lesssim t_{\rm visc}$ (cf.~Tab.~\ref{tab:viscositytab} for a compilation of timescales). Figure \ref{fig:FLASHE20_h} shows snapshots during the evolution of the E20-h model, a representative case for this class.

Detonations typically originate from radii $r_{\rm det}$ close to the inner edge of the initial torus (cf.~Tab.~\ref{tab:torusenergytab}) and then quickly spread radially across the midplane of the torus as well as vertically. In the case of E20-h, $r_{\rm det}\approx 1\times 10^{8}$\,cm. Nuclear burning is evident early on in the evolution ($t>20$\,s) from the production of $^{28}$Si via He, C, and O burning (cf.~Fig.~\ref{fig:FLASHE20_h}). The associated increase in temperature quickly triggers Si burning and production of heavier nuclei up to $^{56}$Ni (cf.~the lower panels of Fig.~\ref{fig:FLASHE20_h}). A prominent self-sustained strong detonation shock develops at around $t\approx 21-22$\,s, which originates close to the midplane and then propagates in the $\pm z$ and lateral directions and quickly develops into a global detonation of the accretion disk. We shall discuss the ignition process in more detail below.

Nuclear burning in the disrupted accretion torus of E20-h continues until roughly $t \sim 2.5\, t_{\rm visc}\approx 75$\,s, when most reactions start to freeze out (Fig.~\ref{fig:energytotal_E20h}). During the explosive phase the temperature and density in the disk midplane reach values above $3\times 10^{8}$\,K and $10^{6}\,\text{g}\,\text{cm}^{-3}$, respectively, with peak values up to $\approx\!3\times 10^9$\,K and $> 10^7\,\text{g}\,\text{cm}^{-3}$, and a cumulative nuclear energy release larger than the internal energy (cf.~Figs.~\ref{fig:FLASHE20_h}, \ref{fig:energytotal_E20h}). The total nuclear energy released during the evolution reaches $\approx\! 10^{49}$\,erg based on the 19-isotope network employed here.

The accretion rate peaks similarly to E20-l at $\dot{M}_{\rm acc}\simeq 2\times10^{-2}\,M_{\odot}\,\text{s}^{-1}$ at $t\approx t_{\rm visc}$ and subsequently transitions into a power-law decay as expected for viscous spreading (cf.~Fig.~\ref{fig:Mdot_E20}). The total material unbound from the BH-disk system as a result of the detonation and viscous outflows combined amounts to $\approx\!6.6\times 10^{-2}\,M_\odot$, roughly $67\%$ of the total cumulative outflows (cf.~Fig.~\ref{fig:CDF_massloss_E20h} and Tab.~\ref{tab:torusenergytab}). The resulting kinetic energies of the outflows are significant, reaching $5\times 10^{50}$\,erg, which may prevent further accretion of envelope material onto the disk (cf.~bottom panel of Fig.~\ref{fig:energytotal_E20h} and Tab.~\ref{tab:torusenergytab}). In general, the kinetic energies of outflows of the `h' models are sufficiently large to unbind the remaining outer envelopes of the collapsing progenitor star (see Tab.~\ref{tab:torusenergytab} for a compilation of binding energies and Sec.~\ref{sec:GRB_SNe} for more detailed discussion).

\begin{figure}[h]
\centering
\includegraphics[width=0.99\linewidth]{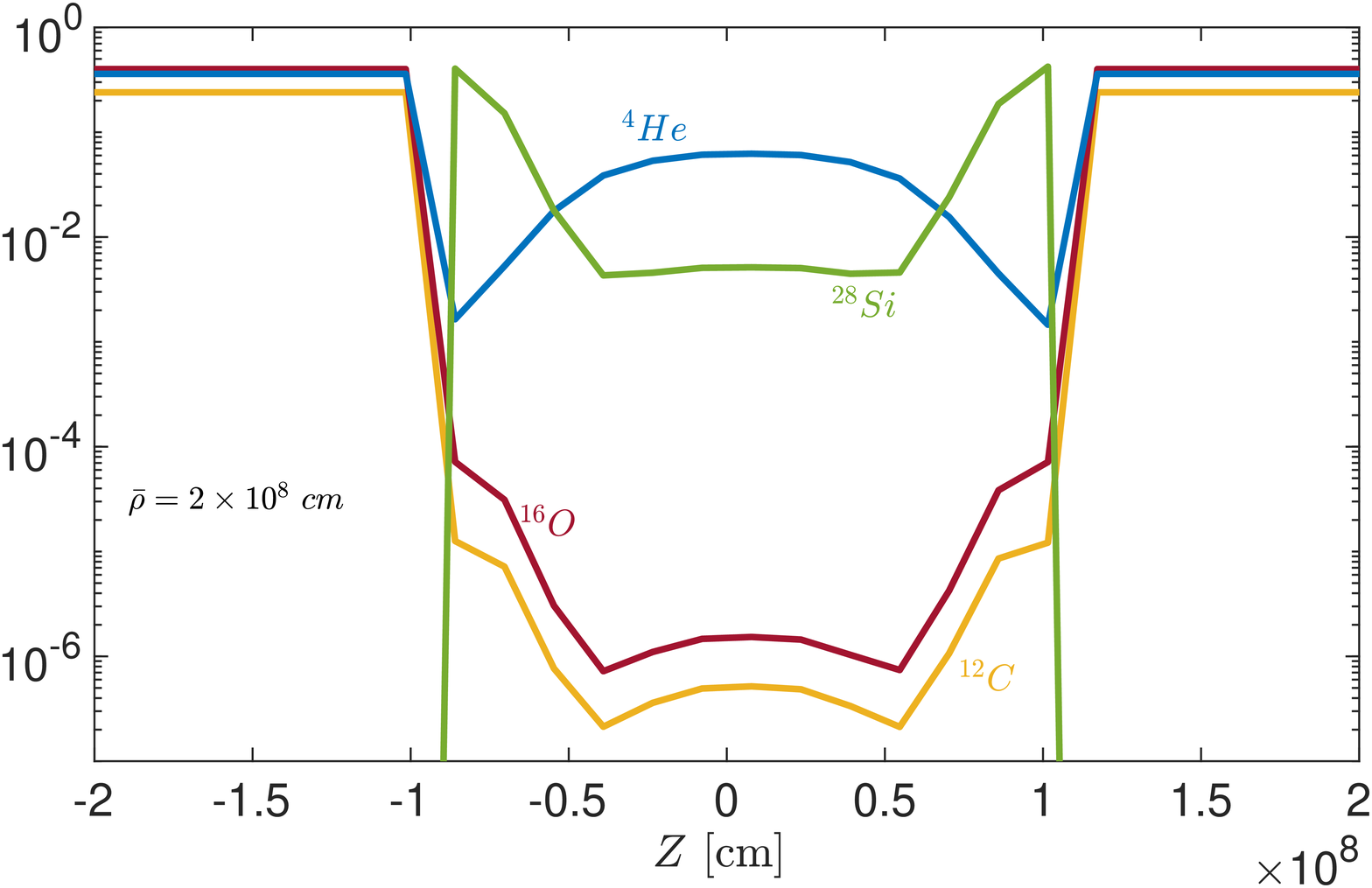}
\includegraphics[width=0.99\linewidth]{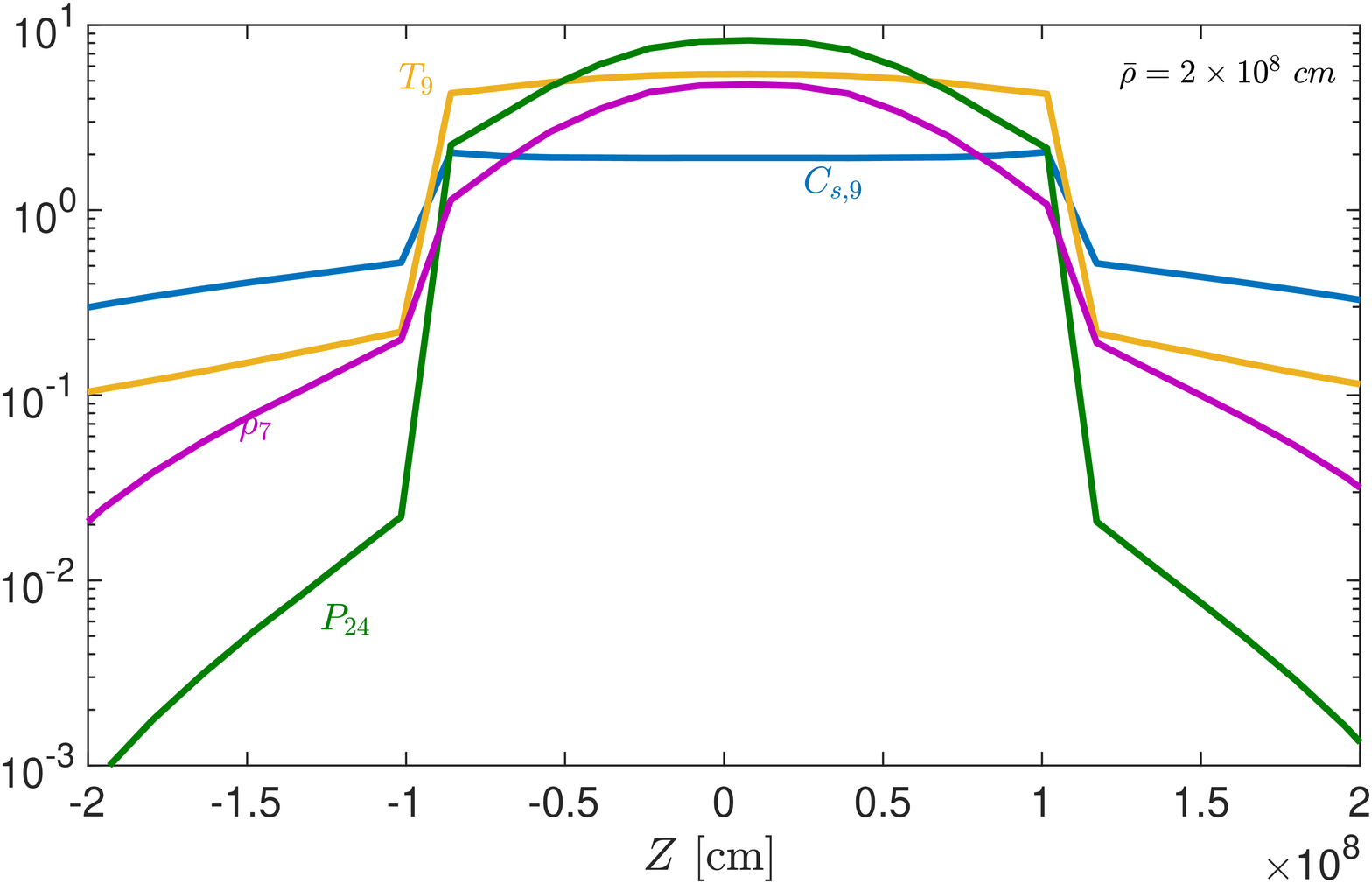}
\caption{Vertical sections through the E20-h accretion disk as a function of height $z$ at fixed radius $\bar{\rho}=2\times 10^{8}$\,cm roughly 1\,ms before the onset of detonation ($t\approx21.63$\,s) that develops around $z\approx \pm 1\times 10^{8}$\,cm. Top:
mass fractions of $^{4}$He, $^{12}$C, $^{16}$O, and $^{28}$Si. Bottom: pressure $p$\,[$10^{24}$\,erg\,cm$^{-3}$], temperature $T$\,[$10^{9}$\,K], density $\rho$ [$10^{7}$\,g\,cm$^{-3}$], and the sound speed $c_{\rm s}$\,[$10^9$\,cm\,s$^{-1}$].}
\label{fig:sections_z_E20h}
\end{figure}

\begin{figure}[h]
\centering
\includegraphics[width=0.99\linewidth]{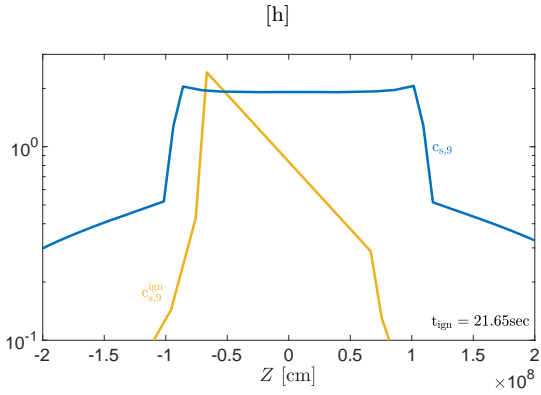}
\caption{Vertical sections through the E20-h accretion disk as a function of height $z$ at fixed radius $\bar{\rho}=2\times 10^{8}$\,cm at the onset of detonation ($t\approx21.65$\,s), showing that the criterion for spontaneous ignition Eq.~\eqref{eq:spontaneous_ignition}, $c_{\rm s}\sim c^{\rm ign}_{\rm s}\equiv |d t_{\rm ign}/dz|^{-1}$ is first satisfied for the burning front at $z\approx - 1\times 10^8$\,cm in this vertical slice. The adiabatic sound speed $c_{\rm s}$ and $c^{\rm ign}_{\rm s}$ are shown in units of $10^9$\,cm\,s$^{-1}$.}
\label{fig:E20h_csign}
\end{figure}

\begin{figure}
\centering
\includegraphics[width=0.99\linewidth]{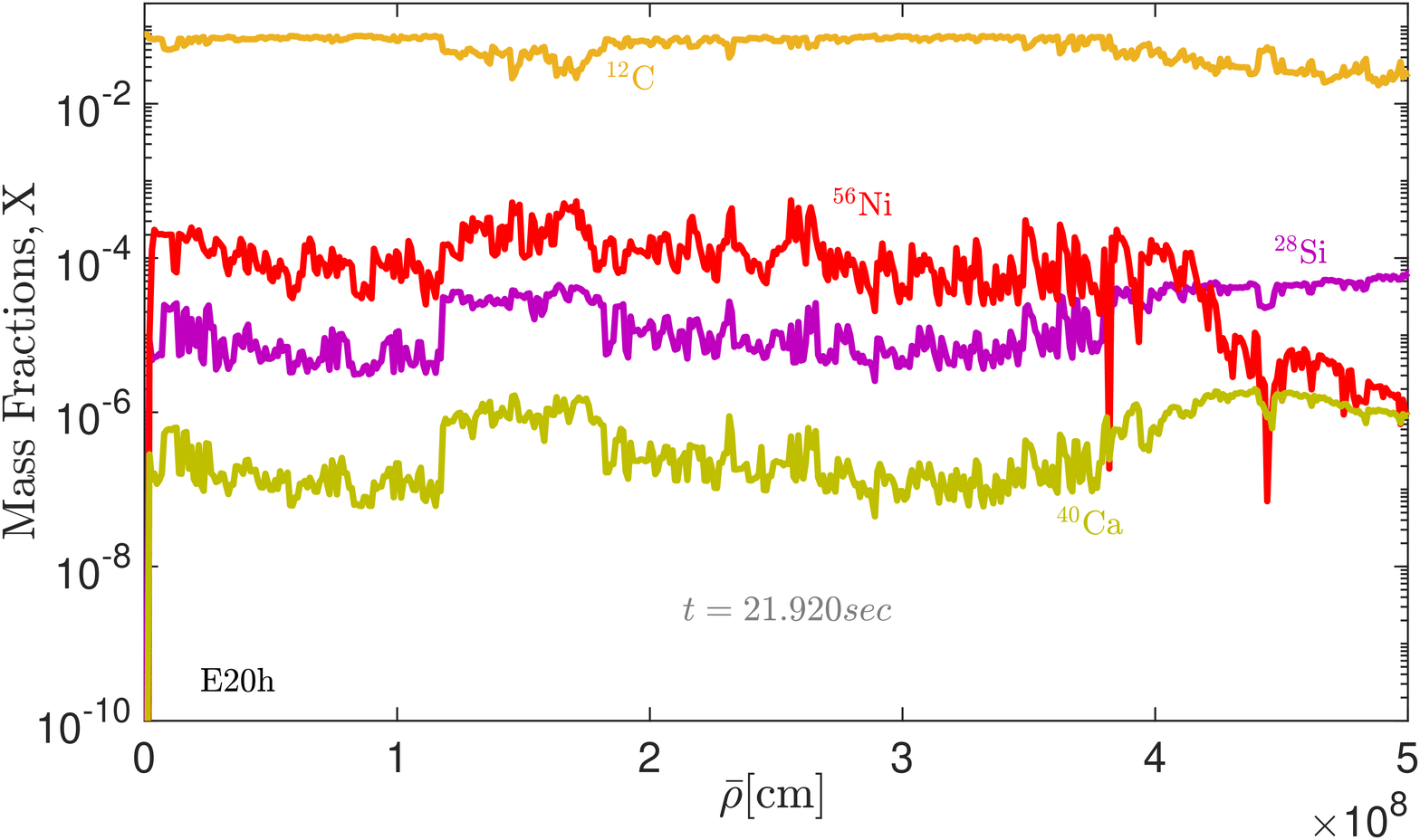}
\includegraphics[width=0.99\linewidth]{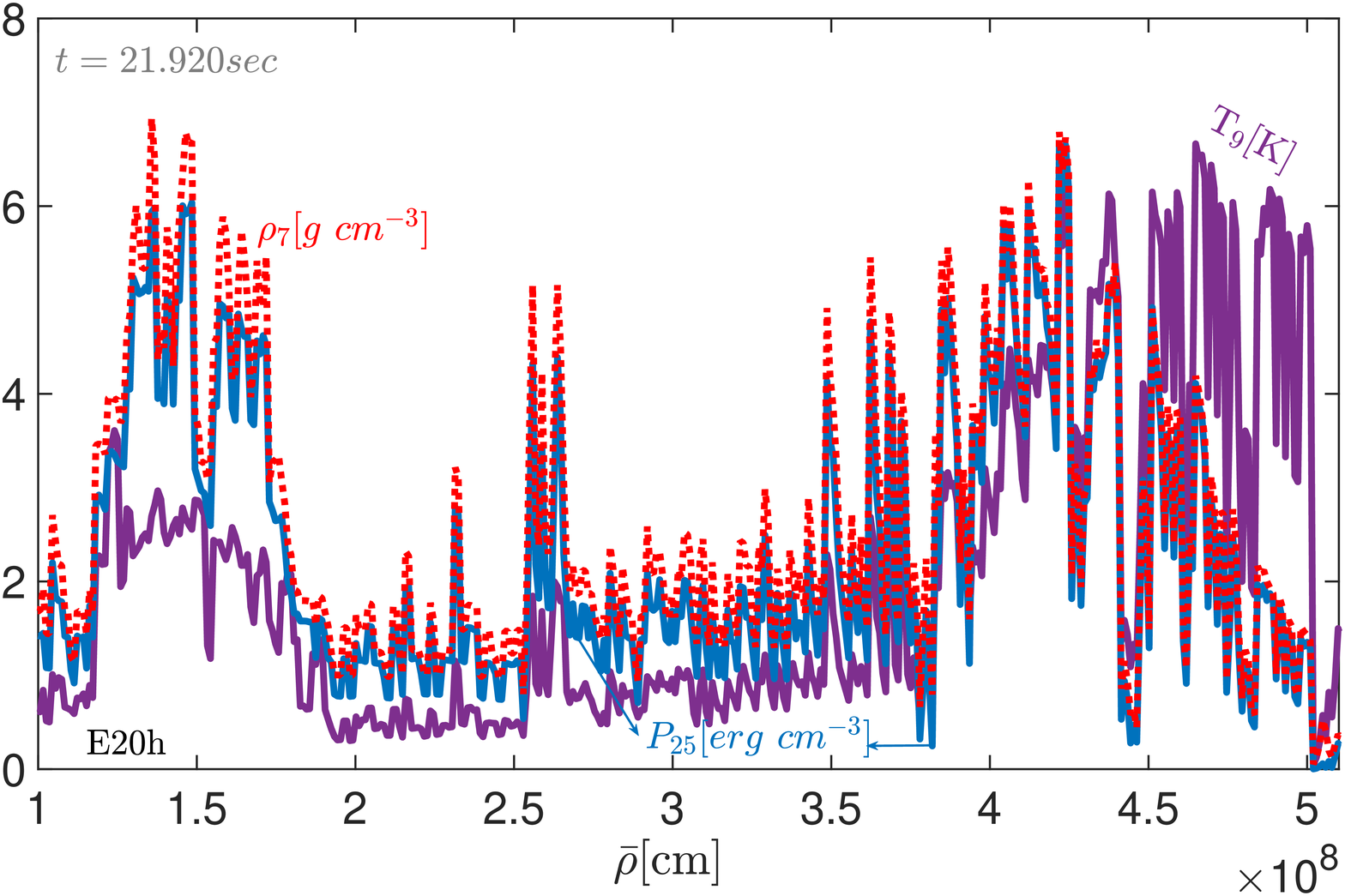}
\caption{Height-averaged radial profiles of various quantities for the E20-h disk at $t=21.920$\,s, just after a the global detonation event has been initiated. Top: mass fractions of $^{12}$C, $^{28}$Si, $^{40}$Ca, and $^{56}$Ni, indicating the synthesis of elements up to $^{56}$Ni behind the shock front. Bottom: pressure $p$\,[$10^{24}$\,erg\,cm$^{-3}$], temperature $T$\,[$10^{9}$\,K], and density $\rho$ [$10^{7}$\,g\,cm$^{-3}$].
}
\label{fig:r-secs_det_E20h}
\end{figure}

\begin{figure}
\centering
\includegraphics[width=0.99\linewidth]{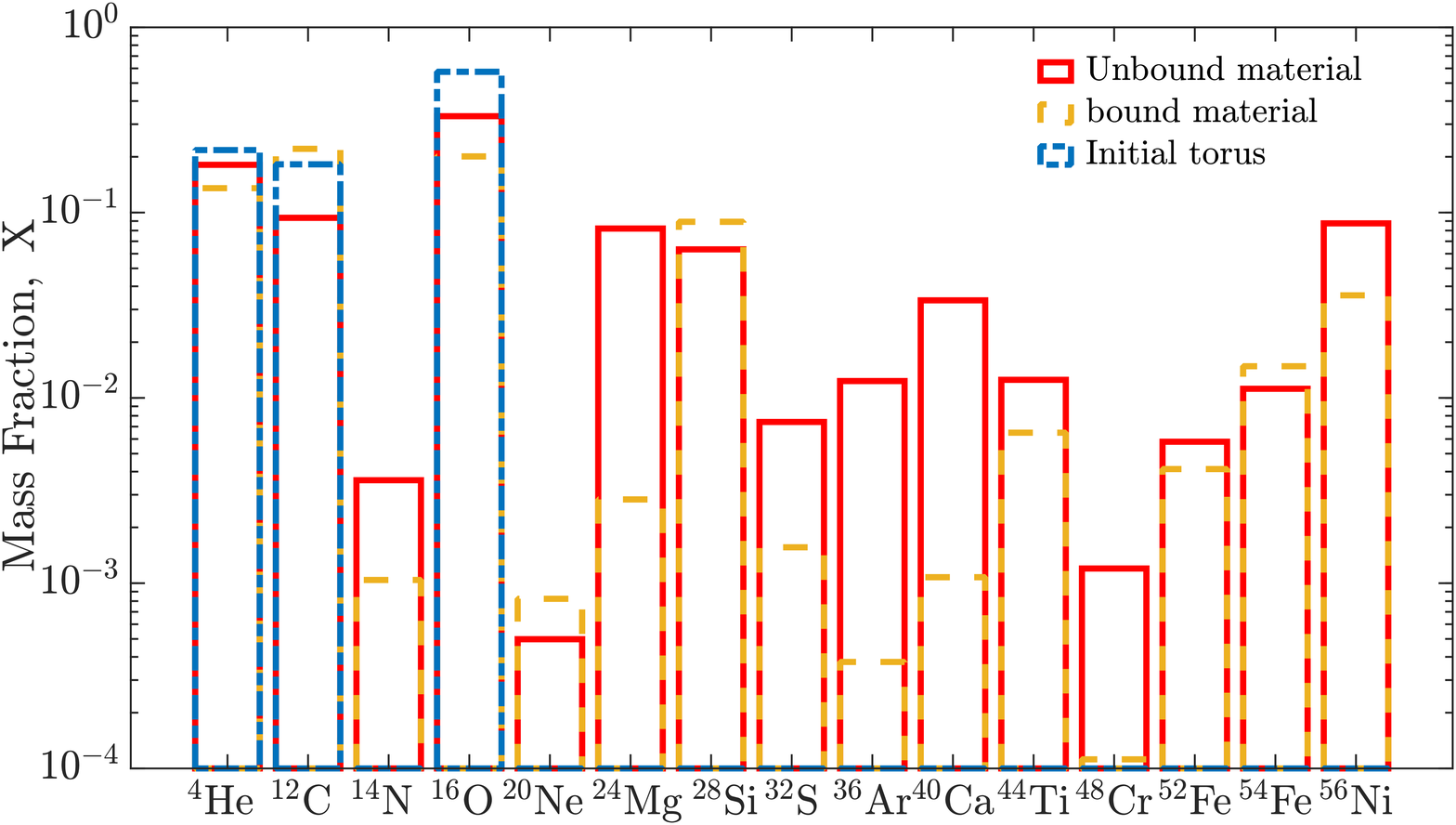}
\caption{Mass fractions of various nuclei for model E20-h extracted at the end of the simulation run and separated into unbound material (ejecta, $M_{\rm ej}$; red) and bound material (yellow), consisting of bound outflows and bound disk material ($M_{{\rm out},\infty}$; cf.~Sec.~\ref{sec:diagnostics}). Shown for reference are also the corresponding mass fractions of the initial disk material before detonation.}
\label{fig:X_elements_E20h}
\end{figure}

\begin{table*}
\begin{centering}
\begin{tabular}{cccccccc}
\hline 
matter type &  & E20-m  & G15B-m  & F15B-m & E20-h  & G15B-h  & F15B-h \tabularnewline

 & & $[M_{\odot}]$ & $[M_{\odot}]$ & $[M_{\odot}]$ & $[M_{\odot}]$ & $[M_{\odot}]$ & $[M_{\odot}]$\tabularnewline
\hline 
 
ejected & $^{48}$Cr & $6.8\times 10^{-5}$  & $2.34\times 10^{-5}$  & $1.96\times 10^{-5}$ & $9.4\times 10^{-5}$  & $8.2\times 10^{-5}$  & $6.7\times 10^{-5}$\tabularnewline
 
 & $^{52}$Fe & $8.4\times 10^{-4}$  & $6.6\times 10^{-4}$  & $6.2\times 10^{-4}$ & $4.3\times 10^{-4}$  & $3.8\times 10^{-4}$  & $2.7\times 10^{-4}$\tabularnewline
 
 & $^{54}$Fe & $2.6\times 10^{-4}$  & $3.2\times 10^{-4}$  & $2.2\times 10^{-4}$ & $7.4\times 10^{-4}$  & $1.65\times 10^{-4}$  & $5.9\times 10^{-4}$ \tabularnewline

 & $^{56}$Ni & $5.4\times 10^{-3}$  & $5.2\times 10^{-3}$  & $3.2\times 10^{-3}$ & $6.9\times 10^{-3}$  & $7.0\times 10^{-3}$  & $4.2\times 10^{-3}$\tabularnewline
\hline 
outflow & $^{56}$Ni & $7.0\times 10^{-3}$  & $5.9\times 10^{-4}$  & $6.2\times 10^{-4}$ & $4.26\times 10^{-3}$  & $2.75\times 10^{-3}$  & $2.84\times 10^{-3}$\tabularnewline
\hline 

$M_{{\rm out}, \infty}$ & $^{56}$Ni & $1.37\times 10^{-1}$  & $3.51\times 10^{-2}$  & $3.58\times 10^{-2}$ & $1.11\times 10^{-1}$  & $2.83\times 10^{-2}$  & $2.91\times 10^{-2}$ \tabularnewline
\hline

\end{tabular}
\par\end{centering}
\caption{Iron-group ejecta from all detonating accretion disk models simulated here, with most of the mass being in $^{56}$Ni. Also listed are the amounts of additional $^{56}$Ni-material in bound outflows and the total amount of $^{56}$Ni synthesized that will eventually be unbound in outflows ($M_{{\rm out}, \infty}$; cf.~Sec.~\ref{sec:diagnostics}).}
\label{tab:X_models}
\end{table*}

The detonation process starts with strong nuclear burning in the disk midplane at radii $r_{\rm det}\approx (1-4) \times 10^8$\,cm. Figure \ref{fig:sections_z_E20h} shows various quantities as a function of height $z$ in this radius regime just before ($\approx\!1$\,ms) detonation. Strong burning behind burning fronts at $z\approx \pm 1 \times 10^{8}$\,cm is evident from Fig.~\ref{fig:sections_z_E20h}, including triple-alpha ($^4$He($2\alpha,\gamma$)$^{12}$C) reactions, carbon burning (via $^{12}$C($\alpha,\gamma$)$^{16}$O, $^{12}$C($^{12}$C,$\gamma$)$^{24}$Mg) and oxygen burning (via $^{16}$O($\alpha,\gamma$)$^{20}$Ne, $^{16}$O($^{16}$O,2$\alpha$)$^{28}$Si, $^{16}$O($^{16}$O, $\gamma$)$^{32}$S etc.), as well as further alpha captures, resulting in the production of $^{28}$Si in the disk midplane.

The burning fronts spontaneously transition into a detonation shock in the $\pm z$-directions once different locations in the burning region become supersonically disconnected in terms of the ignition timescale $t_{\rm ign}$ \citep{Zeldovich1970}. We formulate this condition as \citep{Woosley1990,1997ApJ...475..740N,2013ApJ...763..108F}
\begin{equation}
	\Theta_{\rm ign} \equiv \frac{1}{c_s \nabla t_{\rm ign}} > 1. \label{eq:spontaneous_ignition}
\end{equation}
In order to illustrate the onset of detonation, we apply this criterion to the burning fronts across $z\approx \pm 1 \times 10^{8}$\,cm shown in Fig.~\ref{fig:sections_z_E20h} by approximating
\begin{equation}
	\nabla t_{\rm ign} \approx \frac{d}{dz}\left(\frac{e_{\rm int}}{\beta \dot{Q}_{\rm nuc}}\right).
\end{equation}
Figure \ref{fig:E20h_csign} illustrates that this criterion is approximately satisfied at the onset of the detonation. Since $t_{\rm ign}\propto c_v T^{-(\beta -1)}$ (cf.~Sec.~\ref{sec:diagnostics}) and assuming temperature-sensitive reactions, a sufficiently shallow temperature gradient is required to satisfy the criterion for spontaneous initiation of a detonation Eq.~\eqref{eq:spontaneous_ignition} \citep{Zeldovich1970}. This is mediated here by vigorous turbulence due to viscosity in our accretion disks; shallow temperature gradients are evident from Fig.~\ref{fig:sections_z_E20h}. Model E20-h shows a somewhat asymmetric onset of the detonation with respect to $z=0$ (cf.~Figs.~\ref{fig:FLASHE20_h} and \ref{fig:E20h_csign}). Effects of strong self-gravity can induce small local variations in density and temperature, which, given the high sensitivity of the ignition timescale to these quantities (cf.~Eq.~\eqref{eq:t_ign}), may lead to somewhat asymmetric run-away behavior. For all other models in which effects of self-gravity are expected to be less significant due to smaller disk masses (the detonating models of G15B and F15B), we find symmetric detonation behavior and symmetric production of $^{56}$Ni with respect to the disk midplane.

Behind the newly formed, vertically expanding detonation fronts, the temperature increases up to $\approx\!3\times 10^9$\,K, leading to the production of significant amounts of $^{56}$Ni and intermediate-mass elements. Figure \ref{fig:r-secs_det_E20h} shows height-integrated radial profiles of mass fractions for various elements just after the detonation event has been globally initiated. We report a detailed break-down of total mass fractions for most elements in Fig.~\ref{fig:X_elements_E20h}. Unbound ejecta material typically consists of higher amounts of heavy nuclei (burning products) than bound disk or bound outflow material, as expected. Both bound and unbound material still contain significant amounts of fuel, indicative of overall incomplete burning.

Finally, we point out that very similar detonation dynamics, characteristics, and abundance distributions apply to all other simulated models with prompt detonations. This also applies to our `m' models, with the exception that detonations are initiated somewhat later. For reference, detailed amounts of synthesized iron-group elements are listed in Tabs.~\ref{tab:torusenergytab} and \ref{tab:X_models}. We note that as part of the detonations itself, typically ${\rm few}\times 10^{-3}M_\odot$ of $^{56}$Ni can be directly ejected from the BH-disk system, while an additional ${\rm few}\times 10^{-2}-0.1M_\odot$ of $^{56}$Ni can be unbound later through viscous outflows from the disk.

\subsection{Survey of disk models}
\label{sec:survey_models}

Our simulated disk models (see Tabs.~\ref{tab:models}, \ref{tab:torusenergytab}, \ref{tab:X_models}, and \ref{tab:viscositytab}) for characteristics) are part of a large parameter space of collapsar accretion disks. The detailed outcome in terms of nuclear burning for a given collapsar scenario strongly depends on the progenitor structure, its composition and rotational profile (cf.~Sec.~\ref{sec:initial_models}), in particular. Nevertheless, assuming that the initial disk compositions of fuel found here across various progenitor models (see Tab.~\ref{tab:models}) are roughly representative of a larger population, the question of explodability effectively maps the large parameter space of possible collapsar scenarios to an effectively two-dimensional parameter space for the resulting accretion disks, as we shall argue below.
\begin{figure*}
\centering
\includegraphics[width=0.8\linewidth]{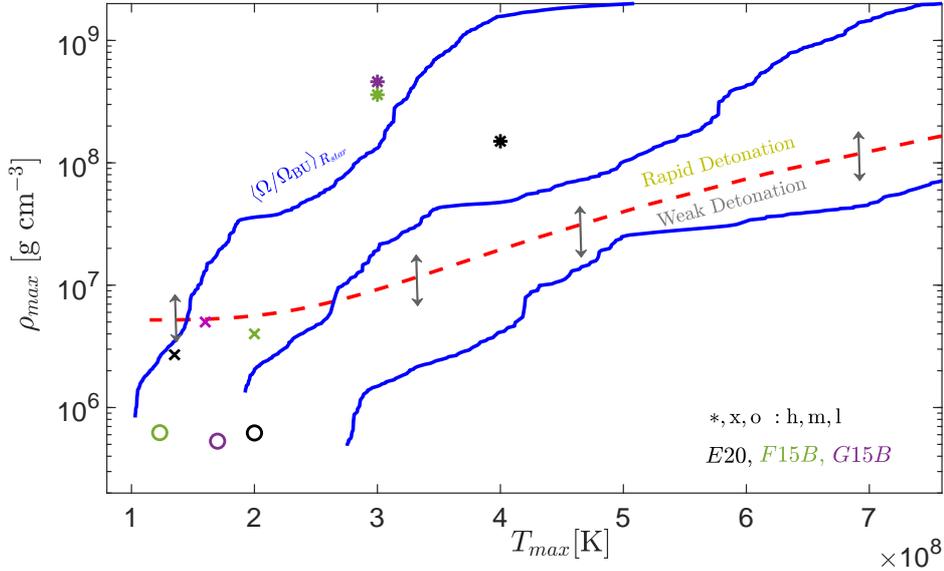}
\caption{Parameter space of collapsar accretion disks mapped onto the plane of maximum temperature and density reached initially. The models numerically evolved here are indicated by individual points. The red line represents an approximate model for the boundary between promptly detonating models and those with delayed or no detonations (see the text for details). This approximate estimate for prompt detonation is remarkably consistent with the detonation behavior of the explicitly evolved accretion disks. Blue lines indicate contours of tori solutions (see Sec.~\ref{sec:initial_models} and Appendix \ref{app:tori_solutions}) obtained by varying the angular momentum power-law parameter $p$ (Eq.~\eqref{eq:j_profile}) for fixed surface angular momentum $\Omega/\Omega_{\rm BU}$ of a fixed progenitor model, illustrating that a large fraction of the $\rho_{\rm max}$-$T_{\rm max}$ parameter space can be obtained from the progenitor models considered here. The blue lines from left to right result from varying models E20-h, E20-m, and E20-l, respectively. These lines do no directly pass through the points indicating the simulation runs, as the one-parameter families of solutions are computed with a simplified EOS and the simulation data is extracted after relaxation under the influence of self-gravity.}

\label{fig:All_detonation}
\end{figure*}

Figure \ref{fig:All_detonation} charts the territory of collapsar accretion disks according to the maximum temperature $T_{\rm max}$ and density $\rho_{\rm max}$ reached in the disk initially. Our simulated models belonging to the `prompt detonation' (`h'), `delayed detonation' (`m') and `no detonation' (`l') categories cluster in distinct regions of this parameter space, respectively. We note that these values can be roughly estimated from the collapsar fallback model and torus initialization method described in Sec.~\ref{sec:initial_models} without performing evolution simulations of the resulting disks (see Tab.~\ref{tab:models} for corresponding values of $T_{\rm max}$ and $\rho_{\rm max}$). Non-detonating models typically occupy the low density $\lesssim\!10^6$\,g\,cm$^{-3}$, delayed detonations the intermediate density $\lesssim\!10^6-10^7$\,g\,cm$^{-3}$, and prompt detonations the high-density $\gtrsim\!10^7$\,g\,cm$^{-3}$ regime, with a weaker temperature dependence. As all collapsar disks are already in a temperature regime $T_{\rm max}>1\times 10^8$\,K conducive to nuclear burning due to the $e_{\rm tot}=0$ condition (Eq.~\eqref{eq:etot}), the main factor in distinguishing between detonating and non-detonating models are the maximum densities (compactnesses) and, to a lesser extend, the detailed compositions of the disks. The blue lines in Fig.~\ref{fig:All_detonation} represent one-parameter families of tori solutions obtained by varying the power-law exponent $p$ of the progenitor's angular momentum profile for fixed surface angular momentum $\Omega/\Omega_{\rm BU}$ of the progenitor star, as described in Sec.~\ref{sec:initial_models} and Appendix \ref{app:tori_solutions}. Changing the power-law exponent $p$ effectively modifies the inner radius $r_{\rm in}$ and thus the compactness of the resulting torus. This is to illustrate that essentially the entire parameter space in terms of $T_{\rm max}$ and $\rho_{\rm max}$ can be obtained from the collapsar progenitor models considered here, depending on the details of the angular momentum profiles.

Given a progenitor model, one can roughly estimate whether or not the resulting accretion disk will lead to a prompt detonation. To this end, let us consider a local detonation wave; across the shock front, the Hugoniot relations of mass, momentum, and energy conservation read
\begin{eqnarray}
\rho_1 D &=& \rho_2 (D-v_2), \label{eq:Hugoniot1} \\
p_1+\rho_1 D^2&=&p_2+\rho_2 (D-v_2)^2, \\
e_{{\rm int}, 1}+\frac{p_1}{\rho_1}+\frac{D^2}{2}&=&e_{{\rm int}, 2}+\frac{p_2}{\rho_2}+\frac{(D-v_2)^2}{2}+Q. \label{eq:Hugoniot3}
\end{eqnarray}
Here, we neglect fluid viscosity, $D$ denotes the velocity of the detonation shock in a reference frame in which unburnt material is at rest; subscripts 1 and 2 refer to upstream (fuel) and downstream (ash) material. In the context of our detonating models, one may take $D$ to be the velocity in $z$-direction of the detonation front, assuming that the detonation is initiated in the disk midplane and expands vertically as well as laterally (cf.~Sec.~\ref{sec:detonation}). Furthermore, $Q$ denotes the nuclear energy release per unit mass due to nuclear reactions behind the shock front. The slowest possible self-sustained steady-state detonation shock $D_{\rm CJ} > c_{\rm 1}$ is determined by the Jouguet condition $D - v_2 = c_{{\rm s},2}$. 
For a polytropic gas with sound speed $c_{\rm s}^2=\gamma p/\rho$ and assuming a strong detonation wave (i.e., $Q\gg e_{{\rm int}, 1}$), the Chapman-Jouguet velocity obtained from Eqs.~\eqref{eq:Hugoniot1}--\eqref{eq:Hugoniot3} simplifies to \citep{Landau1987fluid}
\begin{equation}
    D_{\rm CJ}=\sqrt{2(\gamma^2-1)Q}. \label{eq:D_CJ}
\end{equation}
A necessary condition for a strong detonation is 
\begin{equation}
    \mathcal{M}_{\rm CJ}\simeq \frac{D_{\rm CJ}}{c_{\rm s,1}} > 1, \label{eq:detonation_criterion_approx}
\end{equation}
where $\mathcal{M}$ is the Mach number of the shock in rest frame of the fuel. For given density, temperature, and composition $(\rho,T,X_i)$ of an accretion disk (the fuel), one can directly evaluate Eq.~\eqref{eq:detonation_criterion_approx}.

We employ this criterion as a rough estimate to distinguish between detonating and non-detonating models. To this end, we assume that the detonation is chiefly triggered by helium burning via the triple-alpha reaction. For given $\rho_{\rm max}$ and $T_{\rm max}$, we employ the minimum and maximum $^4$He mass fractions among our models (Tab.~\ref{tab:models}), together with the corresponding energy release $Q$ for a steady-state detonation wave from \cite{Gamezo99}, to estimate upper and lower bounds for $D_{\rm CJ}$. The adiabatic constant $\gamma$ is obtained using the full EOS information. Employing an average estimate of $D_{\rm CJ}$ among different $^4$He mass fractions, and Monte-Carlo sampling over the parameter space in $\rho_{\rm max}$, $T_{\rm max}$, we find an approximate boundary between promptly detonating models and delayed or non-detonating models, shown as a red line in Fig.~\ref{fig:All_detonation}. 

This approximate estimate for prompt detonation is remarkably consistent with the explicitly evolved accretion disks. All `h' models clearly lie in the detonation regime, while the `m' models reside in a `transition zone' close to the rapid detonation boundary. The latter models, given the comparatively lower maximum densities, require a moderate increase in temperature through viscous heating for the ignition criteria to be satisfied. We find that this leads to delayed but otherwise very similar detonation phenomena as compared with the `h' models. We refer to Tabs.~\ref{tab:models} and \ref{tab:X_models} for a summary of burning products.

\section{Discussion}
\label{sec:discussion}

\subsection{Dependence on Viscosity and Resolution}
\label{sec:viscosity}

Since our simulations are hydrodynamic and are conducted in axisymmetry, angular momentum transport is taken into account in a parametrized way through the Shakura-Sunyaev viscosity coefficient $\alpha$ (cf.~Sec.~\ref{sec:code}), with fiducial value $\alpha = 0.1$. We explore the sensitivity of our results with respect to the choice of $\alpha$; see Tab.~\ref{tab:viscositytab} for a summary of all cases simulated. Across a wide range of values for $\alpha$ we find similar evolutionary scenarios for all families of disk models considered here. While `l' models never detonate, regardless of the value of $\alpha$, `h' models detonate for $\alpha \gtrsim 10^{-3}$, while `m' models detonate for $\alpha \gtrsim 10^{-2}$.

\begin{table*}
\begin{centering}
\begin{tabular}{cccccccccccccc}
\hline
$\alpha$-viscosity & $0.5$ & $0.3$ & $0.15$ & $0.1$ & $0.08$ & $0.05$ & $0.01$ & $0.005$ & $0.0001$ & $t_{\rm sim}[{\rm s}]$ & $t_{\rm orb}[{\rm s}]$ & $t_{\rm visc}[{\rm s}]$ & $t_{\rm circ,out}[{\rm s}]$\tabularnewline
\hline 
E20-h & $\surd$ & $\surd$ & $\surd$ & $\surd$ & $\surd$ & $\surd$ & $\surd$ & $-$ & $\surd$ & $225$ & $0.268$ & $30.821$ & $14.306$\tabularnewline 

G15B-h & $-$ & $-$ & $\surd$ & $\surd$ & $\surd$ & $-$ & $\surd$ & $-$ & $-$ & $210 $ & $0.319$ & $33.878$ & $16.241$\tabularnewline
 
F15B-h & $-$ & $\surd$ & $-$ & $\surd$ & $-$ & $\surd$ & $-$ & $\surd$ & $-$  & $215 $ & $0.313$ & $36.435$ & $17.089$\tabularnewline

Detonate? & $\rm Y$ & $\rm Y$ & $\rm Y$ & $\rm Y$ & $\rm Y$ & $\rm Y$ & $\rm Y$ & $\rm Y$ & $\rm N$ & $-$ & $-$ & $-$ & $-$\tabularnewline
\hline 
 
E20-l & $-$ & $-$ & $\surd$ & $\surd$ & $-$ & $-$ & $-$ & $\surd$ & $-$ & $350 $ & $0.359$ & $41.028$ & $20.813$\tabularnewline 

G15B-l & $\surd$ & $-$ & $-$ & $\surd$ & $-$ & $\surd$ & $-$ & $-$ & $-$ & $330 $ & $0.291$ & $41.315$ & $21.557$\tabularnewline

F15B-l & $\surd$ & $-$ & $-$ & $\surd$ & $-$ & $-$ & $\surd$ & $-$ & $-$ & $355 $ & $0.398$ & $44.274$ & $24.046$\tabularnewline 

Detonate? & $\rm N$ & $-$ & $\rm N$ & $\rm N$ & $\rm N$ & $\rm N$ & $\rm N$ & $\rm N$ & $-$ & $-$ & $-$ & $-$ & $-$\tabularnewline
\hline 

E20-m & $\surd$ & $-$ & $\surd$ & $\surd$ & $-$ & $\surd$ & $-$ & $-$ & $-$ & $242 $ & $0.455$ & $39.808$ & $16.182$\tabularnewline 

G15B-m & $-$ & $\surd$ & $-$ & $\surd$ & $-$ & $-$ & $\surd$ & $\surd$ & $-$ & $252 $ & $0.441$ & $43.493$ & $19.760$\tabularnewline 
 
F15B-m & $-$ & $-$ & $-$ & $\surd$ & $-$ & $-$ & $-$ & $\surd$ & $\surd$ & $266 $ & $0.352$ & $44.771$ & $21.893$\tabularnewline 

Detonate? & $\rm Y$ & $\rm Y$ & $\rm Y$ & $\rm Y$ & $-$ & $\rm Y$ & $\rm N$ & $\rm N$ & $\rm N$ & $-$ & $-$ & $-$ & $-$\tabularnewline
\hline 

\end{tabular}
\par\end{centering}
\caption{Simulation runs performed across the $\alpha$-viscosity parameter space. Check-marks refer to setups that were simulated; detonation behavior is indicated by yes/no (Y/N) and applies to all simulated models of the corresponding column and model category (h, l, m). Also listed are the corresponding evolution times, orbital timescales, viscous timescales, and torus formation timescales for our fiducial setups ($\alpha=0.1$).}
\label{tab:viscositytab}
\end{table*}

\begin{figure}
\centering
\includegraphics[width=0.99\linewidth]{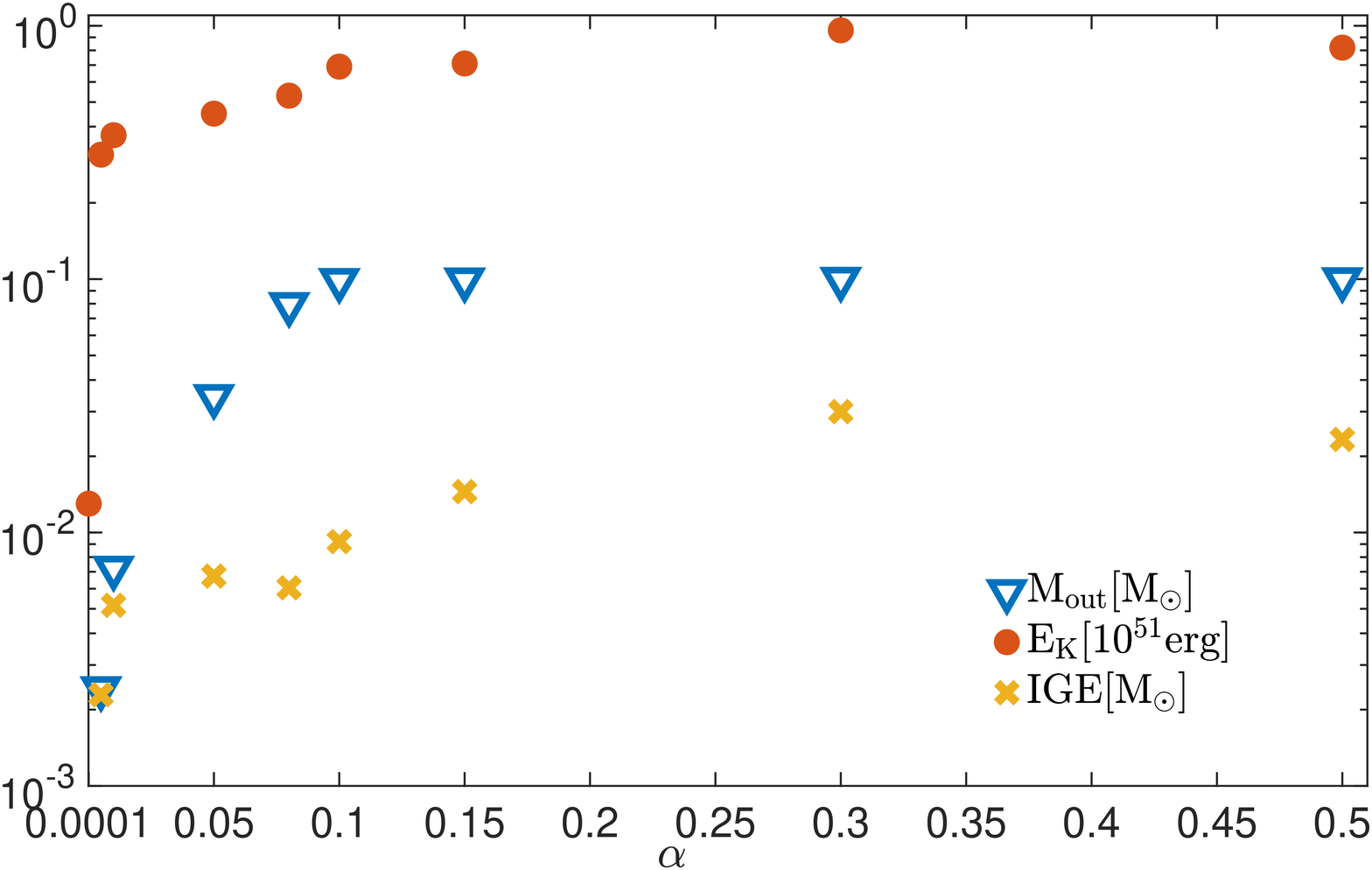}
\caption{Viscosity dependence of key observables for the detonating model E20-h: kinetic energy of outflowing material, total outflow mass, and fraction of iron-group elements thereof as a function of the $\alpha$-viscosity parameter. Data are extracted from various simulations performed with different values for the viscosity parameter but otherwise identical to the fiducial E20-h run.}
\label{fig:visco}
\end{figure}

\begin{figure}
\centering
\includegraphics[width=0.99\linewidth]{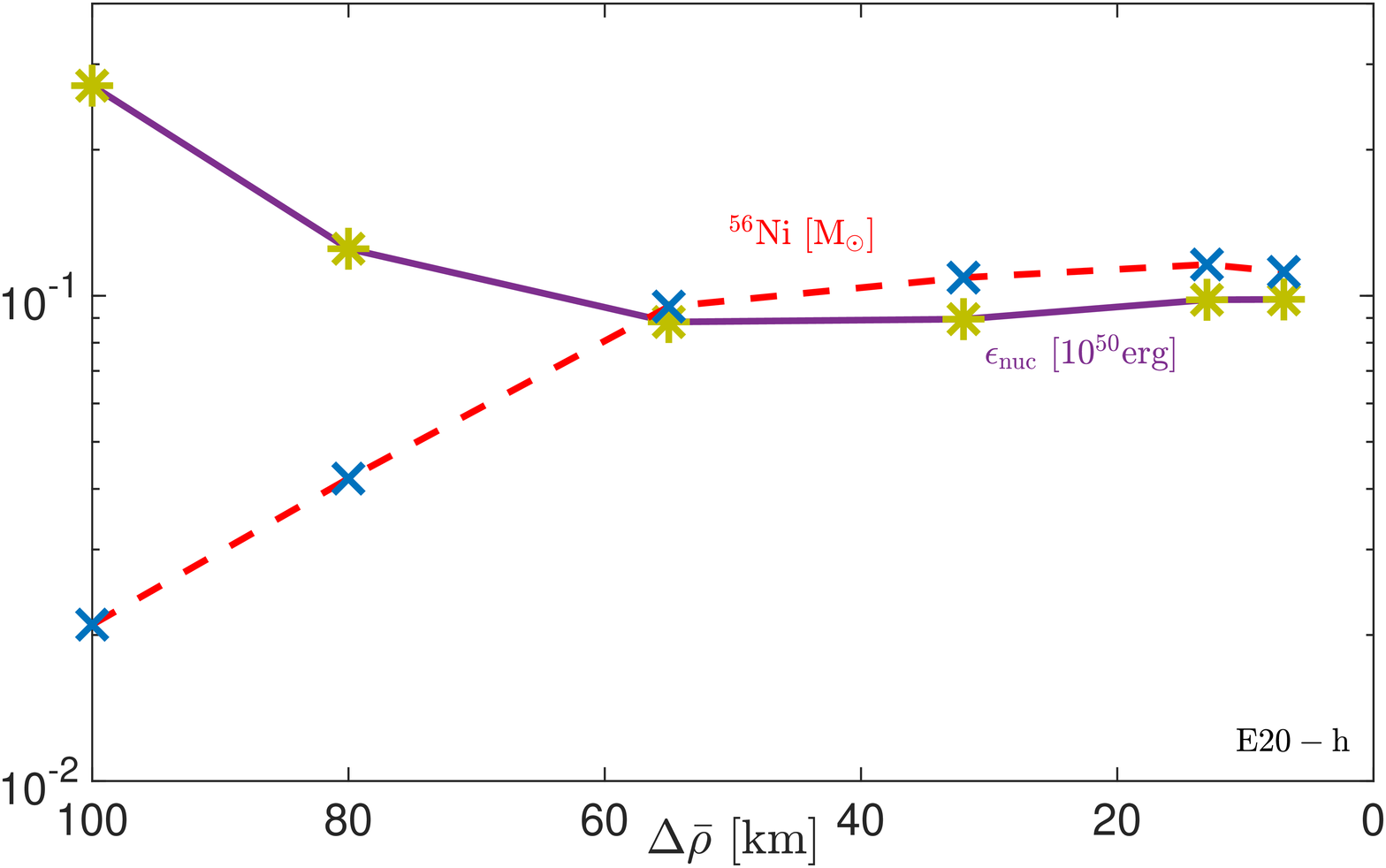}
\caption{Spatial resolution convergence test for model E20-h: total amount of synthesized $^{56}$Ni in units of $[\rm M_{\odot}]$ and the cumulative nuclear energy released in units of $\rm [10^{50}erg]$ as extracted from simulations performed with different spatial resolution $\Delta\bar{\rho}$ but otherwise identical to the fiducial E20-h run (corresponding to the smallest $\Delta\bar{\rho}$ shown here). Global quantities are converged to better than 5\% for the fiducial resolution.}
\label{fig:reso}
\end{figure}

Figure \ref{fig:visco} shows more detailed results from the viscosity study for model E20-h. Key global observables, such as the overall kinetic energy and mass of outflows and the total mass of synthesized iron-group elements remains remarkably constant over a wide range $\alpha\sim 0.05-0.5$ in which this models detonates. Similar findings apply to other models.

The absence of detonations for low values of $\alpha$ can be ascribed to two factors at play. For our `m' models, a moderate increase in temperature through viscous heating is required to mediate nuclear burning. For decreasing $\alpha$, such heat release is distributed over an increasing timescale $t_{\rm visc}\propto \alpha^{-1}$ (cf.~Eq.~\eqref{eq:tvisc}), while the density in the midplane decreases due to viscous spreading, eventually rendering nuclear burning impossible. Another aspect relevant to the promptly detonating models, in particular, is the fact that vigorous turbulence is required to trigger spontaneous detonations. Strong mixing is needed to generate shallow temperature gradients, such that the ignition condition $\Theta_{\rm ign}>1$ (Eq.~\eqref{eq:spontaneous_ignition}) can be satisfied (cf.~Sec.~\ref{sec:detonation}). However, in order to generate such shallow temperature gradients across a burning front, mixing must occur before complete burning of the fuel. This condition can be expressed as (cf.~\citealt{2013ApJ...763..108F})
\begin{equation}
	\frac{t_{\rm burn}}{t_{\rm turb}} = \mathcal{M}_{\rm turb} \Psi \beta \Theta_{\rm ign}^{-1} > 1. \label{eq:mixing_condition}
\end{equation}
Here, $\Psi$ is the ratio of nuclear energy released to the enthalpy of the fluid, $\beta$ is the effective temperature exponent of $\dot{Q}_{\rm nuc}$, and $\mathcal{M}_{\rm turb}= v_{\rm edd}/c_{\rm s}<1$ is the turbulent Mach number. Furthermore, $t_{\rm turb} = l_{\rm edd}/v_{\rm edd}$ is the turbulent eddy turn over time, with $l_{\rm edd}$ and $v_{\rm edd}$ the characteristic eddy length and velocity, respectively. Assuming that the ignition condition Eq.~\eqref{eq:spontaneous_ignition} is marginally satisfied, and given a burning process with energy release $\Psi$, Eq.~\eqref{eq:mixing_condition} shows that still sufficiently vigorous turbulence is required to trigger a detonation.

Although our results obtained with fiducial viscosity coefficient $\alpha=0.1$ apply over a wide range of values for $\alpha$, fully three-dimensional magnetohydrodynamic simulations are required to self-consistently account for angular momentum transport mediated by the magnetorotational instability and thus to assess whether the detonation behaviour found here is indeed a robust outcome of such collapsar accretion disks.

Finally, we assess the robustness of our results against changes in resolution. Figure \ref{fig:reso} reports results from a resolution study on the model E20-h, performed with identical setups except for spatial resolution. We find that the spatial resolution of 6-10\,km employed for our fiducial runs here is sufficient to guarantee convergence of key global quantities such as the total amount of $^{56}$Ni and the total cumulative nuclear energy release to better than $\lesssim 5\%$, with even higher accuracy for some models.

\subsection{$^{56}$Ni production and GRB supernovae}

\begin{figure}
\centering
\includegraphics[width=0.99\linewidth]{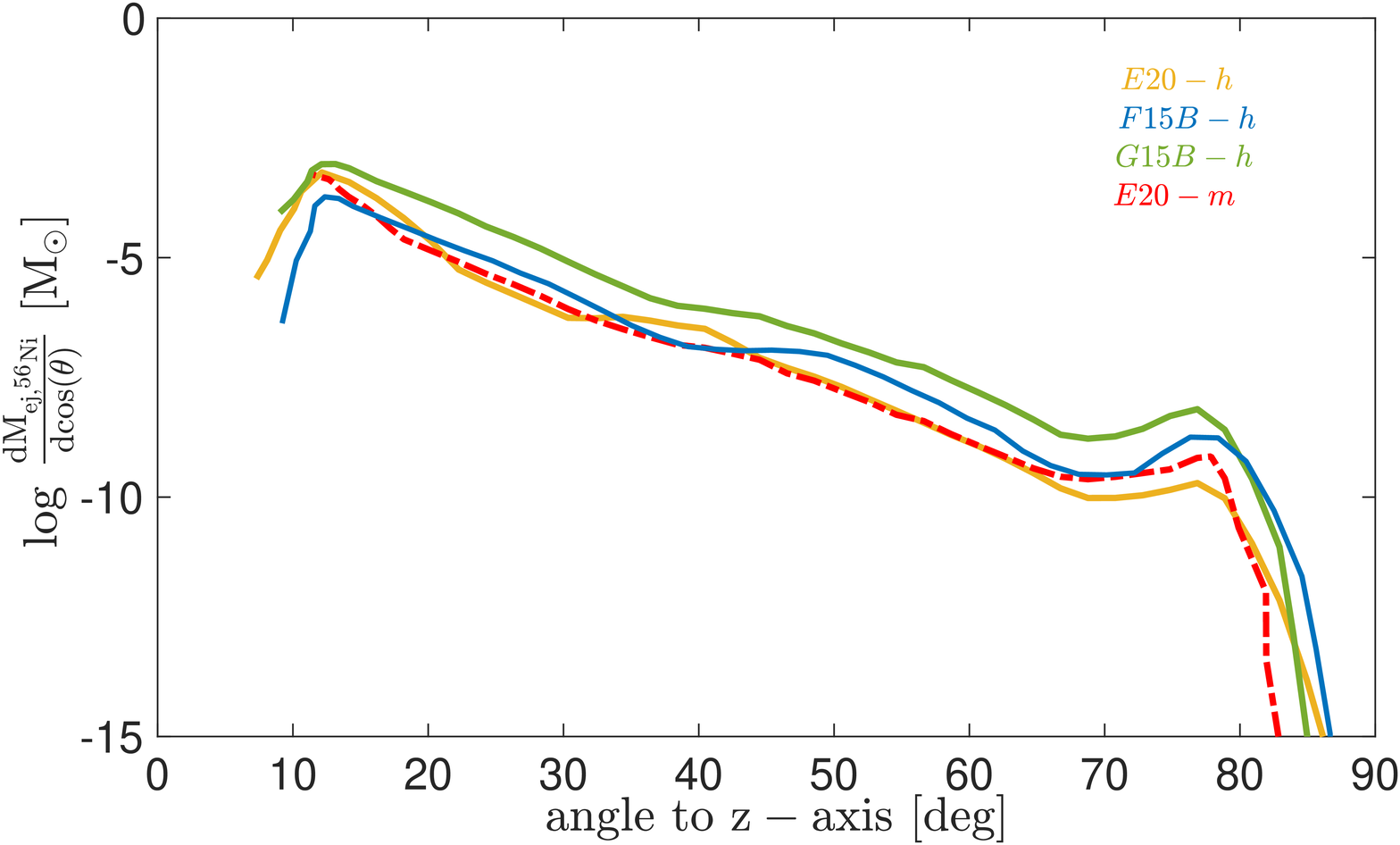}
\includegraphics[width=0.95\linewidth]{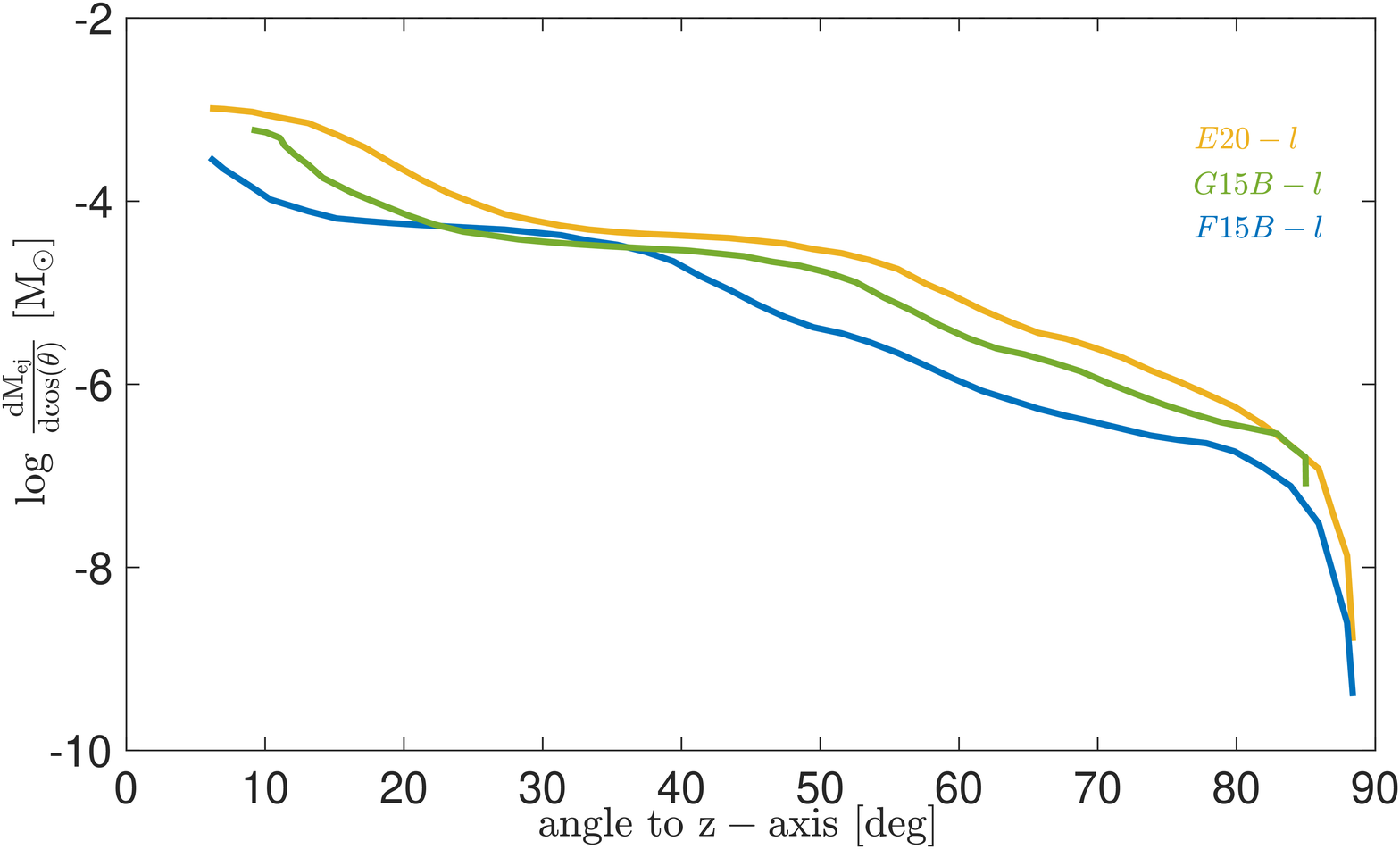}
\caption{Angular distributions of unbound disk outflow mass generated during the time window simulated here. Top: Ejected mass of $^{56}$Ni as a function of polar angle for various detonating models. Most of the unbound $^{56}$Ni is confined to within $\pm 30^{\circ}$ from the polar axis. Bottom: ejected mass including all elements for the non-detonating models as function of polar angle. Unbound matter is again preferentially directed into polar regions.}
\label{fig:Ni56polar}
\end{figure}

An extrapolation of our medium- and high- angular momentum models to late times predict the ejection of $\sim 0.1M_{\odot}$ of $^{56}$Ni over a timescale of several minutes to hours.  Although this amount falls short of the total $\approx 0.3-0.5M_{\odot}$ Nickel production required to explain the optical light curves of GRB supernovae (e.g.~\citealt{Drout+11,Cano+16}), it suggests that late-time disk outflows could be an appreciable contributor in addition to prompt jet-powered explosions from the inner regions close to the black hole (which in the latter case must take place on a very short timescale $\lesssim 1$ s; e.g.~\citealt{Barnes+18}).  This could reduce the tensions placed on black hole models for GRB central engines, which require the initial supernova explosion to ``fail'' in order to form the black hole in the first place.   

Via spectral modeling of GRB-SNe, \citet{Ashall&Mazzali20} find evidence for decreasing $^{56}$Ni abundance as a function of velocity in the ejecta, which they argue is consistent with the $^{56}$Ni being generated at the edges of a bipolar jet, or dredged up to the ejecta surface.  We indeed find that the disk outflows concentrate their kinetic energy in a jet-like structure within $\lesssim 30^{\circ}$ of the angular momentum axis (Fig.~\ref{fig:Ni56polar}).  The kinetic energy of the unbound matter in our `h'- and `m'- simulations generated during the simulated time frame is already much larger than the gravitational binding energy of the outer layers of the star (which we do not model here), but only $30-90\%$ for the `l'-models (cf.~Tab.~\ref{tab:torusenergytab}).  However the latter fractions would also approach or greatly exceed unity if we were to include disk outflows ($M_{{\rm out},\infty}$) that occur over much longer timescales $\gg$ minutes than those we simulate.  Thus, it is likely that most of the $^{56}$Ni synthesized during the early evolution simulated here will ultimately escape to infinity, and plausible that some of this material will be mixed to high velocities in the supernova ejecta on a timescale much shorter than the light curve rise time of weeks.

\subsection{Revealing R-Process Material by Delayed $^{56}$Ni Ejecta}
\label{sec:GRB_SNe}

One timescales of $\lesssim$ minutes following stellar collapse, the very inner regions of collapsar accretion disks are neutron-rich and may eject up to $\gtrsim 0.3M_{\odot}$ material which forms heavy $r$-process elements (e.g.~\citealt{Siegel+19}, see also \citealt{Miller+19}).  If this material contains the heaviest elements (atomic mass number $A \gtrsim 140$) then the high opacity of the lanthanide elements relative to that of ordinary supernova ejecta (e.g.~\citealt{Barnes&Kasen13}) could make these inner $r$-process layers visible as an infrared excess at late times in GRB SNe \citep{Siegel+19}.  Unfortunately, the specific radioactive decay heating rate from $r$-process nuclei is typically an order of magnitude smaller than that of $^{56}$Ni on $\gtrsim$ week timescales when the outer layers of the supernova ejecta are becoming transparent and the inner layers are revealed.  This low ``intrinsic'' luminosity of $r$-process elements make their effect challenging to observe on top of the much higher luminosity of the ordinary supernova material.

Prospects for detecting $r$-process signatures would be improved if the $r$-process enriched layers reside outside a significant quantity of $^{56}$Ni, which can then ``back-light'' the $r$-process layers with a higher luminosity.  Such a ``Nickel inside $r$-process'' geometry may be challenging to obtain if the $^{56}$Ni is produced entirely by the early-time GRB jet (e.g.~\citealt{Barnes+18}) because that earliest material will reside outside the layers of the supernova ejecta polluted by subsequent disk winds.  On the other hand, the $^{56}$Ni which originates from the outer disk, as considered in this paper, will be delayed relative to $r$-process outflows from the inner disk and hence could provide the conditions necessary to make the $r$-process signatures more visible.  Future late-time infrared observations of GRB-SNe could enable a test of this hypothesis.  

\subsection{Implications for GRB Light Curves}  
\label{sec:GRB_lightcurves}

Material from the accretion disk which feeds the inner accretion flow provides a power-source for generating the GRB jet activity (e.g.~\citealt{MacFadyen&Woosley99}).  Indeed, our predicted accretion rates (Fig.~\ref{fig:Mdot_E20}) peak on a timescale of $\sim 1$ minutes, which is similar to the duration of peak gamma-ray emission in long-duration GRBs.  However, on timescales greater than the viscous time, we find a steep decline in the accretion rate $\dot{M} \propto t^{-\beta}$ with $\beta \sim 3.5$ (Fig.~\ref{fig:Mdot_E20}), mainly as a result of mass-loss due to powerful disk winds and convection inhibiting accretion.  

Intriguingly, the prompt gamma-ray/X-ray light curves of long GRBs often exhibit a ``steep decay phase'', whereby the flux at the end of the GRB prompt phase decays rapidly in time as $F \propto t^{-\beta}$ with $\beta \approx 3-5$ \citep{Nousek+06}.  Although many mechanisms have been proposed for the steep decay phase (e.g.~\citealt{Zhang+06} for a review), a cut-off in the feeding rate from the outer disk, due to a combination of disk winds and convection, offers a plausible explanation within our scenario.  \citet{Lindner+10} present a somewhat different physical mechanism for the rapid decay in accretion rate involving outwards propagation of the accretion shock through the progenitor envelope, while convection in the disk as a way to suppress the accretion rate has also been proposed previously \citep{Quataert&Gruzinov00,Milosavljevic+12,Lindner+12}.

\subsection{Comparison to Previous Works}
Although numerous studies have explored core-collapse supernovae, far fewer focused on the nucelosynthetic processes occurring in the inner accretion disk following the formation of a BH in stellar core. 

 Starting with \cite{Bodenheimer&Woosley83} and to these days (e.g. \cite{Kushnir&Katz15} and references therein) various studies suggested the importance on nucleosynthesis in significantly affecting the evolution of CC SNe.  \cite{Bodenheimer&Woosley83}  used 2D simulations to explore the early stages of the collapse of an evolved, massive, rotating  $M=25 M_{\odot}$ star, and include a simplified model for nucleosynthesis. Although followed only for a short time, the simulated evolution already suggested thermonucelar burning and detonation could play an important role in the CC evolution. 

 In their classic paper \cite{MacFadyen&Woosley99} provided a model for long GRBs produced through and  accretion disk around the central BH. They explored a larger region around the central remnant and followed the \emph{early} evolution of the accretion and fallback process, and in particular focused on the innermost region of the disk, rather than the cooler outer part of the disk established at later times, which we study. Their focus was on the GRB production model, and did not explore the detailed nucleosynthesis explored here. A focus on the early evolution of collapsar disks was also taken by \cite{Siegel+19} who studied the production of r-process elements in early outflows from such collapsar accretion disks during the GRB phase (see also \citealt{Miller+19}).

\cite{Lindner+10} ran a spherically-symmetric AMR hydrodynamic simulation of the collapse of a rotating $\rm  M=16\,M_{\odot}$ star. They followed the evolution up to a few hundred seconds, and found the evolution of the accretion rate is consistent with the light curve evolution of long GRBs, and  showed that a thick disk forms around the stellar black hole remnant.
\cite{Lindner+12} showed the kinetic energy exceeds $\sim 5\times10^{50}$\,erg, and  their estimates of the peak net outflow rate at a larger radius are comparable to our FLASH simulation.  Although using similar tools (FLASH) and following the evolution for hundreds of seconds, these simulations did not include nucleosynthesis processes, which are the focus of the current work.

Other studies followed the detailed nucleosynthesis in collapsar accretion disks, but those used highly simplified (and sometimes stationary) disk models, not accounting for the nucleosynthesis feedback on the disk evolution, nor self-consistently considering outflows from the disk \cite[e.g.][and references therein]{Banerjee2013,Tao2015}. Several of these models find considerable production of IME and IGE, and in some cases comparable to our findings (e.g. \cite{Tao2015} up to a factor of a few), however, these models and the regimes in these various works are considerably different than those we explore, and are thus not directly comparable.  
 
Our approach, while not following the early collapse and the formation of an accretion disk, allows us to follow the long-term hydrodynamical evolution of the disk, while incorporating a detailed nucleosynthetic network and self-consistently accounting for nuclear energy and potential thermonuclear detonations. Furthermore, it also allows us to characterize not only the nucleosynthetic products, but also their detailed kinematic evolution, energetic input and material outflow and ejection.

\subsection{Long-term evolution}
Our models explored the long-term evolution of collapsar accretion disks over a few hundred seconds. However, the viscous and dynamical timescales of the outermost regions of the stellar envelope are even longer, and can not be captured by our models. Moreover, our models suggest that in many cases no detonation occurs, and in others in which detonations occur the energy production is insufficient to eject all the material nor the outer envelope, outside the disk regions we studied. It is therefore possible for some material to fall back and potentially give rise to later accretion events and further nuclear burning and energy production in the disk, beyond the simulation time.
These aspects and their importance are, however,  beyond the scope of the current study. Their study requires a consistent modelling of the global large scale structure of the envelope and the longer-term evolution, which are still computationally prohibiting. 

\section{Conclusions}

We have presented simulations of the long-term (many viscous time) evolution of fall-back disks generated by the core collapse of rapidly rotating stars, accounting for the combined effects of viscous angular momentum transport and nuclear burning on the disk dynamics and ejecta.  Our conclusions may be summarized as follows:

\begin{itemize}
    \item We explore disk formation for different stellar progenitor models with a range of angular momentum profiles, the latter put in by hand to account for uncertainty in angular momentum transport in massive stars.  These differences result in 3 classes of models---`l', `m', and `h'---characterized by different amounts (`low', `medium', and `high') of surface angular momentum. We are reasonably justified in evolving a torus of fixed mass because the viscous time of the torus at the circularization radius is typically longer than the mass fall-back time and the torus formation timescale (time for circularization of the outer torus layers) is smaller than the viscous time. Further justification arises from the fact that the kinetic energy of outflows from the disk exceeds the binding energy of the remaining outer layers of the stars.  
    
    \item In our l-models we find no detonations.  Instead, the disk evolves relatively quiescently, losing mass to accretion and outflows.  The total kinetic energy of the disk winds is only $\sim 1-4\times 10^{49}$\,erg (or roughly a factor of a few to ten higher when extrapolated beyond our simulated epoch) and only a tiny amount of iron-group elements are synthesized ($\sim 10^{-8}\,M_\odot$; cf.~Tab.~\ref{tab:torusenergytab}).  Therefore, we expect relatively little impact on the supernova explosion in the low surface angular momentum cases.
    
    \item By contrast, in our m- and h-models, we find strong detonations.  These detonations occur due to spontaneous ignition of a He-C-O mixture in the disk midplane due to high midplane densities and shallow temperature gradients established by vigorous viscous turbulence (see Figs.~\ref{fig:FLASHE20_h}, \ref{fig:sections_z_E20h}, \ref{fig:E20h_csign}).  The occurrence of a detonation is found to be robust to the assumed viscosity of the accretion disk for $\alpha \gtrsim 10^{-3}-10^{-2}$ (cf.~Sec.~\ref{sec:viscosity}).   
    
    \item Based on and verified by our simulation results, we present a detonation criterion and semi-analytical model to predict whether a given stellar progenitor model with given rotation profile leads to an accretion disk that detonates, only weakly detonates, or does not detonate at all (cf.~Sec.~\ref{sec:survey_models}, Fig.~\ref{fig:All_detonation}). This model enables us to scan the configuration space of stellar progenitor models and rotation profiles for detonation behaviour without performing costly multi-dimensional hydrodynamic simulations.
    
    \item Our detonating models result in the production of a significant quantity of iron-group elements, of which we expect $0.05-0.4 M_{\odot}$ will be ejected over a timescale of $\sim$ minutes to hours, including up to $\sim 0.1M_{\odot}$ in $^{56}$Ni (cf.~Tabs.~\ref{tab:torusenergytab} and \ref{tab:X_models}).  The unbound winds carry $\sim 10^{49}-6\times10^{50}$ \,erg of energy (likely a factor of a few higher when extrapolated beyond our simulated epoch) and are mostly concentrated within $\sim 30^{\circ}$ of the rotational axis (Figs.~\ref{fig:energytotal_E20h} and \ref{fig:Ni56polar}).  
    
    \item The delayed $^{56}$Ni produced by detonating collapsar disks could contribute significantly to the optical light of collapsar/GRB supernovae, augmenting that produced during the prompt explosion phase.  The delayed $^{56}$Ni ejection could also provide a radioactive energy source for back-lighting $r$-process elements produced earlier in the event from the inner accretion disk after BH formation, revealing their presence in late-time infrared observations of GRB supernovae.
    
    \item We find a steep decline in the accretion rate $\dot{M} \propto t^{-\beta}$ with $\beta \sim 3.5$ on timescales larger than the viscous timescale (Fig.~\ref{fig:Mdot_E20}), which offers a plausible explanation for the steep decay phase in flux $F \propto t^{-\beta}$, with $\beta\approx 3-5$, observed in many long GRB gamma-ray/X-ray light curves.
    
\end{itemize}

\bigskip{}

DMS acknowledges the support of the Natural Sciences and Engineering Research Council of Canada (NSERC). Research at Perimeter Institute is supported in part by the Government of Canada through the Department of Innovation, Science and Economic Development Canada and by the Province of Ontario through the Ministry of Colleges and Universities.  BDM acknowledges support by NASA through the Astrophysics Theory Program (NNX17AK43G), the National Science Foundation (AST-2002577), and from the Simons Foundation (606260). YZ and HBP acknowledge support for this project from the European Union's Horizon 2020 research and innovation program under grant agreement No 865932-ERC-SNeX.

\appendix

\section{Generation of collapsar disk models}
\label{app:tori_solutions}

In this appendix, we shall briefly outline the individual steps of our scheme to compute initial data for collapsar accretion disks. We distinguish between two usage cases:
\begin{itemize}
    \item[(1)] Computing initial data for simple one-parameter families of models that scan the progenitor parameter space (cf.~Sec.~\ref{sec:survey_models}).
    \item[(2)] Computing approximate initial data for simulation runs.
\end{itemize}

Our scheme proceeds as follows:

\begin{itemize}
	\item[$(i)$] Given a fixed (but arbitrary) stellar model for the progenitor star and modified angular momentum profile (Eq.~\eqref{eq:j_profile}) with given surface rotation rate $\Omega/\Omega_{\rm BU}$ and power-law index $p$, we obtain the inner and outer circularization radius of the disk and its compositional profile employing the fall-back model described in Sec.~\ref{sec:initial_models}:
	\begin{eqnarray}
		r_{\rm circ,in} &=& r_{\rm circ,in}(\Omega/\Omega_{\rm BU},p) \\
		r_{\rm circ,out} &=& r_{\rm circ,out}(\Omega/\Omega_{\rm BU},p) \\
		X_{i} &=& X_i(\bar{\rho}; \Omega/\Omega_{\rm BU},p),
	\end{eqnarray}
	where $X_{i}$ denote the mass fractions of various elements.
	
	\item[$(ii)$] Infalling material on zero-energy orbits requires that the total energy be zero (cf.~Eq.~\eqref{eq:etot})
	\begin{equation}
		e_{\rm tot}(r) = 0. \label{eq:e_tot_appendix}
	\end{equation}
	Constant specific angular momentum is assumed with a value corresponding to the average value between $r_{\rm circ,in}$ and $r_{\rm circ,out}$ predicted by the fall-back model in step $(i)$. Self-gravity of the torus is approximately included by specifying a point mass potential at the center of mass of the torus. For usage case (1), we treat this expression as a fourth-order equation in the independent variables temperature $T$ and density $\rho$, assuming an ideal gas plus radiation. For usage case (2), we employ the Helmholtz EOS and the compositional profiles $X_i(r)=X_i(\bar{\rho})$ determined in step $(i)$.

	\item[$(iii)$] At the location $R_0$ of maximum density $\rho_{\rm max}$ and temperature $T_{\rm max}$ (located in the midplane $z=0$), the pressure gradient must vanish:
	\begin{equation}
		\nabla p = 0 \mskip20mu \text{at} \mskip20mu \bar{\rho} = R_0, \label{eq:grad_press}
	\end{equation}
	where $p = p(\rho,T,X_i)$ is determined either using an ideal gas plus radiation or the Helmholtz EOS, depending on the usage case (1) or (2), respectively. 

    \item[$(iv)$] Conditions \eqref{eq:e_tot_appendix} and \eqref{eq:grad_press} can be solved for $T_{\rm max}$ and $\rho_{\rm max}$.

	\item[$(v)$] Following \citet{1999MNRAS.310.1002S} we normalize the torus density by $\rho_{\rm max}$ from step $(iv)$, resulting in the density distribution
	\begin{equation}
		\frac{\rho}{\rho_{\rm max}} = \left\{ \frac{2d}{d-1}\left[ \frac{R_0}{r}-\frac{1}{2}\left(\frac{R_0}{r\sin\theta}\right)^2 - \frac{1}{2d}\right]\right\}^{\frac{1}{\gamma -1}}. \label{eq:torus_rho_profile}
	\end{equation}
	Here, we set $\gamma=5/3$ in usage case (1) and compute an effective adiabatic index in case (2) using the Helmholtz EOS. Bounded torus configurations require $d>1$. The torus distortion parameter $d$ is related to the internal energy at $R_0$ by
	\begin{equation}
	    e_{\rm int,max} = \frac{GM_{\rm BH}}{R_0} \frac{1}{2\gamma}\frac{d-1}{d}. \label{eq:e_int_d}
	\end{equation}

	\item[$(vi)$] Imposing the requirement $d>1$ for physical solutions, one can obtain $d$
	from condition \eqref{eq:e_int_d} using $T_{\rm max}$ and $\rho_{\rm max}$ from step $(iv)$.

	\item[$(vii)$] With $d$ from step $(vi)$ Eq.~\eqref{eq:torus_rho_profile} yields the density profile.

	\item[$(viii)$] Given the density structure from $(vii)$, Eq.~\eqref{eq:e_tot_appendix} provides the temperature profile and thus all other required quantities.
\end{itemize}

For usage case (1), we employ this procedure to obtain a one-parameter family of solutions by varying the power law index $p$ of the stellar rotation law. Such one-parameter families of models for the stellar models E20-h, E20-m, and E20-l are shown in Fig.~\ref{fig:All_detonation}. For usage case (2), this procedure provides approximate initial data, which once initialized on the grid are evolved and relaxed to an equilibrium state that wipes out initial inconsistencies and serves as the actual initial data for simulation runs (see Secs.~\ref{sec:initial_models} and \ref{sec:results} for more details).

%\newpage

\bibliographystyle{mnras}
%\bibliography{Sne2017edited}

\begin{thebibliography}{}
\makeatletter
\relax
\def\mn@urlcharsother{\let\do\@makeother \do\$\do\&\do\#\do\^\do\_\do\%\do\~}
\def\mn@doi{\begingroup\mn@urlcharsother \@ifnextchar [ {\mn@doi@}
  {\mn@doi@[]}}
\def\mn@doi@[#1]#2{\def\@tempa{#1}\ifx\@tempa\@empty \href
  {http://dx.doi.org/#2} {doi:#2}\else \href {http://dx.doi.org/#2} {#1}\fi
  \endgroup}
\def\mn@eprint#1#2{\mn@eprint@#1:#2::\@nil}
\def\mn@eprint@arXiv#1{\href {http://arxiv.org/abs/#1} {{\tt arXiv:#1}}}
\def\mn@eprint@dblp#1{\href {http://dblp.uni-trier.de/rec/bibtex/#1.xml}
  {dblp:#1}}
\def\mn@eprint@#1:#2:#3:#4\@nil{\def\@tempa {#1}\def\@tempb {#2}\def\@tempc
  {#3}\ifx \@tempc \@empty \let \@tempc \@tempb \let \@tempb \@tempa \fi \ifx
  \@tempb \@empty \def\@tempb {arXiv}\fi \@ifundefined
  {mn@eprint@\@tempb}{\@tempb:\@tempc}{\expandafter \expandafter \csname
  mn@eprint@\@tempb\endcsname \expandafter{\@tempc}}}

\bibitem[\protect\citeauthoryear{{Anderson}}{{Anderson}}{2019}]{Anderson19}
{Anderson} J.~P.,  2019, \mn@doi [\aap] {10.1051/0004-6361/201935027}, \href
  {https://ui.adsabs.harvard.edu/abs/2019A&A...628A...7A} {628, A7}

\bibitem[\protect\citeauthoryear{{Ashall} \& {Mazzali}}{{Ashall} \&
  {Mazzali}}{2020}]{Ashall&Mazzali20}
{Ashall} C.,  {Mazzali} P.,  2020, arXiv e-prints, \href
  {https://ui.adsabs.harvard.edu/abs/2020arXiv200107665A} {p. arXiv:2001.07665}

\bibitem[\protect\citeauthoryear{{Banerjee} \& {Mukhopadhyay}}{{Banerjee} \&
  {Mukhopadhyay}}{2013}]{Banerjee2013}
{Banerjee} I.,  {Mukhopadhyay} B.,  2013, \mn@doi [\apj]
  {10.1088/0004-637X/778/1/8}, \href
  {https://ui.adsabs.harvard.edu/abs/2013ApJ...778....8B} {778, 8}

\bibitem[\protect\citeauthoryear{{Barnes} \& {Kasen}}{{Barnes} \&
  {Kasen}}{2013}]{Barnes&Kasen13}
{Barnes} J.,  {Kasen} D.,  2013, \mn@doi [\apj] {10.1088/0004-637X/775/1/18},
  \href {https://ui.adsabs.harvard.edu/abs/2013ApJ...775...18B} {775, 18}

\bibitem[\protect\citeauthoryear{{Barnes}, {Duffell}, {Liu}, {Modjaz},
  {Bianco}, {Kasen}  \& {MacFadyen}}{{Barnes} et~al.}{2018}]{Barnes+18}
{Barnes} J.,  {Duffell} P.~C.,  {Liu} Y.,  {Modjaz} M.,  {Bianco} F.~B.,
  {Kasen} D.,   {MacFadyen} A.~I.,  2018, \mn@doi [\apj]
  {10.3847/1538-4357/aabf84}, \href
  {https://ui.adsabs.harvard.edu/abs/2018ApJ...860...38B} {860, 38}

\bibitem[\protect\citeauthoryear{{Beloborodov}}{{Beloborodov}}{2003}]{Beloborodov03}
{Beloborodov} A.~M.,  2003, \mn@doi [\apj] {10.1086/374217}, \href
  {https://ui.adsabs.harvard.edu/abs/2003ApJ...588..931B} {588, 931}

\bibitem[\protect\citeauthoryear{{Bodenheimer} \& {Woosley}}{{Bodenheimer} \&
  {Woosley}}{1983}]{Bodenheimer&Woosley83}
{Bodenheimer} P.,  {Woosley} S.~E.,  1983, \mn@doi [\apj] {10.1086/161040},
  \href {https://ui.adsabs.harvard.edu/abs/1983ApJ...269..281B} {269, 281}

\bibitem[\protect\citeauthoryear{{Cannizzo}, {Troja}  \& {Gehrels}}{{Cannizzo}
  et~al.}{2011}]{Cannizzo+11}
{Cannizzo} J.~K.,  {Troja} E.,   {Gehrels} N.,  2011, \mn@doi [\apj]
  {10.1088/0004-637X/734/1/35}, \href
  {https://ui.adsabs.harvard.edu/abs/2011ApJ...734...35C} {734, 35}

\bibitem[\protect\citeauthoryear{{Cano}, {Johansson Andreas}  \&
  {Maeda}}{{Cano} et~al.}{2016a}]{Cano16}
{Cano} Z.,  {Johansson Andreas} K.~G.,   {Maeda} K.,  2016a, \mn@doi [\mnras]
  {10.1093/mnras/stw122}, \href
  {https://ui.adsabs.harvard.edu/abs/2016MNRAS.457.2761C} {457, 2761}

\bibitem[\protect\citeauthoryear{{Cano}, {Johansson Andreas}  \&
  {Maeda}}{{Cano} et~al.}{2016b}]{Cano+16}
{Cano} Z.,  {Johansson Andreas} K.~G.,   {Maeda} K.,  2016b, \mn@doi [\mnras]
  {10.1093/mnras/stw122}, \href
  {https://ui.adsabs.harvard.edu/abs/2016MNRAS.457.2761C} {457, 2761}

\bibitem[\protect\citeauthoryear{{Cantiello}, {Yoon}, {Langer}  \&
  {Livio}}{{Cantiello} et~al.}{2007}]{Cantiello+07}
{Cantiello} M.,  {Yoon} S.~C.,  {Langer} N.,   {Livio} M.,  2007, \mn@doi
  [\aap] {10.1051/0004-6361:20077115}, \href
  {https://ui.adsabs.harvard.edu/abs/2007A&A...465L..29C} {465, L29}

\bibitem[\protect\citeauthoryear{{Chen} \& {Beloborodov}}{{Chen} \&
  {Beloborodov}}{2007}]{Chen&Beloborodov07}
{Chen} W.-X.,  {Beloborodov} A.~M.,  2007, \mn@doi [\apj] {10.1086/508923},
  \href {https://ui.adsabs.harvard.edu/abs/2007ApJ...657..383C} {657, 383}

\bibitem[\protect\citeauthoryear{{Chevalier}}{{Chevalier}}{1989}]{1989ApJ...346..847C}
{Chevalier} R.~A.,  1989, \mn@doi [\apj] {10.1086/168066}, \href
  {http://adsabs.harvard.edu/abs/1989ApJ...346..847C} {346, 847}

\bibitem[\protect\citeauthoryear{{Dessart}, {Burrows}, {Livne}  \&
  {Ott}}{{Dessart} et~al.}{2008}]{Dessart+08}
{Dessart} L.,  {Burrows} A.,  {Livne} E.,   {Ott} C.~D.,  2008, \mn@doi [\apjl]
  {10.1086/527519}, \href
  {https://ui.adsabs.harvard.edu/abs/2008ApJ...673L..43D} {673, L43}

\bibitem[\protect\citeauthoryear{{Drout} et~al.,}{{Drout}
  et~al.}{2011}]{Drout+11}
{Drout} M.~R.,  et~al., 2011, \mn@doi [\apj] {10.1088/0004-637X/741/2/97},
  \href {https://ui.adsabs.harvard.edu/abs/2011ApJ...741...97D} {741, 97}

\bibitem[\protect\citeauthoryear{{Ertl}, {Woosley}, {Sukhbold}  \&
  {Janka}}{{Ertl} et~al.}{2019}]{Ertl+19}
{Ertl} T.,  {Woosley} S.~E.,  {Sukhbold} T.,   {Janka} H.~T.,  2019, arXiv
  e-prints, \href {https://ui.adsabs.harvard.edu/abs/2019arXiv191001641E} {p.
  arXiv:1910.01641}

\bibitem[\protect\citeauthoryear{{Fern{\'a}ndez} \& {Metzger}}{{Fern{\'a}ndez}
  \& {Metzger}}{2013}]{2013ApJ...763..108F}
{Fern{\'a}ndez} R.,  {Metzger} B.~D.,  2013, \mn@doi [\apj]
  {10.1088/0004-637X/763/2/108}, \href
  {http://adsabs.harvard.edu/abs/2013ApJ...763..108F} {763, 108}

\bibitem[\protect\citeauthoryear{{Fryxell}, {Arnett}  \&
  {M{\"u}ller}}{{Fryxell} et~al.}{1989}]{1989BAAS...21.1209F}
{Fryxell} B.~A.,  {Arnett} W.~D.,   {M{\"u}ller} E.,  1989, in Bulletin of the
  American Astronomical Society. p.~1209

\bibitem[\protect\citeauthoryear{{Fryxell} et~al.,}{{Fryxell}
  et~al.}{2000}]{2000ApJS..131..273F}
{Fryxell} B.,  et~al., 2000, \mn@doi [\apjs] {10.1086/317361}, \href
  {http://adsabs.harvard.edu/abs/2000ApJS..131..273F} {131, 273}

\bibitem[\protect\citeauthoryear{{Gamezo}, {Wheeler}, {Khokhlov}  \&
  {Oran}}{{Gamezo} et~al.}{1999}]{Gamezo99}
{Gamezo} V.~N.,  {Wheeler} J.~C.,  {Khokhlov} A.~M.,   {Oran} E.~S.,  1999,
  \mn@doi [\apj] {10.1086/306784}, \href
  {https://ui.adsabs.harvard.edu/abs/1999ApJ...512..827G} {512, 827}

\bibitem[\protect\citeauthoryear{{Gilkis}, {Soker}  \& {Kashi}}{{Gilkis}
  et~al.}{2019}]{Gilkis+19}
{Gilkis} A.,  {Soker} N.,   {Kashi} A.,  2019, \mn@doi [\mnras]
  {10.1093/mnras/sty3008}, \href
  {https://ui.adsabs.harvard.edu/abs/2019MNRAS.482.4233G} {482, 4233}

\bibitem[\protect\citeauthoryear{{Heger}, {Langer}  \& {Woosley}}{{Heger}
  et~al.}{2000}]{Heger+00}
{Heger} A.,  {Langer} N.,   {Woosley} S.~E.,  2000, \mn@doi [\apj]
  {10.1086/308158}, \href
  {https://ui.adsabs.harvard.edu/abs/2000ApJ...528..368H} {528, 368}

\bibitem[\protect\citeauthoryear{{Houck} \& {Chevalier}}{{Houck} \&
  {Chevalier}}{1991}]{1991ApJ...376..234H}
{Houck} J.~C.,  {Chevalier} R.~A.,  1991, \mn@doi [\apj] {10.1086/170272},
  \href {http://adsabs.harvard.edu/abs/1991ApJ...376..234H} {376, 234}

\bibitem[\protect\citeauthoryear{{Hu}}{{Hu}}{2015}]{Tao2015}
{Hu} T.,  2015, \mn@doi [\aap] {10.1051/0004-6361/201425167}, \href
  {https://ui.adsabs.harvard.edu/abs/2015A&A...578A.132H} {578, A132}

\bibitem[\protect\citeauthoryear{{Kasen} \& {Bildsten}}{{Kasen} \&
  {Bildsten}}{2010}]{Kasen&Bildsten10}
{Kasen} D.,  {Bildsten} L.,  2010, \mn@doi [\apj]
  {10.1088/0004-637X/717/1/245}, \href
  {https://ui.adsabs.harvard.edu/abs/2010ApJ...717..245K} {717, 245}

\bibitem[\protect\citeauthoryear{{Kohri}, {Narayan}  \& {Piran}}{{Kohri}
  et~al.}{2005}]{Kohri+05}
{Kohri} K.,  {Narayan} R.,   {Piran} T.,  2005, \mn@doi [\apj]
  {10.1086/431354}, \href
  {https://ui.adsabs.harvard.edu/abs/2005ApJ...629..341K} {629, 341}

\bibitem[\protect\citeauthoryear{{Kumar}, {Narayan}  \& {Johnson}}{{Kumar}
  et~al.}{2008}]{Kumar+08}
{Kumar} P.,  {Narayan} R.,   {Johnson} J.~L.,  2008, \mn@doi [\mnras]
  {10.1111/j.1365-2966.2008.13493.x}, \href
  {https://ui.adsabs.harvard.edu/abs/2008MNRAS.388.1729K} {388, 1729}

\bibitem[\protect\citeauthoryear{{Kushnir}}{{Kushnir}}{2015}]{Kushnir&Katz15}
{Kushnir} D.,  2015, arXiv e-prints, \href
  {https://ui.adsabs.harvard.edu/abs/2015arXiv150203111K} {p. arXiv:1502.03111}

\bibitem[\protect\citeauthoryear{{Kushnir}, {Katz}, {Dong}, {Livne}  \&
  {Fern{\'a}ndez}}{{Kushnir} et~al.}{2013}]{2013ApJ...778L..37K}
{Kushnir} D.,  {Katz} B.,  {Dong} S.,  {Livne} E.,   {Fern{\'a}ndez} R.,  2013,
  \mn@doi [\apjl] {10.1088/2041-8205/778/2/L37}, \href
  {http://adsabs.harvard.edu/abs/2013ApJ...778L..37K} {778, L37}

\bibitem[\protect\citeauthoryear{Landau \& Lifshitz}{Landau \&
  Lifshitz}{1987}]{Landau1987fluid}
Landau L.~D.,  Lifshitz E.~M.,  1987, Fluid Mechanics: Volume 6, 2nd Edition.
Pergamon Press

\bibitem[\protect\citeauthoryear{{Lindner}, {Milosavljevi{\'c}}, {Couch}  \&
  {Kumar}}{{Lindner} et~al.}{2010}]{Lindner+10}
{Lindner} C.~C.,  {Milosavljevi{\'c}} M.,  {Couch} S.~M.,   {Kumar} P.,  2010,
  \mn@doi [\apj] {10.1088/0004-637X/713/2/800}, \href
  {https://ui.adsabs.harvard.edu/abs/2010ApJ...713..800L} {713, 800}

\bibitem[\protect\citeauthoryear{{Lindner}, {Milosavljevi{\'c}}, {Shen}  \&
  {Kumar}}{{Lindner} et~al.}{2012}]{Lindner+12}
{Lindner} C.~C.,  {Milosavljevi{\'c}} M.,  {Shen} R.,   {Kumar} P.,  2012,
  \mn@doi [\apj] {10.1088/0004-637X/750/2/163}, \href
  {https://ui.adsabs.harvard.edu/abs/2012ApJ...750..163L} {750, 163}

\bibitem[\protect\citeauthoryear{{Ma} \& {Fuller}}{{Ma} \&
  {Fuller}}{2019}]{Ma&Fuller19}
{Ma} L.,  {Fuller} J.,  2019, \mn@doi [\mnras] {10.1093/mnras/stz2009}, \href
  {https://ui.adsabs.harvard.edu/abs/2019MNRAS.488.4338M} {488, 4338}

\bibitem[\protect\citeauthoryear{{MacFadyen} \& {Woosley}}{{MacFadyen} \&
  {Woosley}}{1999}]{MacFadyen&Woosley99}
{MacFadyen} A.~I.,  {Woosley} S.~E.,  1999, \mn@doi [\apj] {10.1086/307790},
  \href {https://ui.adsabs.harvard.edu/abs/1999ApJ...524..262M} {524, 262}

\bibitem[\protect\citeauthoryear{{Maeda} et~al.,}{{Maeda}
  et~al.}{2007}]{Maeda+07}
{Maeda} K.,  et~al., 2007, \mn@doi [\apj] {10.1086/520054}, \href
  {https://ui.adsabs.harvard.edu/abs/2007ApJ...666.1069M} {666, 1069}

\bibitem[\protect\citeauthoryear{{Maeder}}{{Maeder}}{1987}]{Maeder87}
{Maeder} A.,  1987, \aap, \href
  {https://ui.adsabs.harvard.edu/abs/1987A&A...178..159M} {178, 159}

\bibitem[\protect\citeauthoryear{{Meakin}, {Seitenzahl}, {Townsley}, {Jordan},
  {Truran}  \& {Lamb}}{{Meakin} et~al.}{2009}]{Mea+09}
{Meakin} C.~A.,  {Seitenzahl} I.,  {Townsley} D.,  {Jordan} IV G.~C.,  {Truran}
  J.,   {Lamb} D.,  2009, \mn@doi [\apj] {10.1088/0004-637X/693/2/1188}, \href
  {http://adsabs.harvard.edu/abs/2009ApJ...693.1188M} {693, 1188}

\bibitem[\protect\citeauthoryear{{Metzger}}{{Metzger}}{2012}]{Metzger12}
{Metzger} B.~D.,  2012, \mn@doi [\mnras] {10.1111/j.1365-2966.2011.19747.x},
  \href {https://ui.adsabs.harvard.edu/abs/2012MNRAS.419..827M} {419, 827}

\bibitem[\protect\citeauthoryear{{Metzger}, {Piro}  \& {Quataert}}{{Metzger}
  et~al.}{2008}]{Metzger+08}
{Metzger} B.~D.,  {Piro} A.~L.,   {Quataert} E.,  2008, \mn@doi [\mnras]
  {10.1111/j.1365-2966.2008.13789.x}, \href
  {https://ui.adsabs.harvard.edu/abs/2008MNRAS.390..781M} {390, 781}

\bibitem[\protect\citeauthoryear{{Metzger}, {Giannios}, {Thompson},
  {Bucciantini}  \& {Quataert}}{{Metzger} et~al.}{2011}]{Metzger+11}
{Metzger} B.~D.,  {Giannios} D.,  {Thompson} T.~A.,  {Bucciantini} N.,
  {Quataert} E.,  2011, \mn@doi [\mnras] {10.1111/j.1365-2966.2011.18280.x},
  \href {https://ui.adsabs.harvard.edu/abs/2011MNRAS.413.2031M} {413, 2031}

\bibitem[\protect\citeauthoryear{{Miller}, {Sprouse}, {Fryer}, {Ryan},
  {Dolence}, {Mumpower}  \& {Surman}}{{Miller} et~al.}{2019}]{Miller+19}
{Miller} J.~M.,  {Sprouse} T.~M.,  {Fryer} C.~L.,  {Ryan} B.~R.,  {Dolence}
  J.~C.,  {Mumpower} M.~R.,   {Surman} R.,  2019, arXiv e-prints, \href
  {https://ui.adsabs.harvard.edu/abs/2019arXiv191203378M} {p. arXiv:1912.03378}

\bibitem[\protect\citeauthoryear{{Milosavljevi{\'c}}, {Lindner}, {Shen}  \&
  {Kumar}}{{Milosavljevi{\'c}} et~al.}{2012}]{Milosavljevic+12}
{Milosavljevi{\'c}} M.,  {Lindner} C.~C.,  {Shen} R.,   {Kumar} P.,  2012,
  \mn@doi [\apj] {10.1088/0004-637X/744/2/103}, \href
  {https://ui.adsabs.harvard.edu/abs/2012ApJ...744..103M} {744, 103}

\bibitem[\protect\citeauthoryear{{Modjaz} et~al.,}{{Modjaz}
  et~al.}{2019}]{Modjaz+19}
{Modjaz} M.,  et~al., 2019, arXiv e-prints, \href
  {https://ui.adsabs.harvard.edu/abs/2019arXiv190100872M} {p. arXiv:1901.00872}

\bibitem[\protect\citeauthoryear{{M{\"o}sta} et~al.,}{{M{\"o}sta}
  et~al.}{2014}]{Mosta+14}
{M{\"o}sta} P.,  et~al., 2014, \mn@doi [\apjl] {10.1088/2041-8205/785/2/L29},
  \href {https://ui.adsabs.harvard.edu/abs/2014ApJ...785L..29M} {785, L29}

\bibitem[\protect\citeauthoryear{{Nagataki}}{{Nagataki}}{2018}]{Nagataki+18}
{Nagataki} S.,  2018, \mn@doi [Reports on Progress in Physics]
  {10.1088/1361-6633/aa97a8}, \href
  {https://ui.adsabs.harvard.edu/abs/2018RPPh...81b6901N} {81, 026901}

\bibitem[\protect\citeauthoryear{{Narayan}, {Piran}  \& {Kumar}}{{Narayan}
  et~al.}{2001}]{Narayan+01}
{Narayan} R.,  {Piran} T.,   {Kumar} P.,  2001, \mn@doi [\apj]
  {10.1086/322267}, \href
  {https://ui.adsabs.harvard.edu/abs/2001ApJ...557..949N} {557, 949}

\bibitem[\protect\citeauthoryear{{Niemeyer} \& {Woosley}}{{Niemeyer} \&
  {Woosley}}{1997}]{1997ApJ...475..740N}
{Niemeyer} J.~C.,  {Woosley} S.~E.,  1997, \mn@doi [\apj] {10.1086/303544},
  \href {http://adsabs.harvard.edu/abs/1997ApJ...475..740N} {475, 740}

\bibitem[\protect\citeauthoryear{{Nousek} et~al.,}{{Nousek}
  et~al.}{2006}]{Nousek+06}
{Nousek} J.~A.,  et~al., 2006, \mn@doi [\apj] {10.1086/500724}, \href
  {https://ui.adsabs.harvard.edu/abs/2006ApJ...642..389N} {642, 389}

\bibitem[\protect\citeauthoryear{Papaloizou \& Pringle}{Papaloizou \&
  Pringle}{1984}]{PapaloizouPringle}
Papaloizou J. C.~B.,  Pringle J.~E.,  1984, \mn@doi [Monthly Notices of the
  Royal Astronomical Society] {10.1093/mnras/208.4.721}, 208, 721

\bibitem[\protect\citeauthoryear{{Perets}, {Li}, {Lombardi}  \&
  {Milcarek}}{{Perets} et~al.}{2016}]{Per+16}
{Perets} H.~B.,  {Li} Z.,  {Lombardi} James~C. J.,   {Milcarek} Stephen~R. J.,
  2016, \mn@doi [\apj] {10.3847/0004-637X/823/2/113}, \href
  {https://ui.adsabs.harvard.edu/abs/2016ApJ...823..113P} {823, 113}

\bibitem[\protect\citeauthoryear{{Quataert} \& {Gruzinov}}{{Quataert} \&
  {Gruzinov}}{2000}]{Quataert&Gruzinov00}
{Quataert} E.,  {Gruzinov} A.,  2000, \mn@doi [\apj] {10.1086/309267}, \href
  {https://ui.adsabs.harvard.edu/abs/2000ApJ...539..809Q} {539, 809}

\bibitem[\protect\citeauthoryear{{Sekiguchi} \& {Shibata}}{{Sekiguchi} \&
  {Shibata}}{2011}]{Sekiguchi&Shibata11}
{Sekiguchi} Y.,  {Shibata} M.,  2011, \mn@doi [\apj]
  {10.1088/0004-637X/737/1/6}, \href
  {https://ui.adsabs.harvard.edu/abs/2011ApJ...737....6S} {737, 6}

\bibitem[\protect\citeauthoryear{{Shakura} \& {Sunyaev}}{{Shakura} \&
  {Sunyaev}}{1973}]{Shakura&Sunyaev73}
{Shakura} N.~I.,  {Sunyaev} R.~A.,  1973, \aap, \href
  {https://ui.adsabs.harvard.edu/abs/1973A&A....24..337S} {500, 33}

\bibitem[\protect\citeauthoryear{{Siegel}, {Barnes}  \& {Metzger}}{{Siegel}
  et~al.}{2019}]{Siegel+19}
{Siegel} D.~M.,  {Barnes} J.,   {Metzger} B.~D.,  2019, \mn@doi [\nat]
  {10.1038/s41586-019-1136-0}, \href
  {https://ui.adsabs.harvard.edu/abs/2019Natur.569..241S} {569, 241}

\bibitem[\protect\citeauthoryear{{Stone}, {Pringle}  \& {Begelman}}{{Stone}
  et~al.}{1999}]{1999MNRAS.310.1002S}
{Stone} J.~M.,  {Pringle} J.~E.,   {Begelman} M.~C.,  1999, \mn@doi [\mnras]
  {10.1046/j.1365-8711.1999.03024.x}, \href
  {http://adsabs.harvard.edu/abs/1999MNRAS.310.1002S} {310, 1002}

\bibitem[\protect\citeauthoryear{{Surman}, {McLaughlin}  \&
  {Sabbatino}}{{Surman} et~al.}{2011}]{Surman+11}
{Surman} R.,  {McLaughlin} G.~C.,   {Sabbatino} N.,  2011, \mn@doi [\apj]
  {10.1088/0004-637X/743/2/155}, \href
  {https://ui.adsabs.harvard.edu/abs/2011ApJ...743..155S} {743, 155}

\bibitem[\protect\citeauthoryear{{Takiwaki}, {Kotake}  \& {Suwa}}{{Takiwaki}
  et~al.}{2016}]{Takiwaki+16}
{Takiwaki} T.,  {Kotake} K.,   {Suwa} Y.,  2016, \mn@doi [\mnras]
  {10.1093/mnrasl/slw105}, \href
  {https://ui.adsabs.harvard.edu/abs/2016MNRAS.461L.112T} {461, L112}

\bibitem[\protect\citeauthoryear{{Timmes} \& {Swesty}}{{Timmes} \&
  {Swesty}}{2000}]{2000ApJS..126..501T}
{Timmes} F.~X.,  {Swesty} F.~D.,  2000, \mn@doi [\apjs] {10.1086/313304}, \href
  {http://adsabs.harvard.edu/abs/2000ApJS..126..501T} {126, 501}

\bibitem[\protect\citeauthoryear{{Woosley}}{{Woosley}}{1990}]{Woosley1990}
{Woosley} S.~E.,  1990, in Supernovae. pp 182--212

\bibitem[\protect\citeauthoryear{{Woosley}}{{Woosley}}{2010}]{Woosley10}
{Woosley} S.~E.,  2010, \mn@doi [\apjl] {10.1088/2041-8205/719/2/L204}, \href
  {https://ui.adsabs.harvard.edu/abs/2010ApJ...719L.204W} {719, L204}

\bibitem[\protect\citeauthoryear{{Woosley} \& {Bloom}}{{Woosley} \&
  {Bloom}}{2006}]{Woosley&Bloom06}
{Woosley} S.~E.,  {Bloom} J.~S.,  2006, \mn@doi [\araa]
  {10.1146/annurev.astro.43.072103.150558}, \href
  {https://ui.adsabs.harvard.edu/abs/2006ARA&A..44..507W} {44, 507}

\bibitem[\protect\citeauthoryear{{Woosley}, {Wunsch}  \& {Kuhlen}}{{Woosley}
  et~al.}{2004}]{Woosley2004}
{Woosley} S.~E.,  {Wunsch} S.,   {Kuhlen} M.,  2004, \mn@doi [\apj]
  {10.1086/383530}, \href
  {https://ui.adsabs.harvard.edu/abs/2004ApJ...607..921W} {607, 921}

\bibitem[\protect\citeauthoryear{Zel'dovich, Librovich, Makhviladze  \&
  Sivashinskil}{Zel'dovich et~al.}{1970}]{Zeldovich1970}
Zel'dovich Y.~B.,  Librovich V.~B.,  Makhviladze G.~M.,   Sivashinskil G.~I.,
  1970, \mn@doi [Journal of Applied Mechanics and Technical Physics]
  {10.1007/BF00908106}, 11, 264

\bibitem[\protect\citeauthoryear{{Zenati}, {Perets}  \& {Toonen}}{{Zenati}
  et~al.}{2019}]{Zen+19a}
{Zenati} Y.,  {Perets} H.~B.,   {Toonen} S.,  2019, \mn@doi [\mnras]
  {10.1093/mnras/stz316}, \href
  {https://ui.adsabs.harvard.edu/abs/2019MNRAS.486.1805Z} {486, 1805}

\bibitem[\protect\citeauthoryear{{Zenati}, {Bobrick}  \& {Perets}}{{Zenati}
  et~al.}{2020}]{Zen+19b}
{Zenati} Y.,  {Bobrick} A.,   {Perets} H.~B.,  2020, \mn@doi [\mnras]
  {10.1093/mnras/staa507}, \href
  {https://ui.adsabs.harvard.edu/abs/2020MNRAS.493.3956Z} {493, 3956}

\bibitem[\protect\citeauthoryear{{Zhang}, {Fan}, {Dyks}, {Kobayashi},
  {M{\'e}sz{\'a}ros}, {Burrows}, {Nousek}  \& {Gehrels}}{{Zhang}
  et~al.}{2006}]{Zhang+06}
{Zhang} B.,  {Fan} Y.~Z.,  {Dyks} J.,  {Kobayashi} S.,  {M{\'e}sz{\'a}ros} P.,
  {Burrows} D.~N.,  {Nousek} J.~A.,   {Gehrels} N.,  2006, \mn@doi [\apj]
  {10.1086/500723}, \href
  {https://ui.adsabs.harvard.edu/abs/2006ApJ...642..354Z} {642, 354}

\makeatother
\end{thebibliography}

%\bibliographystyle{aasjournal}

%%%%%%%%%%%%%%%%%%%%%%%%%%%%%%%%%%%%%%%%%%%%%%%%%%

%%%%%%%%%%%%%%%%% APPENDICES %%%%%%%%%%%%%%%%%%%%%%

% Don't change these lines
\bsp	% typesetting comment
\label{lastpage}
\end{document}